\documentclass[aps,prx,twocolumn,amsmath,amssymb,floatfix,superscriptaddress]{revtex4}
\usepackage{bbold,rotating}
\usepackage{comment}
\usepackage{physics}
\usepackage[dvipsnames]{xcolor}
\usepackage{amsmath,amsthm,amssymb,amsfonts,stmaryrd,wasysym,graphicx,multirow,color,textcomp}
\usepackage{hyperref}\usepackage{url}

\usepackage[utf8x]{inputenc}

\begin{document}

\title{Partial fillings of the bosonic \texorpdfstring{$E_8$}{E8} quantum Hall state}
\author{Pak Kau Lim} 
\author{Michael Mulligan}
\affiliation{Department of Physics and Astronomy, University of California, Riverside, California 92511, USA.}
\author{Jeffrey C. Y. Teo}
\affiliation{Department of Physics, University of Virginia, Charlottesville, Virginia 22904, USA.}

\begin{abstract}
We study bosonic topological phases constructed from electrons. In addition to a bulk excitation energy gap, these bosonic phases also have a fermion energy gap, below which all local excitations in the bulk and on the edge are even combinations of electrons. We focus on chiral phases, in which all low-energy edge excitations move in the same direction, that arise from the short-range entangled $E_8$ quantum Hall state, the bosonic analog of the filled lowest Landau level of electrons. The $E_8$ edge-state theory features an $E_8$ Kac-Moody symmetry that can be decomposed into ${\cal G}_A \times {\cal G}_B$ subalgebras, such as $SU(3) \times E_6$, $SO(M) \times SO(16-M)$, and $G_2 \times F_4$. (Here, $\{SO(M) \}$, $\{SU(N)\}$, and $\{E_8, G_2, F_4 \}$ denote orthogonal, unitary, and exceptional Lie algebras.) Using these symmetry decompositions, we construct exactly solvable coupled-wire model Hamiltonians for families of long-range entangled ${\cal G}_A$ or ${\cal G}_B$ bosonic fractional quantum Hall states that ``partially fill" the $E_8$ state and are pairwise related by a generalized particle-hole symmetry. These long-range entangled states feature either Abelian or non-Abelian topological order. Some support the emergence of non-local Dirac and Majorana fermions, Ising anyons, metaplectic anyons, Fibonacci anyons, as well as deconfined $\mathbb{Z}_2$ gauge fluxes and charges. 
\end{abstract}

\maketitle
\tableofcontents
\section{Introduction}

Of all the topological phases of matter in two spatial dimensions, the integer quantum Hall effect (IQHE) of electrons is particularly special \cite{prangegirvin1990, sarma2008perspectives}.
The IQHE was the first topological phase to be recognized experimentally; it furthermore provides a basis for understanding more general topological states, such as the fractional quantum Hall effect (FQHE).
The IQHE is observed when all single-electron states of an integer number of Landau levels are occupied; the fractional effect occurs when a Landau level is partially filled.
Due to the extensive degeneracy of a partially-filled Landau level, the FQHE ground state is determined by the underlying microscopic electron interactions (and disorder in realistic systems).
Empirically, the most ``attractive" states are found in the lowest Landau level at filling fractions $\nu = p/(2p+1)$ (with integer $p$) and the particle-hole conjugate filling fractions $1 - \nu$.
Both the IQHE and FQHE are stable against weak external perturbations.

To what extent is this picture unique to electron systems?
In particular, what are the analogous families of states composed out of fundamental bosons?
In this paper, we report progress on answering these questions.

To begin, we first need to understand what it means for a bosonic system to exhibit an integer quantum Hall effect.
The IQHE of electrons is characterized by the absence of any topological order, i.e., there are no anyonic quasiparticle excitations~\cite{Wilczekbook} and the topological entanglement entropy~\cite{KitaevPreskill06} vanishes.
Furthermore, the IQHE of electrons has robust gapless edge modes that do not require symmetry for their stabilization.
These properties can only be reproduced by bosonic states with $8k$ chiral edge modes ($k$ is a non-negative integer) \cite{kitaev11lecture, PhysRevB.86.125119}, the simplest example occurring when $k=1$---the so-called $E_8$ state.
We will therefore identify the $E_8$ state as the bosonic analog of the $\nu = 1$ IQHE of electrons.
(The primary distinction between the $E_8$ state and the bosonic QH state proposed in \cite{PhysRevLett.110.046801} is that the edge modes of the $E_8$ state do not require symmetry for their stability.)
As its name suggests, the $E_8$ state (reviewed in \S \ref{sec:E8review}) is based on the $E_8$ Lie algebra; for instance, the $K$-matrix in its Chern-Simons~\cite{WenZee92} field theory description is the $E_8$ Cartan matrix.

Unlike the IQHE of electrons, the $E_8$ state does not have a single-particle interpretation. The bosonic state is inherently interacting. For example, the primitive low-energy excitations, the $E_8$ bosons, are all even combinations of electrons. This makes the general description, let alone a specific construction, of a bosonic analog of the FQHE of electrons less clear.
We attack this problem following the earlier work of Lopes et al.~\cite{PhysRevB.100.085116}, where it was shown: (1) how the $E_8$ state can be constructed from interacting electrons in a coupled-wire model (an anisotropic 2d array of coupled Luttinger liquids that serves as the normal state for a large variety of distinct topological phases, e.g., \cite{PhysRevLett.88.036401, PhysRevB.89.085101}, among many others); and (2) within this construction, the $E_8$ state is the parent state of states with $G_2$ or $F_4$ Fibonacci topological order.
(Other constructions of a state with Fibonacci topological order are given in \cite{mongg2,PhysRevB.91.235112, PhysRevB.103.235118}.)
These and the other phases considered in this paper are bosonic in the sense that there is a fermion energy gap, which can be made arbitrarily large, below which all local excitations in the bulk and on the edge are even combinations of electrons.

Within this construction \cite{PhysRevB.100.085116}, the $G_2$ and $F_4$ symmetries arise from the Lie algebraic factorization $G_2 \times F_4 \subset E_8$.
The theory of conformal embeddings \cite{francesco2012conformal} allows this symmetry factorization to be realized within the $E_8$ edge-state theory.
This factorization allows the $E_8$ quasiparticle excitations (created, for instance, along any boundary by operators in the edge-state theory) to be decomposed into $G_2$ and $F_4$ Fibonacci components, in such a way that the product is a boson.
This is a generalization of spin-charge separation in the theory of a 1D spin-1/2 electron \cite{giamarchi2003quantum}.

The bulk-boundary correspondence \cite{elitzur1989remarks, wen1992theory} implies this edge-state symmetry factorization can be used to construct states with either $G_2$ or $F_4$ topological order as follows.
The $E_8$ state is equivalent to a collection of wires---each hosting nonchiral conformal field theories with $E_8$ symmetry---that are glued together via interactions that gap out nearby counter-propagating modes.
The $G_2 \times F_4 \subset E_8$ conformal embedding allows for an alternative set of interactions that gap out the nonchiral $E_8$ wires in such a way that at low energies only chiral $G_2$ or $F_4$ edge states remain.
The relative dominance of the different sets of interactions can be tuned by an external magnetic field (proportional to $\nu^{-1}$) transverse to the plane of the wires and/or short-ranged density-density interactions between modes on a finite number of nearby wires.
Importantly, the various gap-generating interactions are local, being constructed from products of the fundamental $E_8$ boson creation/annihilation operators.

Here, we show how this picture can be generalized to a large class of conformal embeddings ${\cal G}_A \times {\cal G}_B \subset E_8$ \cite{Bais:1986zs, PhysRevD.34.3092}: 
\begin{align}
\label{E8embeddings0}
\mathcal{G}_A\times\mathcal{G}_B =
    \begin{cases}
    & SU(3)\times E_6 \\
    & SU(2)\times E_7\\
    & SU(5)\times SU(5)\\
    & SO(M)\times SO(16-M) \\
    & G_2\times F_4\\
    & U(1)_8\times SU(8) \\
    & SU(2)_4\times Sp(8) 
    \end{cases}
\end{align}
for $M = 1, \ldots, 8$.
Above, $SO(1)$ refers to a state---described in detail in the main text---with Ising topological order. The levels of the Wess-Zumino-Witten (WZW)~\cite{WessZumino71,WittenWZW,witten1984} theories are $k=1$ unless specified otherwise.
By the mechanism outlined in the previous paragraph, we show how the above embeddings give rise to states with either ${\cal G}_A$ or ${\cal G}_B$ topological order.
We refer to the ${\cal G}_A$ or ${\cal G}_B$ topological states as bosonic fractional $E_8$ states. 

\begin{figure}[htbp]
\includegraphics[width=0.48\textwidth]{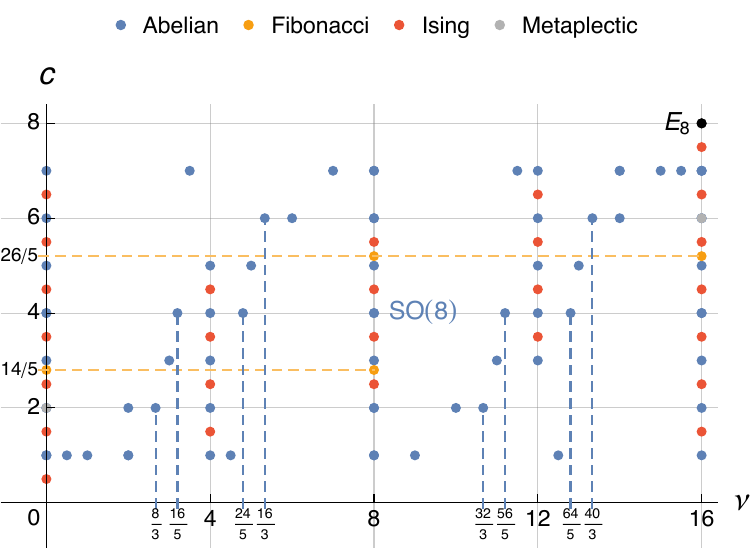}
\caption{Bosonic fractional quantum Hall (bFQH) states with filling number $\nu$ and chiral central charge $c$. 
Particle-hole conjugation flips $(\nu,c)\leftrightarrow(16-\nu,8-c)$. The particle-hole symmetric point lies at $(\nu,c)=(8,4)$. The metaplectic states and the Abelian states are overlapping at $(\nu,c)=(0,2), (16,6)$.}\label{fig:fillingcentralchargeplot}
\end{figure}

We catalog the possible bosonic fractional $E_8$ states in Fig.~\ref{fig:fillingcentralchargeplot} according to their filling number $\nu$ and chiral central charge $c$.
Recall that $\nu$ and $c$ determine the electrical ($\sigma_{xy}$) and thermal ($\kappa_{xy}$) \cite{PhysRevB.55.15832, 2002NuPhB.636..568C} Hall conductance via the relations:
\begin{align}
\label{electricthermalresponse}
\sigma_{xy} = \nu\frac{e^2}{h},\quad \kappa_{xy} = c\frac{\pi^2k_B^2}{3h}T,
\end{align}
where $T$ is the temperature.
In general, the chiral central charges $c_A$ and $c_B$ of the ${\cal G}_A$ and ${\cal G}_B$ states are not integers; this is the case for the $G_2$ and $F_4$ states, for instance. 
When this occurs, the ${\cal G}_A$ and ${\cal G}_B$ states have non-Abelian topological order.  

The filling numbers of the ${\cal G}_A$ and ${\cal G}_B$ states associated with the conformal embedding ${\cal G}_A \times {\cal G}_B \subset E_8$ satisfy $\nu_A + \nu_B = \nu_{E_8} = 16$.
Similarly, the chiral central charges $c_A + c_B = c_{E_8} = 8$.
These constraints are reminiscent to the relations between the filling numbers and chiral central charges of particle-hole conjugate states in the lowest Landau level, upon the replacements $16 \rightarrow 1$ and $8 \rightarrow 1$.
(The relative factor of 2 between $\nu_{E_8}$ and $c_{E_8}$ arises from the construction of the $E_8$ state from a collection of electron wires, in which the $E_8$ bosons are formed from electrons of unit charge.)
To emphasize this analogy, we refer to ${\cal G}_A$ and ${\cal G}_B$ topological states---associated with a particular embedding ${\cal G}_A \times {\cal G}_B \subset E_8$---as particle-hole conjugates.
Indeed, this identification can be made precise with the particle-hole symmetry operation given in \cite{PhysRevB.100.085116}.
This mapping---reviewed in Appendix \ref{PHconjugationappendix}---relates ${\cal G}_A$ and ${\cal G}_B$ degrees of freedom in such a way that particle-hole conjugate states are obtained from Hamiltonians that are conjugate with respect to the particle-hole symmetry operation.

\begin{figure}[htbp]
\centering\includegraphics[width=0.45\textwidth]{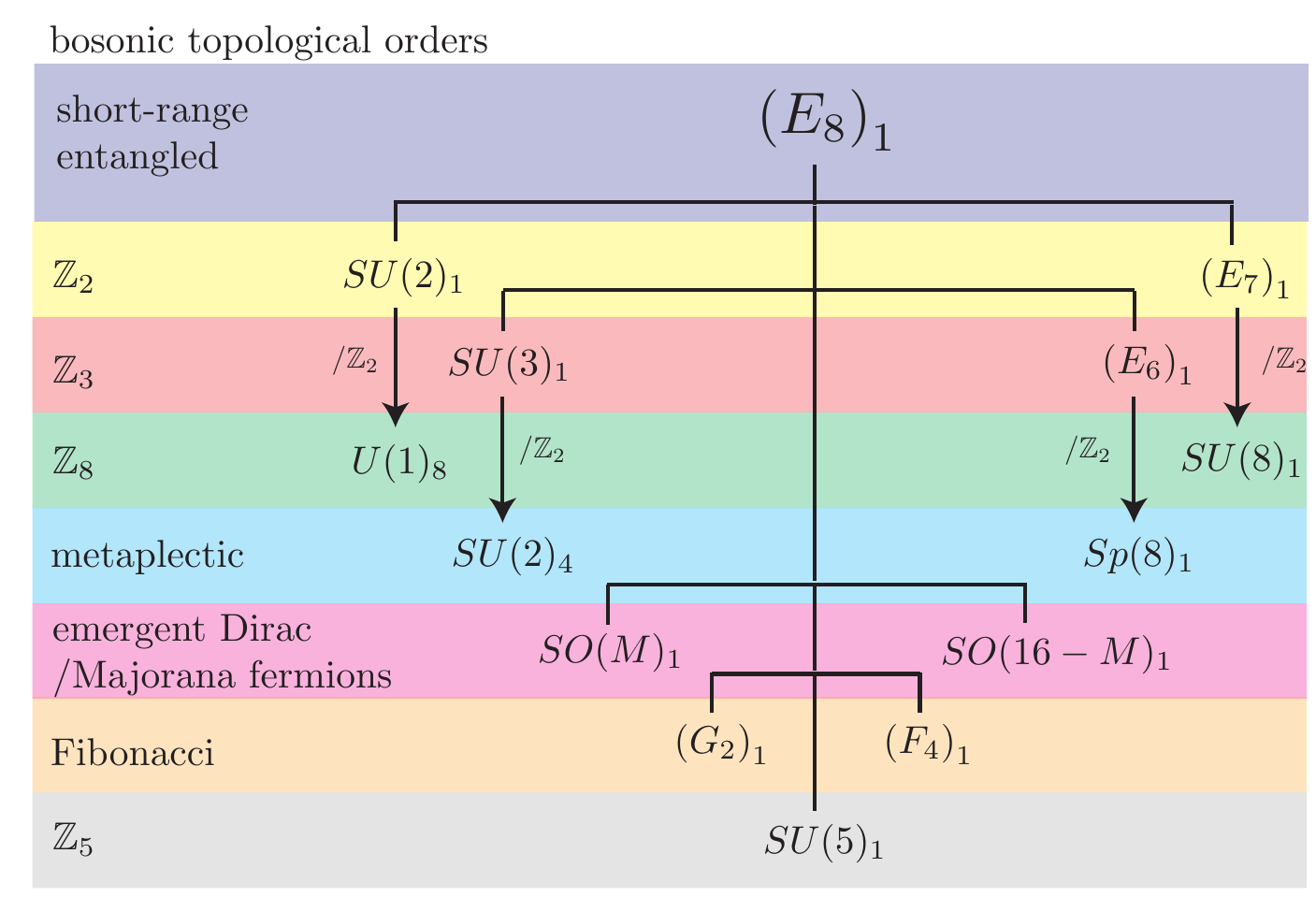}
\caption{Family tree of bosonic fractional quantum Hall states that descend from the $E_8$ state, and their topological orders. States on the same line are pairwise related by particle-hole conjugation. Vertical down arrows represent $\mathbb{Z}_2$ gauging (or orbifolding).}\label{fig:familytree}
\end{figure}

It is worth noting that, for a fixed conjugate pair $\mathcal{G}_A$ and $\mathcal{G}_B$, there are, in general, multiple inequivalent ways they can be embedded in $E_8$, with distinct electrical responses. 
In other words, a
bosonic fractional quantum state with $\mathcal{G}_{A/B}$ topological order can occur at different filling numbers (see Fig.~\ref{fig:fillingcentralchargeplot}) and have distinct quasiparticle charge assignments. In this paper, we exhaust these charged phases with topological orders appearing in \eqref{E8embeddings0} (perhaps except the metaplectic phases $SU(2)_4$ and $Sp(8)_1$). We present the electric charges carried by the quasiparticle primary fields on the edge in all these phases in Appendix~\ref{topologicaldataappendix}. The various bosonic fractional quantum Hall phases encountered in this paper can be summarized by the family tree in Fig~\ref{fig:familytree}.

The remainder of this paper is organized as follows.
In \S \ref{sec:E8review} we review the topological properties of the $E_8$ state and its coupled-wire construction \cite{PhysRevB.100.085116}.
This section introduces the technical tools we later apply to construct various fractional $E_8$ states.
In \S \ref{fractionalbosonicstatessection} we show how the conformal embedding approach leads to Abelian (\S \ref{sec:AbelianStates}) and non-Abelian (\S \ref{nonabeliansection}) fractional $E_8$ states.
In \S \ref{discussionsection} we summarize and discuss possible directions of future work.
There are a number of appendices that describe details used in the main text. 
Appendix \ref{Appendix:qIChargeParity} exhausts all non-local Dirac fermion presentations of the $E_8$ WZW algebra, and consequently accounts for all the fermion charge vectors used in the construction of the $E_8$ state.
In Appendix \ref{app:momentum}, we derive an explicit formula for electron Fermi momenta that ensures translation invariance is preserved in the various coupled-wire constructions of the ${\cal G}_A$ and ${\cal G}_B$ states.
In Appendix \ref{PHconjugationappendix} we review the particle-hole symmetry with respect to the $E_8$ state.
Appendix \ref{topologicaldataSU8} details the topological order of the $SU(8)_1$ state.
In Appendix \ref{app:Sp8}, we discuss the current and primary operators of the metaplectic $Sp(8)_1$ and $SU(2)_4$ conformal field theories.
Appendix \ref{topologicaldataappendix} contains charge assignment tables for the quasiparticle primary fields of the bosonic fractional quantum Hall states in \eqref{E8embeddings0} and Fig.~\ref{fig:familytree}

\section{The bosonic \texorpdfstring{$E_8$}{E8} quantum Hall state}\label{sec:E8review}

In this section, we review the coupled-wire construction of the $E_8$ quantum Hall state~\cite{PhysRevB.100.085116} at filling fraction $\nu=16$.

\subsection{Review of the $E_8$ state}

The $E_8$ state is a bosonic topological state of matter.
``Bosonic" means that its fundamental excitations have bosonic self and mutual statistics.
The state is insulating, possessing a finite bulk excitation gap $E_g^{(0)} > 0$.
The topological order of the $E_8$ state is trivial: All bulk and boundary excitations are local integral combination of the fundamental bosons.
One consequence of this is that the ground state is non-degenerate.
These properties are summarized by saying that the $E_8$ state is a short-range entangled bosonic topological state.
The state supports eight gapless, chiral edge modes.
These edge states are responsible for the quantized (in appropriate units) electric $\sigma_{xy}$ and thermal $\kappa_{xy}$ Hall conductivities $\sigma_{xy} = \nu_{E_8} {e^2 \over h}$ and $\kappa_{xy} = c_{E_8} \frac{\pi^2k_B^2}{3h}T$,
where $\nu_{E_8} = 16$ and $c_{E_8} = 8$.
These conductivities distinguish the $E_8$ state from the topologically trivial insulator, for which both $\sigma_{xy}$ and $\kappa_{xy}$ vanish.
In contrast to the IQHE of electrons, for which both the filling number and the chiral central charge equal one, the $E_8$ state has an unconventional Wiedemann-Franz law 
\cite{FranzWiedemann1853} since $c_{E_8} \neq \nu_{E_8}$.
The $E_8$ state is adiabaticlly connected by stable equivalence to eight copies of the IQHE of electrons \cite{plamadeala2013short, cano2014bulk}.

In general, the chiral central charge $c$
of a bosonic topological phase is related to its anyon quasiparticle content through the Gauss-Milgram formula \cite{kitaev2006anyons}, 
\begin{align}e^{2\pi ic/8}=\frac{1}{\mathcal{D}}\sum_xd_x^2e^{2\pi ih_x}.
\end{align} 
Here, the sum is over anyon classes $x$, with quantum dimensions $d_x$ and spins $h_x$, and is normalized by the total quantum dimension $\mathcal{D}=\sqrt{\sum_xd_x^2}$. 
For trivial topological order, the chiral central charge of a bosonic short-range entangled state must be $c\equiv0$ modulo 8.
% For bosonic states with trivial topological order, $d_x = 1$ and $h_x = 1$ for all $x$.
% Therefore the chiral central charge of a bosonic short-range entangled state must be divisible by 8, i.e., $c\equiv0$ modulo 8.
Up to the addition of unprotected nonchiral edge modes, the $E_8$ state
is therefore the simplest fully-chiral short-range entangled topological state.

The Chern-Simons theory for the $E_8$ state is \cite{PhysRevB.86.125119}
 \begin{align}
 S_{\mathrm{bulk}}=\frac{1}{4\pi}\int_{2+1}(K_{E_8})_{IJ}a^I\wedge da^J+\frac{e}{2\pi}\int_{2+1}t_IA\wedge da^I.\end{align}
Here, $a_I$ ($I = 1, \ldots, 8$) are dynamical $U(1)$ gauge fields;
$A$ is the external electromagnetic gauge potential;
$(K_{E_8})_{IJ}$ is the Cartan matrix of the $E_8$ Lie algebra:
\begin{align}
K_{E_8}=\left(\begin{array}{cccccccc}
2 & -1\\
-1 & 2 & -1\\
 & -1 & 2 & -1\\
 &  & -1 & 2 & -1\\
 &  &  & -1 & 2 & -1 &  & -1\\
 &  &  &  & -1 & 2 & -1\\
 &  &  &  &  & -1 & 2\\
 &  &  &  & -1 &  &  & 2
\end{array}\right),\label{KE8}
\end{align}
with elements not shown equal to zero;
$t_I$ is the charge vector (in the electron basis);
repeated indices are summed over.
The wedge product $a^I\wedge da^J = \epsilon^{\mu \nu \rho} a^I_\mu \partial_\nu a^J_\rho$, where $\epsilon^{\mu \nu \rho}$ is the totally antisymmetric symbol and $\mu, \nu, \rho \in \{0,1, 2 \} = \{t,x,y \}$.

Positive-definiteness of $K_{E_8}$ implies the topological state is fully chiral, i.e., all 8 edge modes move in the same direction.
The state is bosonic since there are only even entries along the diagonal.
Excitations are defined by 8-dimensional integer vectors $l_I, l'_I$: the self and mutual statistics, $\pi l_I (K^{-1}_{E_8})^{IJ} l_I$ and $2 \pi l_I (K^{-1}_{E_8})^{IJ} l'_I$, are integer multiples of $2 \pi$.
Unimodularity of $K_{E_8}$, i.e., $|\det K_{E_8}| = 1$, ensures $E_8$ is short-range entangled.
$K_{E_8}$ is the unique 8-dimensional matrix (up to equivalence by $GL(8, \mathbb{Z})$ similarity transformation) with these properties.
All excitations of the $E_8$ state are even combinations of electrons and therefore carry even electric charge.
Given an 8-dimensional integral vector $l_I$, representing an $E_8$ excitation, its charge is $l_I \tilde{q}^I \equiv l_I(K_{E_8}^{-1})^{IJ}t_J$ (in units of $e$).
The charge vector and $K$-matrix together determine the filling fraction $\nu = \tilde q^I (K^{-1}_{E_8})_{IJ} \tilde q^J$.
These considerations imply that $\nu$ must be an integral multiple of 8.

\subsection{Coupled-wire construction of the \texorpdfstring{$E_8$}{E8} state}

The coupled-wire construction of the $E_8$ state given in Ref.~\onlinecite{PhysRevB.100.085116} begins with a 2d array of metallic wires, each consisting of 11 channels of nonrelativistic fermions (electrons). 
Near the Fermi level, the electron operator $c^\sigma_{ya}(\mathsf{x})$ decomposes in terms of left ($L$) and right ($R$) propagating chiral Dirac electrons that can be represented by vertex operators of bosonized variables $\Phi^\sigma_{ya}(\mathsf{x})$ \cite{giamarchi2003quantum}: \begin{align}c^\sigma_{ya}(\mathsf{x})\sim\exp\left[i\left(\Phi^\sigma_{ya}(\mathsf{x})+k^\sigma_{ya}\mathsf{x}\right)\right],\label{electrons}\end{align} where $a=1,\ldots,11$; $y$ is an integer that labels a given wire; $\mathsf{x}$ is the continuous spatial coordinate along the wire; $\sigma=R,L=+1,-1$ signifies the direction of propagation.
A dimensionful factor $1/\sqrt{l_0}$ (where $l_0$ is a microscopic length scale) that ensures the electron operators have the correct engineering dimensions is suppressed under the proportionality sign $\sim$ in \eqref{electrons}.
The Fermi momentum in the $\mathsf{x}$-direction is
\begin{align}
k^\sigma_{ya}=\frac{eB}{\hbar c}\mathsf{y}+\sigma k_{F,a},
\label{electronmomentum}
\end{align} 
where $\mathsf{y}=yd$ is the vertical location of the $y^{\mathrm{th}}$ wire and $d$ is the separation between adjacent wires. The first term in \eqref{electronmomentum} shifts the electron momentum in the presence of an out-of-plane magnetic field ${\bf B}=B\hat{\bf z}$, written in Landau gauge $A_x=-B\mathsf{y}$. The bare momentum $k_{F,a}$ is model dependent and will be determined by momentum conservation of backscattering interactions presented below. The filling number, which counts the electron number per magnetic flux quantum $\phi_0=hc/e$, is \begin{align}\nu=\frac{N_e}{N_B}=\frac{\frac{l}{2\pi}\sum_a2k_{F,a}}{Bdl/\phi_0}=\frac{\hbar c}{eBd}\sum_a2k_{F,a},\label{fillingnumber}\end{align} where $l$ is the length of each wire. 

The bosonized variables $\Phi^\sigma_{ya}$ are described by the Luttinger liquid Lagrangian density, \begin{align}
\mathcal{L}=\sum_y\sum_{a=1}^{11}\sum_{\sigma=\pm}\frac{1}{4\pi}\sigma\partial_{\mathsf{x}}\Phi^\sigma_{ya}\partial_{\mathsf{t}}\Phi^\sigma_{ya}-\mathcal{H},
\label{L0}
\end{align} 
and obey the equal-time commutation relations, \begin{align}\left[\Phi_{ya}^\sigma(\mathsf{x}),\partial_{\mathsf{x}'}\Phi_{y'a'}^{\sigma'}(\mathsf{x}')\right]=2\pi i\sigma\delta^{\sigma\sigma'}\delta_{yy'}\delta_{aa'}\delta(\mathsf{x}-\mathsf{x}').\label{ETCR0}\end{align} 
Before introducing any backscattering interactions, the Hamiltonian ${\cal H}$ equals the bare Hamiltonian density, \begin{align}\mathcal{H}_0=\sum_y\sum_{a,a'=1}^{11}\sum_{\sigma,\sigma'=\pm}v^{aa'}_{\sigma\sigma'}\partial_{\mathsf{x}}\Phi^\sigma_{ya}\partial_{\mathsf{x}}\Phi^{\sigma'}_{ya'}.\label{H0}\end{align} 
$\mathcal{H}_0$ includes the single-body massless Dirac Hamiltonian $\mathcal{H}_{\mathrm{Dirac}}=iv\sigma{c^\sigma_{ya}}^\dagger\partial_{\mathsf{x}}c^\sigma_{ya}=\frac{v}{4\pi}\left(\partial_{\mathsf{x}}\Phi^\sigma_{ya}\right)^2$, as well as the intra-wire density-density interactions $\mathcal{H}_{\mathrm{int}}=u^{aa'}_{\sigma\sigma'}n^\sigma_{ya}n^{\sigma'}_{ya'}$, where $n^\sigma_{ya}={c^\sigma_{ya}}^\dagger c^\sigma_{ya}=\sigma\partial_{\mathsf{x}}\Phi^\sigma_{ya}/(2\pi)$ is the electron number density for given $y,a,\sigma$.

On each wire, a non-chiral bosonic $E_8$ Wess-Zumino-Witten (WZW) conformal field theory (CFT) \cite{francesco2012conformal} can be singled out by introducing a set of many-body backscattering interactions within the wire that gaps out all fermionic excitations with odd fermion parity. 
To do this, we first perform a basis transformation that decomposes the 11 counter-propagating pairs of Dirac electrons into $E_8$ bosons and three decoupled ``integrated" Dirac fermions $f^\sigma_{yn}$, for $n=1,2,3$.
This corresponds to the symmetry decomposition:
\begin{align}U(11)_1=\left(E_8\right)_1\otimes U(3)_1.\end{align} 
The basis transformation from the electron to the ``Chevalley" basis, \begin{align}\tilde{\Phi}_{yI}^{\sigma}=\sum_{a,\sigma'}U_{Ia}^{\sigma\sigma'}\Phi_{ya}^{\sigma'},\quad\tilde{k}_{yI}^{\sigma}=\sum_{a,\sigma'}U_{Ia}^{\sigma\sigma'}k_{ya}^{\sigma'},
\label{eq:electronChevalleyConversion}
\end{align} uses the integral unimodular $U$ matrix, 
\begin{gather}
U=\begin{pmatrix}U^{++}&U^{+-}\\U^{-+}&U^{--}\end{pmatrix},\label{Umatrix}\\
\begin{split}
 & (U^{++}|U^{+-})=(U^{--}|U^{-+})=\\
 & \left(\resizebox{7.8cm}{!}{$
 \begin{array}{ccccccccccc|ccccccccccc}
-1 & -1 & -1 &  &  &  &  &  &  &  &  &  &  &  &  &  &  &  &  &  &  & -1\\
 &  & 1 & 1 &  &  &  &  &  &  &        &  &  &  &  &  &  &  &  &  &  & \\
 &  &  & -1 & 1 &  &  &  &  &  &       &  &  &  &  &  &  &  &  &  &  & \\
 &  &  &  & -1 & 1 &  &  &  &  &       &  &  &  &  &  &  &  &  &  &  & \\
 &  &  &  &  & -1 & -1 &  &  &  &       &  &  &  &  &  &  &  &  &  &  & \\
 &  &  &  &  &  & 1 & 1 &  &  &       &  &  &  &  &  &  &  &  &  &  & \\
 &  &  &  &  &  &  & -1 & 1 &  &       &  &  &  &  &  &  &  &  &  &  & \\
 &  & -1 & 1 & 1 & 1 &  &  &  &  &  &  &  &  &  &  &  &  &  &  & 1 & -1\\
 &  &  &  &  &  &  &  &  & 1 & 1 & 1 &&&&&&&&&&\\
 &  &  &  &  &  &  &  &  & 3 & -5 & -2 & -1 & -2 & 2 & 2 & 2 & -2 & 2 & 2\\
2 &  & 1 & -1 & -1 & -1 & 1 & -1 & -1 &  &  &  &  &  &  &  &  &  &  &  & -1 & 3
\end{array}$}
\right),
\end{split}\nonumber 
\end{gather}
where the rows and columns of $U^{\sigma\sigma'}$ are labeled by $I,a=1,\ldots,11$. 
The Lagrangian density \eqref{L0} in the ``Chevalley" basis is \begin{align}\begin{split}\mathcal{L}=\sum_y&\left(\sum_{I,J=1}^{8}\sum_{\sigma=\pm}\frac{1}{4\pi}\sigma\left(K_{E_8}^{-1}\right)^{IJ}\partial_{\mathsf{x}}\tilde\Phi^\sigma_{yI}\partial_{\mathsf{t}}\tilde\Phi^\sigma_{yJ}\right.\\&+\left.\sum_{n=1}^3\sum_{\sigma=\pm}\frac{1}{4\pi}\sigma\partial_{\mathsf{x}}\tilde\Phi^\sigma_{y,n+8}\partial_{\mathsf{t}}\tilde\Phi^\sigma_{y,n+8}\right)-\mathcal{H},\end{split}\label{Lc}\end{align}
where $K_{E_8}$ is the Cartan matrix defined in \eqref{KE8}.
Since $K_{E_8}$ has unit determinant, its inverse $K_{E_8}^{-1}$ is also integral. 

For each chiral sector $\sigma=R,L=+,-$, the first eight bosonized variables $\tilde\Phi^\sigma_{yI}$, $I=1,\ldots,8$, generate the $E_8$ WZW CFT at level 1. 
The vertex operators, \begin{align}
\left[\mathtt{E}_{E_8}(\mathsf{x})\right]^\sigma_{yI}=\exp\left[i(\tilde\Phi_{yI}^{\sigma}(\mathsf{x})+\tilde{k}_{yI}^{\sigma}\mathsf{x})\right],\label{E8simpleroots}\end{align} 
correspond to the simple roots of the $E_8$ algebra.
These operators all have spin $|h|=1$ and are (each the exponential of) integral linear combinations of the bosonized electrons \eqref{electrons}. The electric charges $\tilde{q}_J$ (in units of $e$) of the simple roots $\left[\mathtt{E}_{E_8}(\mathsf{x})\right]^\sigma_{yI}$ can be read off by summing over the entries in each of the first 8 rows of the $U$ matrix: \begin{align}\begin{split}\tilde{q}_{J=1,\ldots,8}&=\sum_{a=1}^{11}\left(U^{++}_{Ja}+U^{+-}_{Ja}\right)\\&=(-4,2,0,0,-2,2,0,2).\end{split}\label{E8simlerootscharge}\end{align} 
The last three elements of $\tilde\Phi^\sigma_{y,I+8+n}$ define the ``integrated" Dirac fermions: $f_{yn}^{\sigma}=\exp\left[i(\tilde\Phi_{y,8+n}^{\sigma}+\tilde{k}_{y,n+8}^{\sigma}\mathsf{x})\right]$, for $n=1,2,3$. 
These bosons generate the $U(3)_1$ symmetry, which is decoupled from the $(E_8)_1$ symmetry. 
The ``integrated" Dirac fermions have spin $|h|=1/2$ and carry electric charges $\tilde{q}_{n=1,2,3}=(3,1,1)$.

The electron density-density interactions $u^{aa'}_{\sigma\sigma'}$--- equivalently, the velocity matrix $v^{aa'}_{\sigma\sigma'}$ in \eqref{H0}---can be tuned so that the bare Hamiltonian density in the ``Chevalley" basis is \begin{align}\begin{split}\mathcal{H}_0=\frac{\tilde{v}}{4\pi}\sum_y\sum_{\sigma=\pm}&\left[\sum_{I,J=1}^8\left(K_{E_8}^{-1}\right)^{IJ}\partial_{\mathsf{x}}\tilde\Phi^\sigma_{yI}\partial_{\mathsf{x}}\tilde\Phi^\sigma_{yJ}\right.\\&\left.+\sum_{n=1}^3\left(\partial_{\mathsf{x}}\tilde\Phi^\sigma_{y,n+8}\right)^2\right].\end{split}\label{E8conformalH}\end{align} 
At this fixed point, the $E_8$ bosons and the three integrated Dirac fermions are completely decoupled and their theories are conformally symmetric. 
The fine tuning of the density interactions can be relaxed after a bulk excitation energy gap is established by backscattering terms. 
The 2D topological phase thus constructed is robust against all gap preserving perturbations, including small deviations of the density interactions away from their fine-tuned values. 

The three integrated Dirac fermions can be gapped out by the intra-wire backscattering interactions \begin{align}\begin{split}\mathcal{H}^f_{\mathrm{intra}}&=u_{\mathrm{intra}}\sum_y\sum_{n=1}^3{f^R_{yn}}^\dagger f^L_{yn}+h.c.\\&=2u_{\mathrm{intra}}\sum_y\sum_{n=1}^3\cos\left(\tilde\Phi^L_{y,n+8}-\tilde\Phi^R_{y,n+8}\right).\end{split}\label{E8Hintra}\end{align} 
These interactions conserve charge because $f^R_{yn}$ and $f^L_{yn}$ carry the same charge. In addition, if the $L$ and $R$ ``integrated" Dirac fermions have equal momentum, $\tilde{k}^R_{y,8+n}=\tilde{k}^L_{y,8+n}$, then the $\mathsf{x}$-dependent oscillation factors $e^{ik\mathsf{x}}$ in \eqref{E8Hintra} cancel. 
The intra-wire interaction strength  sets the fermion gap energy scale $E_g^1\sim u_{\mathrm{intra}}$. 
We take this to be the largest energy scale for all of the bosonic topological states constructed using the coupled-wire models in this paper.

The $(E_8)_1$ Kac-Moody (KM)~\cite{Kac68,Moody68} current algebra is spanned by the eight Cartan generators $\left[\mathtt{H}_{E_8}\right]_I$ and 240 roots $\left[\mathtt{E}_{E_8}\right]_{\boldsymbol\alpha}$ (eight of which have already been given in \eqref{E8simpleroots}). 
On any given wire $y$ and in chiral sector $\sigma=R,L$, the Cartan generators are \begin{align}\left[\mathtt{H}_{E_8}(\mathsf{x})\right]^\sigma_{yI}=\partial_{\mathsf{x}}\tilde\Phi_{yI}^{\sigma}(\mathsf{x}),\label{E8CartangeneratorsChevalley}\end{align} where $I=1,\ldots,8$. 
These operators, which are proportional to the normal ordered product $\left[\mathtt{E}_{E_8}\right]_I^\dagger\left[\mathtt{E}_{E_8}\right]_I$, are bosonic combination of electrons. 
The roots are spin-1 vertex operators of integral combinations of $\tilde\Phi_I$: \begin{align}\left[\mathtt{E}_{E_8}(\mathsf{x})\right]^\sigma_{y,\boldsymbol\alpha}=\exp\left[i\tilde\alpha^I(\tilde\Phi_{yI}^{\sigma}(\mathsf{x})+\tilde{k}_{yI}^{\sigma}\mathsf{x})\right],\label{E8rootsChevalley}\end{align} where the root vector $\tilde{\boldsymbol\alpha}=(\tilde\alpha^1,\ldots,\tilde\alpha^8)$ (in the ``Chevalley" basis) has integral entries and length $|\boldsymbol\alpha|=\sqrt{K_{IJ}\tilde\alpha^I\tilde\alpha^J}=\sqrt{2}$. 
It will become clear below that there are 240 root vectors. 
Since the eight simple roots in \eqref{E8simpleroots} are bosonic combinations of electrons, so are all the 240 roots. 

Introducing the complex Euclidean spacetime parameters $z=e^{2\pi(\tilde{v}\tau+i\mathsf{x})/l}$ and $\bar{z}=e^{2\pi(\tilde{v}\tau-i\mathsf{x})/l}$ (i.e., mapping spacetime to the cylinder), where $\tau=i\mathsf{t}$ is the Wick rotated time and $l$ is the length of a wire, the chiral operators evolve as holomorphic and anti-holomorphic fields $\tilde\Phi^L(x,\tau)=\tilde\Phi^L(z)$ and $\tilde\Phi^R(x,\tau)=\tilde\Phi^R(\bar{z})$ at the conformal fixed point $\mathcal{H}_0+\mathcal{H}^f_{\mathrm{intra}}$. Focusing on the $L$ sector on any given wire, the Chevalley bosonized operators obey the operator product expansion (OPE):
\begin{align}\left\langle\tilde\Phi_I(z)\tilde\Phi_J(w)\right\rangle=-(K_{E_8})_{IJ}\log(z-w)+\ldots,\label{E8bosonizedKMalgebra}
\end{align} up to finite non-singular factors, including those that are responsible for the anticommutation between electron operators of different channels. The $(E_8)_1$ KM currents obey the OPEs (up to finite non-singular terms):
\begin{align}\begin{split}
\left[\mathtt{H}(z)\right]_I\left[\mathtt{H}(w)\right]_J&=\frac{(K_{E_8})_{IJ}}{(z-w)^2}+\ldots,\\
\left[\mathtt{H}(z)\right]_I\left[\mathtt{E}(w)\right]_{\boldsymbol\alpha}&=\frac{(K_{E_8})_{IJ}\tilde\alpha^J}{z-w}\left[\mathtt{E}(w)\right]_{\boldsymbol\alpha}+\ldots,\\
\left[\mathtt{E}(z)\right]_{\boldsymbol\alpha}\left[\mathtt{E}(w)\right]_{-\boldsymbol\alpha}&=\frac{1}{(z-w)^2}+\frac{\tilde\alpha^I}{z-w}\left[\mathtt{H}(w)\right]_I+\ldots,\\
\left[\mathtt{E}(z)\right]_{\boldsymbol\alpha}\left[\mathtt{E}(w)\right]_{\boldsymbol\beta}&=\frac{Z_{\boldsymbol\alpha\boldsymbol\beta}}{z-w}\left[\mathtt{E}(w)\right]_{\boldsymbol\alpha+\boldsymbol\beta}+\ldots,\end{split}\label{E8KMalgebra}\end{align} if $(K_{E_8})_{IJ}\alpha^I\beta^J=-1$. The cocycle coefficients $Z_{\boldsymbol\alpha\boldsymbol\beta}$ ensure the last equality is unchanged under the exchange $z\leftrightarrow w$ and $\boldsymbol\alpha\leftrightarrow\boldsymbol\beta$. They can be determined from the non-singular pieces in \eqref{E8bosonizedKMalgebra} that guarantee electron mutual anticommutation. The exact form of $Z_{\boldsymbol\alpha\boldsymbol\beta}$ is inconsequential to this paper and will not be presented.

The $E_8$ current algebra, as well as its subalgebras discussed in the following sections, is sometimes more conveniently presented using a {\em non-local} Dirac fermion basis \begin{align}
\label{nonlocaldiracdef}
d_{yj}^\sigma(\mathsf{x})=\frac{\psi_{2j-1}(\mathsf{x})+i\psi_{2j}(\mathsf{x})}{\sqrt{2}}\sim e^{i\phi^\sigma_{yj}(\mathsf{x})+ik^\sigma_{yj}\mathsf{x}},
\end{align}
for $j=1,\ldots,8$, where the ``Cartan-Weyl" bosons $\phi^\sigma_{yj}$ and Fermi momenta $k^\sigma_{yj}$ are defined by the (non-unimodular) basis transformation:
\begin{align}\tilde\Phi^\sigma_{yI}=R_I^j\phi^\sigma_{yj},\quad\tilde{k}_I=R_I^jk_j.\label{diracRmatrix}
\end{align}
The electric charge $q_j$ carried by $d_j$ is related to the charge $\tilde{q}_I$ of the $E_8$ simple roots in \eqref{E8simlerootscharge} using the same transformation: $\tilde{q}_I=R_I^jq_j$ (see also \eqref{appchargevector}). 
The $R$-transformation obeys $R^j_IR^l_J\delta_{jl}=(K_{E_8})_{IJ}$. 
The vectors $\boldsymbol\alpha_I=(R^1_I,\ldots,R^8_I)$ represent the 8 simple roots of $E_8$ in Euclidean space, and their entries $R^j_I$ are integers or half-integers. 
(Explicit examples are presented in appendix~\ref{Appendix:qIChargeParity}.) It can be shown that, depending on the choice of the $R$-matrix, the electric charge vector ${\bf q}=(q_1,\ldots,q_8)$ of the non-local Dirac fermions are entry-wise permutations of one of the following vectors: (i) $(\pm4,0,0,0,0,0,0,0)$, (ii) $(2s_1,2s_2,2s_3,2s_4,0,0,0,0)$ for $s_{1,2,3,4}=\pm 1$, or (iii) $(3s_1,s_2,s_3,s_4,s_5,s_6,s_7,s_8)$ for $s_{1,\ldots,8}=\pm1$ and $\prod_{j=1}^8s_j=-1$. 
The possible ${\bf q}$ have the same length squared: \begin{align}\nu_{E_8}=|{\bf q}|^2=\sum_{J,J'=1}^8\tilde{q}_J\left(K_{E_8}^{-1}\right)^{JJ'}\tilde{q}_{J'}=16.
\label{E8fillingnumber}\end{align} 
In cases (i) and (ii), the non-local Dirac fermions all have even electric charges, and in case (iii), the fermions all have odd charges. 
The condition on the product of the $s_j$ in case (iii) ensures the $E_8$ current operators, which are even combinations of electrons, have even charge. 
Ignoring the integrated fermions that are gapped out by $\mathcal{H}^f_{\mathrm{intra}}$ \eqref{E8Hintra}, the low-energy parts of the Lagrangian \eqref{Lc} and bare Hamiltonian \eqref{E8conformalH} densities, which only retain the $E_8$ degrees of freedom, become in the ``Cartan-Weyl" basis: \begin{align}\begin{split}\mathcal{L}^{E_8}&=\sum_y\sum_{j=1}^8\sum_{\sigma=\pm}\frac{1}{4\pi}\sigma\partial_{\mathsf{x}}\phi^\sigma_{yj}\partial_{\mathsf{t}}\phi^\sigma_{yj}-\mathcal{H},\\\mathcal{H}^{E_8}_0&=\frac{\tilde{v}}{4\pi}\sum_y\sum_{j=1}^8\sum_{\sigma=\pm}\left(\partial_{\mathsf{x}}\phi^\sigma_{yj}\right)^2.\end{split}\label{E8CartanWeylLH}\end{align} 
(Recall the full Hamiltonian ${\cal H}$ includes ${\cal H}^{E_8}_0$ and various to-be-discussed backscattering interactions.)
Although this looks identical to the theory of 8 free Dirac fermions, the fermion non-locality dictates that the Hilbert space must be changed so that only the bosonic $E_8$ current operators (and their combinations) are integral and local. Below we express the 248 $E_8$ currents ${\bf J}_{E_8}$ as fermion bilinears or spinor combinations.

On each wire $y$ and in chiral sector $\sigma=R,L$, the 8 Cartan generators of $(E_8)_1$ in \eqref{E8CartangeneratorsChevalley} are equal to the fermion densities, \begin{align}\left[\mathtt{H}_{E_8}\right]_j=\partial_{\mathsf{x}}\phi_j\sim d_j^\dagger d_j,\quad j=1,\ldots,8.\label{E8Cartangenerators}\end{align} 
Each one of these is a linear combination of Cartan generators in \eqref{E8CartangeneratorsChevalley}. The 240 $(E_8)_1$ roots in \eqref{E8rootsChevalley} can be expressed in the Cartan-Weyl basis as \begin{align}\left[\mathtt{E}_{E_8}(\mathsf{x})\right]_{\boldsymbol\alpha}=e^{i\tilde\alpha^I(\tilde\Phi_I(\mathsf{x})+\tilde{k}_I\mathsf{x})}=e^{i\alpha^j(\phi_j(\mathsf{x})+k_j\mathsf{x})},\label{E8rootsCartanWeyl}\end{align} where the Chevalley and Cartan-Weyl root vectors and momenta are related by $\alpha^j=\tilde\alpha^IR^j_I$ and $\tilde{k}_I=R_I^jk_j$. 
The roots consist of (i) the 112 $SO(16)_1$ roots and (ii) the 128 $SO(16)_1$ even spinors. 
The $SO(16)_1$ roots are the fermion bilinears $d_{j_1}d_{j_2}$, $d_{j_1}^\dagger d_{j_2}$, $d_{j_1}d_{j_2}^\dagger$, or $d_{j_1}^\dagger d_{j_2}^\dagger$, for $1\leq j_1<j_2\leq8$:
\begin{align}\left[\mathtt{E}_{E_8}(\mathsf{x})\right]_{\boldsymbol\alpha}=e^{i(\pm\phi_{j_1}(\mathsf{x})\pm\phi_{j_2}(\mathsf{x}))+i\alpha^jk_j\mathsf{x}},\label{SO16roots}\end{align} where the $SO(16)$ root vectors in the Cartan-Weyl basis are the integral vectors $\boldsymbol\alpha=\pm{\bf e}_{j_1}\pm{\bf e}_{j_2}$.
Here, ${\bf e}_{j}$ is the unit 8-vector with a 1 in the $j$-th entry and 0 elsewhere.
Together with the Cartan generators, they form the $SO(16)_1$ WZW subalgebra, whose current operators are 
$J_{pq}=i\psi_p\psi_q$, $1\leq p<q\leq16$ (see \eqref{nonlocaldiracdef}). The $(E_8)_1$ theory extends $SO(16)_1$ by including its even spinors, \begin{align}\left[\mathtt{E}_{E_8}(\mathsf{x})\right]_{\boldsymbol\alpha}=e^{i\varepsilon^j(\phi_j(\mathsf{x})+k_j\mathsf{x})/2}.\label{SO16evenspinors}\end{align} The $E_8$ root vectors $\boldsymbol\alpha=\boldsymbol\varepsilon/2$ here have half-integer entries $\varepsilon^j/2=\pm1/2$, where $\prod_{j=1}^8\varepsilon^j=+1$. Since the root vectors in \eqref{SO16roots} and \eqref{SO16evenspinors} both have length $|\boldsymbol\alpha|=\sqrt{2}$, they all correspond to bosonic vertex operators $e^{i\alpha^j\phi_j}$ with spin $|h|=1$. Moreover, \eqref{SO16roots} and \eqref{SO16evenspinors} are both local integral operators by construction since they both originate from \eqref{E8simpleroots} and \eqref{E8rootsChevalley}, which are even combinations of electrons. The distinction of $SO(16)$ roots and even spinors is artificial and depends on the choice of the Cartan-Weyl basis $R^a_I$. The $SO(16)$ embedding in $E_8$ is not unique. In the physical theory, all $E_8$ bosons, including both $SO(16)$ roots and even spinors, can be rotated into one another by the $E_8$ symmetry and should be treated impartially. 

The ``Chevalley" bosons $\tilde{\boldsymbol\Phi}=(\tilde\Phi_1,\ldots,\tilde\Phi_8)$ take values in the torus $\mathbb{R}^8/2\pi\mathbb{Z}^8$: $\tilde\Phi_I\equiv\tilde\Phi_I+2\pi n_I$, for any integer $n_I$.
This equivalence may alternatively be described as an invariance (of operators constructed out of the ``Chevalley" bosons) under large gauge transformations. 
Using the $R$-transformation, the ``Cartan-Weyl" bosons $\boldsymbol\phi=(\phi_1,\ldots,\phi_8)$ have the equivalence $\phi_j\equiv\phi_j+2\pi(R^{-1})_j^In_I$, where the vector ${\bf r}=R^{-1}{\bf n}$ lives inside the lattice $\mathcal{R}$ that contains the $E_8$ root vectors $\boldsymbol\alpha$ as primitive lattice vectors. $\mathcal{R}$ consists of vectors ${\bf r}=(r_1,\ldots,r_8)$ with all integral or all half-integral entries and even trace $\sum_{j=1}^8r_j$. A vertex operator $e^{im^j\phi_j}$ is local if and only if it is invariant under all large gauge transformations, i.e., ${\bf m}\cdot{\bf r}$ is integral. 
Since the lattice $\mathcal{R}$ is self-dual, $e^{im^j\phi_j}$ is local and integral if and only if ${\bf m}$ lives in $\mathcal{R}$. 
Therefore the ``Cartan-Weyl" bosons $\boldsymbol\phi=(\phi_1,\ldots,\phi_8)$ live in the compactified torus $T_{E_8}=\mathbb{R}^8/2\pi\mathcal{R}$. 
If the theory \eqref{E8CartanWeylLH} were 8 free local Dirac fermions, $\boldsymbol\phi$ would live in a different torus $T_{U(1)^8}=\mathbb{R}^8/2\pi\mathbb{Z}^8$. 
$T_{E_8}$ and $T_{U(1)^8}$ are not related by unimodular transformation and therefore the corresponding Hilbert spaces are different.
Consequently, the bosonic $(E_8)_1$ theory is inequivalent to the theory of 8 free (local) Dirac fermions, and the $(E_8)_1$ Hilbert space is spanned by fermion bilinear or even spinor excitations.

The coupled-wire model of the $E_8$ state is completed by including inter-wire $E_8$ current backscattering interactions ${\cal H}_{\rm inter}$. 
Summarizing, the full Hamiltonian density (including for completeness the ``integrated" fermion modes) $\mathcal{H}=\mathcal{H}_0 + {\cal H}_{\rm intra} +\mathcal{H}_{\mathrm{inter}}$ 
consists of the bare Hamiltonian $\mathcal{H}_0$ \eqref{E8conformalH}, the ``integrated" fermion gap-generating term $\mathcal{H}^f_{\mathrm{intra}}$ \eqref{E8Hintra} that removes all local fermion excitations below the energy scale $E_g^1\sim u_{\mathrm{intra}}$, and the $E_8$ backscattering interactions, \begin{widetext}\begin{align}\mathcal{H}_{\mathrm{inter}}&=u_{\mathrm{inter}}\sum_y{\bf J}_y^R\cdot{\bf J}_{y+1}^L\label{E8Hinter}\\
&=u_{\mathrm{inter}}\sum_y\left(\sum_{j=1}^8{\left[\mathtt{H}_{E_8}\right]^R_{y,j}}^\dagger\left[\mathtt{H}_{E_8}\right]^L_{y+1,j} \, + \, \sum_{\boldsymbol\alpha}{\left[\mathtt{E}_{E_8}\right]^R_{y,\boldsymbol\alpha}}^\dagger\left[\mathtt{E}_{E_8}\right]^L_{y+1,\boldsymbol\alpha}\right)\nonumber\\
&=u_{\mathrm{inter}}\sum_y\left[\sum_{j=1}^8\partial_{\mathsf{x}}\phi^R_{y,j}\partial_{\mathsf{x}}\phi^L_{y+1,j}\, - \, \sum_{\boldsymbol\alpha}\cos\left(\boldsymbol\alpha\cdot\boldsymbol\theta_{y+1/2}\right)\right].\nonumber\end{align}
\end{widetext}
The interactions in \eqref{E8Hinter} are analogous to those in the $O(N)$ Gross-Neveu model \cite{GrossNeveu1974,PhysRevB.94.165142}.
The entries of $\boldsymbol\theta_{y+1/2}=(\theta_{y+1/2,1},\ldots,\theta_{y+1/2,8})$ are the sine-Gordon angle variables $\theta_{y+1/2,j}=\phi^R_{y,j}-\phi^L_{y+1,j}$, and the sum over $\boldsymbol\alpha$ ranges over all 240 roots of $E_8$. Like $\mathcal{H}_0$ and $\mathcal{H}^f_{\mathrm{intra}}$, the inter-wire interactions are combinations of integral products of electron operators because $\left[\mathtt{H}_{E_8}\right]^\sigma_{ya}$ and $\left[\mathtt{E}_{E_8}\right]^\sigma_{y\boldsymbol\alpha}$ are. 
Because the roots ${E^R_{\boldsymbol\alpha}}^\dagger$ and $E^L_{\boldsymbol\alpha}$ have opposite electric charge, $\mathcal{H}_{\mathrm{inter}}$ conserves charge. 
The last identity of \eqref{E8Hinter} holds when the Fermi momenta appearing in the $E_8$ roots \eqref{E8rootsChevalley}, \eqref{SO16roots} and \eqref{SO16evenspinors} cancel, i.e., $k^R_{y,j}=k^L_{y+1,j}$ in the ``Cartan-Weyl" basis (or equivalently $\tilde{k}^R_{y,I}=\tilde{k}^L_{y+1,I}$ in the ``Chevalley" basis). 
These momentum conservation conditions of $\mathcal{H}_{\mathrm{inter}}$ and the ones from $\mathcal{H}^f_{\mathrm{intra}}$, $\tilde{k}^R_{y,n+8}=\tilde{k}^L_{y,n+8}$ for $n=1,2,3$, require the electron bare Fermi momenta $k_{F,a}$ (see \eqref{electronmomentum}) to take a particular form: \begin{align}
k_{F,a}=\frac{1}{2}\frac{eBd}{\hbar c}\sum_{J,J'=1}^8\left(U^{++}_{Ja}+U^{+-}_{Ja}\right)
\left(K_{E8}^{-1}\right)^{JJ'}
\tilde{q}_{J'},
\label{E8momentum}\end{align} where $U$ is the unimodular matrix \eqref{Umatrix}
that defines the $E_8$ simple roots, and $\tilde{q}_J$ are the electric charges \eqref{E8simlerootscharge} of the simple roots.
The Fermi momentum solution \eqref{E8momentum} is a special case discussed in appendix~\ref{app:momentum} (see \eqref{appmomentumAbelian}). The momentum conservation condition \eqref{E8momentum} also guarantees that the filling number \eqref{fillingnumber} agrees with \eqref{E8fillingnumber}.

With $u_{\mathrm{inter}}>0$, the sine-Gordon potentials in \eqref{E8Hinter} are marginally relevant in the renormalization group sense at the 1-loop level due to the $E_8$ current OPE \eqref{E8KMalgebra} and generate a mass gap \cite{PhysRevB.100.085116}. 
The potentials simultaneously pin the angle variables $\boldsymbol\theta_{y+1/2}=(\theta_{y+1/2,1},\ldots,\theta_{y+1/2,8})$. The ground state expectation value $\langle\boldsymbol\theta_{y+1/2}\rangle$ sits inside the lattice $2\pi\mathcal{R}$ so that $\langle\boldsymbol\alpha\cdot\boldsymbol\theta_{y+1/2}\rangle$ are integers multiple of $2\pi$, for all $E_8$ roots, and minimize the sine-gordon potentials. The angle variables can be shifted by the large gauge transformation $\phi^R_{y,j}\to\phi^R_{y,j}+2\pi r_j$ for any lattice vector ${\bf r}$ in $\mathcal{R}$. Therefore, up to the large gauge transformation, there is a unique potential minimum and ground state. The inter-wire $E_8$ current backscattering strength sets the finite energy scale $E_g^0\sim u_{\mathrm{inter}}$ of bulk excitations of the $E_8$ state. Throughout this paper, we assume the excitation energy gap is much smaller than the fermion gap $E_g^1\sim u_{\mathrm{intra}}$. For the $E_8$ quantum Hall state, all excitations between $E_g^0$ and $E_g^1$ are bosonic even combinations of electrons. 

If the coupled-wire model is defined on a closed torus where $\mathsf{x}\equiv\mathsf{x}+l$ and $y\equiv y+\mathsf{L}$ are both periodic, the summation of the inter-wire interactions in \eqref{E8Hinter} runs over all wires $y=1,\ldots,\mathsf{L}$. The unique ground state is separated from all excitations by the bulk gap $E_g^0\sim u_{\mathrm{inter}}$. On the other hand, if the model is defined on a cylinder where the $y$ direction is open, the summation in \eqref{E8Hinter} runs from $y=1,\ldots,\mathsf{L}-1$. $\mathcal{H}_{\mathrm{inter}}$ leaves behind the gapless left (right) propagating chiral $E_8$ level 1 WZW CFTs on the open boundaries at $y=1$ (resp.~$y=\mathsf{L}$). All low-energy edge excitations are bosonic and have even electric charge. This can in principle be experimentally verified by shot-noise tunneling at a quantum point contact.

Lastly, we remark that certain sub-collections of sine-Gordon potentials in the inter-wire interaction \eqref{E8Hinter} are sufficient to introduce the bulk excitation energy gap. For example, instead of backscattering the entire $E_8$ current algebra, a gap opens if only the eight $E_8$ simple roots are back-scattered, $\sum_{I=1}^8\cos(\boldsymbol\alpha_I\cdot\boldsymbol\theta_{y+1/2})$. This is because the eight independent terms obey Haldane's nullity condition~\cite{PhysRevLett.74.2090}, $\left[\boldsymbol\alpha_I\cdot\boldsymbol\theta_{y+1/2},\boldsymbol\alpha_J\cdot\boldsymbol\theta_{y+1/2}\right]=0$, and completely gap the eight-component boson theory. Alternatively, instead of summing over all the 240 $E_8$ roots $\boldsymbol\alpha$ in \eqref{E8Hinter}, one can restrict the sum to include only the 112 $SO(16)$ roots. Despite only involving the $SO(16)_1$ currents, the model will still leave behind the chiral $E_8$ level 1 WZW CFTs on the edges, and the quantum Hall state constructed will still carry the same short-ranged entangled $(E_8)_1$ topological order, i.e., an absence of fractionalization. This is because the even spinors of $SO(16)_1$ are still local operators that are integral combinations of electrons and extend $SO(16)_1$ to $(E_8)_1$. In other words, the even spinors are ``condensed" in the anyon condensation picture~\cite{PhysRevB.79.045316, 2018ARCMP...9..307B}. The $SO(16)_1$ topological order is killed. This is because the odd spinor and odd fermion excitations, which carry mutual semionic braiding statistics with the even spinor, are confined by the locality of the even spinor. Moreover, the $SO(16)_1$ sine-Gordon potentials pin the angle variables $\theta_{y+1/2,j}$ to the same minimum since the large gauge transformation $\theta_{y+1/2,j}\equiv\theta_{y+1/2,j}+2\pi r_j$, for any ${\bf r}$ in $\mathcal{R}$, are still set by the locality of the $(E_8)_1$ current operators. Therefore, the ground state remains unique and is identical to that of the $(E_8)_1$ state. 

\section{Fractional bosonic states}
\label{fractionalbosonicstatessection}

In the previous section, we reviewed the sense in which the $E_8$ state is the bosonic analogue of the completely filled lowest Landau level of electrons. 
In this section, we present a family of bosonic fractional quantum Hall (bFQH) states that ``partially fill" this $E_8$ state. 
These bFQH states can be viewed as bosonic analogues of the fermionic fractional quantum Hall (fFQH) states that occur when the Landau level is partially filled. 
Examples of fFQH states include the Abelian Laughlin states~\cite{Laughlin83} at filling $\nu=1/m$, for $m$ odd, and their particle-hole conjugates at filling $\nu=1-1/m$, as well as the non-Abelian Moore-Read Pfaffian state~\cite{MooreRead,GreiterWenWilczekPRL91} at filling $\nu=1/2$ and its particle-hole conjugate, the anti-Pfaffian state~\cite{LevinHalperinRosenow07,LeeRyuNayakFisher07}. 
These states are topologically ordered and long-range entangled. 
They support fractional quasiparticle excitations that are not local integral combinations of electrons and that must exist non-locally in conjugate pairs (or multiplets). 
These excitations can carry fractional electric charge and exhibit anyonic (i.e., not bosonic or fermionic) statistics. 

The bFQH states constructed in this section are also topologically ordered and support fractionalization. 
Like the parent $E_8$ state, the bFQH states differ from the fFQH states by the presence of a fermionic energy gap $E_g^1$, below which all local bulk and edge excitations are even (i.e., bosonic) combinations of electrons. 
Excitations of the bFQH states have a minimum energy gap $0 \leq E_g^0 < E_g^1$ above the ground state energy (taken here to be zero).
As long as the energy of an excited state is below the fermion gap $E_g^1$, the excitation must have the same fermion parity as the ground state. 
This requirement excludes the filled Landau level, the Laughlin states, and the Pfaffian states from our constructions because the gapless edge of these fermionic states (where $E_g^0 =0$) support odd-electron excitations with arbitrarily small energy in the thermodynamic limit.

There are two related senses in which the bFQH states considered in this section ``partially fill" the $E_8$ state.
The first is simply that filling fractions of the bFQH states are less than or equal to the $E_8$ filling fraction $\nu_{E_8} = 16$.
The second is that the bFQH edge states are described by WZW theories with a KM symmetry $\mathcal{G}$ that is a subalgebra of $(E_8)_1$ symmetry. 
This $\mathcal{G}$ symmetry is generated by KM current operators $J_{\mathcal{G}}^i$, which are even combinations of electrons and together form a subcollection (of linear combinations) of the $E_8$ currents.
Similar to the $E_8$ algebra (c.f.~\ref{E8KMalgebra}), the $\mathcal{G}$ subalgebra has a closed OPE (up to non-singular terms), \begin{align}J_{\mathcal{G}}^i(z)J_{\mathcal{G}}^j(w)=\frac{k\delta^{ij}}{(z-w)^2}+\frac{if^{ijk}}{z-w}J_{\mathcal{G}}^k(w)+\ldots,\label{WZWcurrentOPE}\end{align} where the integer $k$, known as the level of the WZW algebra, is 1 unless specified otherwise, $f^{ijk}$ are the structure constants of the Lie algebra of $\mathcal{G}$, and $z=e^{2\pi(\tilde{v}\tau+i\mathsf{x})/l}$ (similarly for $w$) is the complex spacetime coordinate along the edge. 
Since $\mathcal{G}$ sits inside $E_8$, the filling number and central charge that determine the electric and thermal transport \eqref{electricthermalresponse} of the corresponding bFQH state must be less than or equal to the $E_8$ values: 
\begin{align}
\nu\leq\nu_{E_8}=16,\quad c\leq c_{E_8}=8.
\end{align} 
Moreover, we restrict our focus to subalgebras $\mathcal{G}$ whose ``particle-hole conjugate" coset $E_8/\mathcal{G}$ is also a subalgebra of $E_8$. 
In other words, the edge WZW theory $\mathcal{G}$ of the bFQH state must be one of the two components of a bipartite conformal embedding into $E_8$, 
\begin{align}\mathcal{G}_A\times\mathcal{G}_B\subseteq E_8.
\label{E8bipartition}
\end{align} 
This implies the energy momentum tensors combine as $T_{\mathcal{G}_A}+T_{\mathcal{G}_B}=T_{E_8}$, and the two sectors decouple from each other so that the OPE of currents belonging to the $A$ and $B$ subalgebras $J_A(z)J_B(w)$ are non-singular. 
Eq.~\eqref{E8bipartition} implies the particle-hole pair have conjugate electrical and thermal transport: 
\begin{align}\nu_A+\nu_B=16,\quad c_A+c_B=8.
\end{align} 
These relations generalize those in the lowest Landau level of electrons, $\nu_X+\nu_{\bar{X}}=1$ and $c_X+c_{\bar{X}}=1$, where $X$ is a fFQH state, $\bar{X}$ is its particle-hole conjugate, and the filling number and central charge of the filled lowest Landau level are $\nu_{\mathrm{LLL}}=c_{\mathrm{LLL}}=1$. 

In this paper, we consider the following (bipartite) conformal embeddings of $E_8$ at level 1:
\begin{align}
\label{E8embeddings}
\mathcal{G}_A\times\mathcal{G}_B =
    \begin{cases}
    & SU(3)\times E_6, \\
    & SU(2)\times E_7,\\
    & SU(5)\times SU(5),\\
    & SO(M)\times SO(16-M), \\
    & G_2\times F_4,\\
    & U(1)_8\times SU(8), \\
    & SU(2)_4\times Sp(8),
    \end{cases}
\end{align}
for $M = 1, \ldots, 8$, where all the KM algebras are level 1 except the two orbifold theories, $U(1)_8=SU(2)_1/\mathbb{Z}_2$ and $SU(2)_4=SU(3)_1/\mathbb{Z}_2$. 
The exceptional $E_6$ and $E_7$ algebras at level 1 and their particle-hole conjugates, $SU(3)_1$ and $SU(2)_1$, the $D$-series $D_r=SO(2r)$ at level 1 for $r=1,\ldots,7$, and the orbifold theory $SU(8)_1=(E_7)_1/\mathbb{Z}_2$ and its particle-hole conjugate $U(1)_8=SU(2)_1/\mathbb{Z}_2$ are Abelian states, where all anyonic excitations have quantum dimension $d=1$ and exhibit single-channel fusion rules. 
The $B$-series $B_r=SO(2r+1)$ at level 1, for $r=1,\ldots,6$, have non-Abelian Ising topological orders and support Ising anyons~\cite{NayakWilczek96} $\sigma$ with quantum dimension $d_\sigma=\sqrt{2}$ and the two-channel fusion rule $\sigma\times\sigma=1+\psi$. 
The $Sp(8)_1=(E_6)_1/\mathbb{Z}_2$ orbifold theory and its particle-hole conjugate $SU(2)_4=SU(3)_1/\mathbb{Z}_2$ have non-Abelian topological order and host metaplectic anyons \cite{PhysRevB.87.165421} $\Sigma$ with quantum dimension $d_\Sigma=\sqrt{3}$ and the multi-channel fusion rule $\Sigma\times\Sigma=1+E+E^2$. 
The exceptional $G_2$ and $F_4$ algebras at level 1 host non-Abelian Fibonacci anyons~\cite{SlingerlandBais01}, whose fusion rule $\tau\times\tau=1+\tau$ and non-Abelian braiding allow for universal topological quantum computation \cite{nayak2008non}.

In \S\ref{sec:E8review}, we reviewed the coupled-wire construction of the $E_8$ state, based on a 2d array of wires each carrying 11 Dirac electron channels with kinetic Hamiltonian $\mathcal{H}_0$ in \eqref{H0}. 
The fermion gap $E_g^1$ was introduced by the intra-wire many-body backscattering $\mathcal{H}^f_{\mathrm{intra}}$ in \eqref{E8Hintra}, while the bulk excitation energy gap was introduced by the inter-wire $E_8$ current backscattering $\mathcal{H}_{\mathrm{inter}}$ in \eqref{E8Hinter}. 
Given a conformal embedding $\mathcal{G}_A\times\mathcal{G}_B\subseteq E_8$, the $\mathcal{G}_A$ bFQH state can be constructed using the Hamiltonian, \begin{align}\mathcal{H}[\mathcal{G}_A]&=\mathcal{H}_0+\mathcal{H}^f_{\mathrm{intra}}+\mathcal{H}^B_{\mathrm{intra}}+\mathcal{H}^A_{\mathrm{inter}},\label{HA}\end{align} that only back-scatters the $\mathcal{G}_A$ currents between adjacent wires by \begin{align}\mathcal{H}^A_{\mathrm{inter}}&=u_{\mathrm{inter}}\sum_y{\bf J}_{y,A}^R\cdot{\bf J}_{y+1,A}^L,\label{HinterA}\end{align} and backscatters the $\mathcal{G}_B$ currents within each wire by \begin{align}\mathcal{H}^B_{\mathrm{intra}}&=u_{\mathrm{intra}}\sum_y{\bf J}_{y,B}^R\cdot{\bf J}_{y,B}^L.\label{HintraB}\end{align} 
In an open cylinder geometry of $\mathsf{L}$ wires, while \eqref{HintraB} sums over all wires, the inter-wire interaction sum in \eqref{HinterA} only involves $y=1,\ldots,\mathsf{L}-1$. 
The model gaps all bulk excitations, but leaves behind chiral $\mathcal{G}_A$ WZW CFTs
on the boundary edges. The particle-hole conjugate $\mathcal{G}_B$ phase can be constructed similarly by exchanging the $A$ and $B$ sectors.

In the following, we construct (i) the conformal embeddings in \eqref{E8embeddings} and present the (ii) the charge assignments of the KM currents and primary fields of the respective WZW theories. 
Similar to the $E_8$ state, the coupled wire models are exactly-solvable when the backscattering interactions preserve momentum conservation. 
In each model studied below, we detail (iii) the the electron Fermi momenta that ensures momentum along the wire is conserved and connect the edge electric transport $\sigma_{xy}=\nu e^2/h$ to the bulk electron filling number $\nu=N_e/N_B$. 
We begin with the $SU(3)\times E_6$ decomposition of $E_8$; this construction is then straightforwardly generalized to all other the Abelian states in \eqref{E8embeddings}. 
After this, we then construct non-Abelian bFQH states of $E_8$ with Ising, Fibonacci, and metaplectic topological orders.

\subsection{Abelian states}\label{sec:AbelianStates}

\subsubsection{The Abelian \texorpdfstring{$\mathbb{Z}_3$}{Z3} \texorpdfstring{$SU(3)$}{SU(3)} and \texorpdfstring{$E_6$}{E6} states}\label{sec:SU3E6}

We begin with the $\mathcal{G}_A\times\mathcal{G}_B=SU(3)\times E_6$ decomposition of the $E_8$ WZW algebra at level 1. We present in detail the inequivalent $SU(3)\times E_6$ conformal embeddings in $E_8$ and the corresponding $SU(3)_1$ and $(E_6)_1$ bosonic fractional quantum Hall states. This decomposition demonstrates most features that appear in general Abelian bipartitions $\mathcal{G}_A\times\mathcal{G}_B\subseteq E_8$ using simpy-laced WZW algebras $\mathcal{G}_{A/B}$. We rely on the Cartan-Weyl representation of $E_8$ using the eight non-local Dirac fermions $d^\sigma_{yj}(\mathsf{x})=e^{i\phi^\sigma_{yj}(\mathsf{x})+ik^\sigma_{yj}\mathsf{x}}$, for $j=1,\ldots,8$, defined in \eqref{nonlocaldiracdef}. Recall the chiral $E_8$ algebra is generated by the eight Cartan generators $\partial_{\mathsf{x}}\phi_j$ in \eqref{E8Cartangenerators} and the 240 roots $e^{i\alpha^j(\phi_j(\mathsf{x})+k_j\mathsf{x})}$ in \eqref{E8rootsCartanWeyl}. (The wire index $y$ and chiral sector label $\sigma=R,L$ are fixed and suppressed.) The $E_8$ root vectors $\boldsymbol\alpha=(\alpha^1,\ldots,\alpha^8)$ all have length $|\boldsymbol\alpha|=\sqrt{2}$. The set of $E_8$ root vectors $\Delta_{E_8}$ consists of 112 $SO(16)$ root vectors, where $\alpha^j=0,\pm1$, and 128 $SO(16)$ even spinors, where $\alpha^j=\varepsilon^j/2=\pm1/2$ with $\prod_{j=1}^8\varepsilon^j=+1$. An $SU(3)\times E_6$ embedding in $E_8$ is a particular assignment of $SU(3)$ and $E_6$ current operators inside the $E_8$ algebra. 

\begin{figure}[htbp]
\includegraphics[width=.47\textwidth]{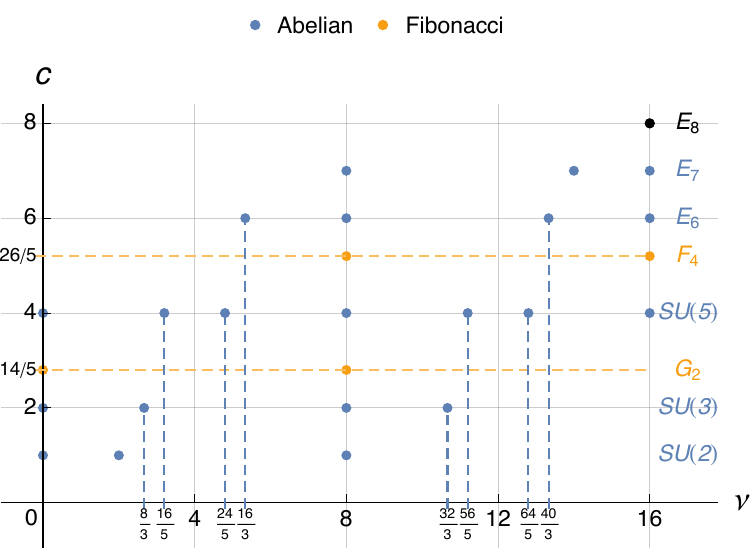}
\caption{The filling numbers $\nu$ and central charges $c$ of bosonic fractional quantum Hall (bFQH) states involving the exceptional Lie algebras $G_2$, $F_4$, $E_{6,7,8}$ as well as $A_4=SU(5)$. 
}
\label{fig:exceptionclassesstates}\end{figure}

We start by choosing a particular subset of $E_8$ current operators that generate an $SU(3)$ subalgebra. 
The $SU(3)$ Lie algebra has rank 2 and dimension 8. It is generated by 2 Cartan generators and 6 roots. It suffices to choose, from the $E_8$ root lattice, 2 simple root vectors $\boldsymbol\alpha_1$ and $\boldsymbol\alpha_2$ of $SU(3)$, so that the inner products $K_{IJ}=\boldsymbol\alpha_I\cdot\boldsymbol\alpha_J$ are the entries of the $SU(3)$ Cartan matrix, \begin{align}K_{SU(3)}=\left(K_{IJ}\right)_{2\times2}=\begin{pmatrix}2&-1\\-1&2\end{pmatrix}.\label{eq:CartanMatrixSU3}\end{align} 
Roots $\boldsymbol \alpha_1$ and $\boldsymbol \alpha_2$ generate the $SU(3)$ root lattice $\Delta_{SU(3)}$, which consists of the 6 root vectors $\boldsymbol\alpha_{SU(3)}=\pm\boldsymbol\alpha_1$, $\pm\boldsymbol\alpha_2$ and $\pm(\boldsymbol\alpha_1+\boldsymbol\alpha_2)$. It can be shown from properties of a root system that each of these roots is an $E_8$ root vector. For example, $\boldsymbol\alpha_1+\boldsymbol\alpha_2=\boldsymbol\alpha_1-2\left(\frac{\boldsymbol\alpha_2\cdot\boldsymbol\alpha_1}{\boldsymbol\alpha_2\cdot\boldsymbol\alpha_2}\right)\boldsymbol\alpha_2$ is the reflection of $\boldsymbol\alpha_1$ on the hyperplane normal to $\boldsymbol\alpha_2$. 
It lies in the $E_8$ root lattice because a root system, in general, is invariant under the reflection about the hyperplane normal to any root vector. 

In general, the 6 root operators of $SU(3)$ are the vertex operators $\left[\mathtt{E}_{SU(3)}(\mathsf{x})\right]_{\boldsymbol\alpha_{SU(3)}}=e^{i\alpha_{SU(3)}^j(\phi_j(\mathsf{x})+k_j\mathsf{x})}$, corresponding to the 6 $SU(3)$ root vectors. 
The 2 Cartan generators of $SU(3)$ are $\left[\mathtt{H}_{SU(3)}(\mathsf{x})\right]_a=\beta_a^j\partial_{\mathsf{x}}\phi_j$, for $a=1,2$, where $\boldsymbol\beta_1,\boldsymbol\beta_2$ is an orthonormal basis of the space spanned by $\boldsymbol\alpha_1,\boldsymbol\alpha_2$.
We take $\boldsymbol\beta_1=\boldsymbol\alpha_1/\sqrt{2}$ and $\boldsymbol\beta_2=(\boldsymbol\alpha_1+2\boldsymbol\alpha_2)/\sqrt{6}$. The Cartan generators and the real and imaginary parts of the root operators are the current operators $J_{SU(3)}$ that generate the (real) $SU(3)$ WZW algebra at level 1 and obey the OPE \eqref{WZWcurrentOPE}.

We choose the 2 simple root vectors of $SU(3)$ to be \begin{align}\boldsymbol\alpha_1={\bf e}_1-{\bf e}_2,\quad\boldsymbol\alpha_2={\bf e}_2-{\bf e}_3.\label{SU3simpleroots}\end{align} 
These simple roots define the rows of a $2\times8$ matrix $A_{SU(3)}$. 
The root operators of $SU(3)$ are then \begin{align}\left[\mathtt{E}_{SU(3)}(\mathsf{x})\right]_{\boldsymbol\alpha_{SU(3)}}=e^{i(\phi_a(\mathsf{x})-\phi_b(\mathsf{x}))+i(k_a-k_b)\mathsf{x}},
\label{SU3roots}\end{align} where $\boldsymbol\alpha_{SU(3)}={\bf e}_a-{\bf e}_b$ are the 6 $SU(3)$ roots, and $a,b$ are distinct integers that range from 1 to 3. Given the simple roots \eqref{SU3simpleroots}, the Cartan generators are \begin{align}\begin{split}\left[\mathtt{H}_{SU(3)}(\mathsf{x})\right]_1&=\frac{\partial_{\mathsf{x}}\phi_1-\partial_{\mathsf{x}}\phi_2}{\sqrt{2}},\\\left[\mathtt{H}_{SU(3)}(\mathsf{x})\right]_2&=\frac{\partial_{\mathsf{x}}\phi_1+\partial_{\mathsf{x}}\phi_2-2\partial_{\mathsf{x}}\phi_3}{\sqrt{6}}.\end{split}\label{SU3Cartangenerators}\end{align} 
It can be shown that any alternative choice of simple root vectors $\boldsymbol\alpha'_{1,2}$ of $SU(3)$ inside the $E_8$ root lattice is related to the one above, $\boldsymbol\alpha'_J=w\boldsymbol\alpha_J$, by a rotation or reflection $w$ inside the automorphism group of the $E_8$ root lattice $\Delta_{E_8}$, \begin{align}\mathrm{Aut}(E_8)=\left\{w\in O(8):w(\Delta_{E_8})=\Delta_{E_8}\right\}.\label{E8Weylgroup}\end{align} 
This automorphism group is identical to the Weyl group \cite{conway2013sphere}, $W(E_8)$ of $E_8$ that is generated by reflections about hyperplanes perpendicular to the $E_8$ root vectors. 
Consequently, all $SU(3)$ root embeddings $\Delta_{SU(3)}\subseteq\Delta_{E_8}$ are equivalent up to the Weyl symmetry. 

The $E_6$ sector is the orthogonal complement of $SU(3)$ in $E_8$. The $E_6$ Lie algebra has rank 6 and dimension 78. It is generated by 6 Cartan generators and 72 roots. The Cartan generators $\left[\mathtt{H}_{E_6}(\mathsf{x})\right]_b=\gamma_b^j\partial_{\mathsf{x}}\phi_j$, for $b=1,\ldots,6$, can be chosen using an orthonormal basis $\boldsymbol\gamma_1,\ldots,\boldsymbol\gamma_6$ of the subspace perpendicular to the $SU(3)$ root system $\Delta_{SU(3)}$. 
The $E_6$ roots, \begin{align}\left[\mathtt{E}_{E_6}(\mathsf{x})\right]_{\boldsymbol\alpha_{E_6}}=e^{i\alpha_{E_6}^j(\phi_j(\mathsf{x})+k_j\mathsf{x})},
\label{E6roots}\end{align} are the $E_8$ roots whose root vectors are perpendicular to those in $SU(3)$: $\boldsymbol\alpha_{E_6}\cdot\boldsymbol\alpha_{SU(3)} = 0$. Specifically, the set of $E_6$ root vectors $\Delta_{E_6}$ consists of (i) 40 integral vectors of the form $\boldsymbol\alpha_{E_6}=(0,0,0,\alpha^4,\ldots,\alpha^8)$, where two of $\alpha^{4,\ldots,8}$ are $\pm1$ and the rest are 0, and (ii) 32 half integral vectors $\boldsymbol\alpha_{E_6}=(\varepsilon^1,\ldots,\varepsilon^8)/2$ where $\varepsilon^j=\pm1$, $\varepsilon^1=\varepsilon^2=\varepsilon^3$ and $\prod_{j=1}^8\varepsilon^j=+1$. The $E_6$ root system $\Delta_{E_6}$ can be generated by the 6 simple root vectors $\boldsymbol\alpha_{J=1,\ldots,6}$ that are the rows of the following matrix, 
\begin{align}
A_{E_6} = 
\begin{pmatrix}
 &  &  & 1 & -1 &  &  &  \\
  &  &  &  & 1 & -1 &  &  \\
  &  &  &  &  & 1 & -1 &  \\
  &  &  &  &  &  & 1 & -1 \\
 \frac{1}{2} & \frac{1}{2} & \frac{1}{2} & -\frac{1}{2} &
   -\frac{1}{2} & -\frac{1}{2} & -\frac{1}{2} & \frac{1}{2} \\
  &  &  &  &  &  & 1 & 1 
\end{pmatrix}.
\label{E6simpleroots}
\end{align} 
The inner products $K_{IJ}=\boldsymbol\alpha_I\cdot\boldsymbol\alpha_J$ are identical to the entries of the Cartan matrix of $E_6$ \begin{align}
K_{E_6} = A_{E_6}A_{E_6}^T=
\begin{pmatrix}
2 & -1 & & & &\\
 -1 & 2 & -1 & & &\\
 & -1 & 2 & -1 & & -1 \\
 & & -1 & 2 & -1 &\\
 & & & -1 & 2 &\\
 & & -1 &  &  & 2
\end{pmatrix}.
\label{E6CartanMatrix}
\end{align} The Cartan generators $[\mathtt{H}_{E_6}]_{b=1,\ldots,6}$ and the real and imaginary parts of the root operators $[\mathtt{E}_{E_6}]_{\boldsymbol\alpha_{E_6}}$ are the current operators $J_{E_6}$ that generate the (real) $E_6$ WZW algebra at level 1 and obey the OPE \eqref{WZWcurrentOPE}. Since the $SU(3)$ and $E_6$ root systems are orthogonal, the current operators have non-singular mutual OPE $J_{SU(3)}(z)J_{E_6}(w)$ and the current modes mutually commute. Because the $SU(3)$ ($E_6$) theory is generated by 2 (resp.~6) independent bosonized variables, their central charges are $c_{SU(3)_1}=2$ and $c_{(E_6)_1}=6$. 
Not only do the central charges add up to $c_{(E_8)_1}=8$, the Sugawara energy-momentum tensors obey $T_{SU(3)_1}+T_{(E_6)_1}=T_{(E_8)_1}$. 
Equivalently, the two CFTs are $(E_8)_1$ coset duals of each other, \begin{align}SU(3)_1=\frac{(E_8)_1}{(E_6)_1},\quad (E_6)_1=\frac{(E_8)_1}{SU(3)_1}.\label{SU3E6cosets}\end{align} Moreover, as all $SU(3)$ root embeddings $\Delta_{SU(3)}\subseteq\Delta_{E_8}$ are equivalent up to the Weyl symmetry \eqref{E8Weylgroup}, the same equivalence holds for the $E_6$ root embeddings $\Delta_{E_6}=\Delta_{SU(3)}^\perp$.

Having completed the definition of the conformal embedding $SU(3)\times E_6\subseteq E_8$, the $SU(3)$ and $E_6$ level 1 quantum Hall states can be constructed using the coupled-wire Hamiltonians in \eqref{HA}. For example, the $SU(3)_1$ model Hamiltonian $\mathcal{H}[SU(3)]$ contains the inter-wire (intra-wire) backscattering of $SU(3)$ (resp.~$E_6$) currents: \begin{widetext}\begin{align}\begin{split}\mathcal{H}^{SU(3)}_{\mathrm{inter}}&=u_{\mathrm{inter}}\sum_y\left[\sum_{a=1}^2{\left[\mathtt{H}_{SU(3)}\right]^R_{y,a}}^\dagger\left[\mathtt{H}_{SU(3)}\right]^L_{y+1,a}-\sum_{\boldsymbol\alpha_{SU(3)}}\cos\left(\boldsymbol\alpha_{SU(3)}\cdot\boldsymbol\theta_{y+1/2}\right)\right],\\\mathcal{H}^{E_6}_{\mathrm{intra}}&=u_{\mathrm{intra}}\sum_y\left[\sum_{b=1}^6{\left[\mathtt{H}_{E_6}\right]^R_{y,b}}^\dagger\left[\mathtt{H}_{E_6}\right]^L_{y,b}-\sum_{\boldsymbol\alpha_{E_6}}\cos\left(\boldsymbol\alpha_{E_6}\cdot\boldsymbol\theta_y\right)\right],\end{split}\label{SU3E6sineGordon}\end{align}\end{widetext} where the entries of $\boldsymbol\theta_{y+1/2}=(\theta_{y+1/2,1},\ldots,\theta_{y+1/2,8})$ and $\boldsymbol\theta_y=(\theta_{y,1},\ldots,\theta_{y,8})$ are the inter-wire and intra-wire sine-Gordon variables $\theta_{y+1/2,j}=\phi^R_{y,j}-\phi^L_{y+1,j}$ and $\theta_{y,j}=\phi^R_{y,j}-\phi^L_{y,j}$, respectively. With the density interactions in $\mathcal{H}_0+\mathcal{H}^f_{\mathrm{intra}}$ that give rise to the conformal $E_8$ fixed point, the current backscattering interactions in \eqref{SU3E6sineGordon} are marginally relevant in the RG sense if $u_{\mathrm{inter}/\mathrm{intra}}>0$. The sums of roots $\boldsymbol\alpha_{SU(3)}$ and $\boldsymbol\alpha_{E_6}$ in the sine-Gordon potentials in \eqref{SU3E6sineGordon} can be restricted to only include the simple roots $\boldsymbol\alpha^{SU(3)}_{1,2}$ in \eqref{SU3simpleroots} for the $SU(3)$ sector and $\boldsymbol\alpha^{E_6}_{J=1,\ldots,6}$ in \eqref{E6simpleroots} for the $E_6$ sector. This restricted set of sine-Gordon potentials pins the ground state expectation values of the angle variables $\Theta^{SU(3)}_{y+1/2,I=1,2}=\boldsymbol\alpha^{SU(3)}_I\cdot\boldsymbol\theta_{y+1/2}$ and $\Theta^{E_6}_{y,J=1,\ldots,6}=\boldsymbol\alpha^{E_6}_J\cdot\boldsymbol\theta_y$ to lie at an integral multiple of $2\pi$. Since all roots are integral combinations of simple ones, the other sine-Gordon potentials in \eqref{SU3E6sineGordon} with non-simple roots do not compete, as they share the same minima. On a closed torus geometry, the angle variables $\Theta^{SU(3)}_{y+1/2,I=1,2}$ and $\Theta^{E_6}_{y,J=1,\ldots,6}$ form a maximal set of independent commuting bosonized operators and therefore the sine-Gordon potentials in \eqref{SU3E6sineGordon} introduce a finite bulk excitation energy gap. 

Unlike the $E_8$ quantum Hall state, $\mathcal{H}[SU(3)]$ has three degenerate ground states on a torus. To see this, we recall from \S\ref{sec:E8review} that the bosonized variables are compactified and identified by large gauge transformations $\phi^\sigma_{y,j}\equiv\phi^\sigma_{y,j}+2\pi r^\sigma_{y,j}$, where ${\bf r}^\sigma_y=(r^\sigma_{y,1},\ldots,r^\sigma_{y,8})$ is a vector inside the lattice $\mathcal{R}=\mathrm{span}_{\mathbb{Z}}(\Delta_{E_8})$ generated by the $E_8$ roots. 
Up to large gauge transformations, all ground state expectation values $\left\langle\Theta^{E_6}_{y,J=1,\ldots,6}\right\rangle$ that minimize $\mathcal{H}^{E_6}_{\mathrm{intra}}$ are equivalent. On the other hand, there are three gauge inequivalent sets of ground state expectation values $\left\langle\Theta^{SU(3)}_{y+1/2,I=1,2}\right\rangle$ that minimize $\mathcal{H}^{SU(3)}_{\mathrm{inter}}$. They correspond to the three anyon classes of bulk quasiparticle excitations, \begin{align}\mathbb{Z}_3=\left\{1,\mathcal{E},\overline{\mathcal{E}}\right\},\end{align} and the three primary fields in the chiral edge CFTs when the model is defined on an open cylinder with boundaries. Non-trivial primary fields and anyons $\mathcal{E}$ and $\overline{\mathcal{E}}$ cannot appear alone and must come in conjugate pairs $\mathcal{E}\times\overline{\mathcal{E}}=1$, triplets $\mathcal{E}\times\mathcal{E}\times\mathcal{E}=\overline{\mathcal{E}}\times\overline{\mathcal{E}}\times\overline{\mathcal{E}}=1$, or any multiplets that fuse to the trivial sector 1. On the chiral edge, each non-trivial primary field $\mathcal{E}$ is a super-selection sector of vertex operators $\mathcal{E}^p$ that rotate irreducibly under the $SU(3)$ WZW algebra. Specifically, $\mathcal{E}$ is spanned by three non-local fields \begin{align}\mathcal{E}=\mathrm{span}\left\{
\resizebox{6.65cm}{!}{$
    e^{i(\phi_1+\phi_2-2\phi_3)/3},  e^{i(\phi_1-2\phi_2+\phi_3)/3}, e^{i(-2\phi_1+\phi_2+\phi_3)/3}
   $}
\right\}.\label{SU3primary}\end{align} Fields inside the same super-selection sector obey the current OPE \begin{align}J_g(z)\mathcal{E}^p(w)=\frac{\rho(g)^p_q}{z-w}\mathcal{E}^q(w)+\ldots,\label{currentirrep}\end{align} where $g$ is a Lie algebra element, $J_g(z)$ is its corresponding WZW current operator, and $\rho(g)$ is a matrix representation of $g$. For example, the $SU(3)$ root operator $[\mathtt{E}_{SU(3)}(z)]_{\boldsymbol\alpha_1}=e^{i(\phi_1(z)-\phi_2(z))}$ rotates $e^{i{\bf m}\cdot\boldsymbol\phi}=e^{i(-2\phi_1+\phi_2+\phi_3)/3}$ into \begin{align}\begin{split}&\left[\mathtt{E}_{SU(3)}(z)\right]_{\boldsymbol\alpha_1}e^{i{\bf m}\cdot\boldsymbol\phi(w)}\\&=e^{i(\boldsymbol\alpha_1+{\bf m})\cdot\boldsymbol\phi(w)-\langle\boldsymbol\alpha_1\cdot\boldsymbol\phi(z),{\bf m}\cdot\phi(w)\rangle+\ldots}\\&=e^{i(\phi_1(w)-2\phi_2(w)+\phi_3(w))/3-\log(z-w)+\ldots}\\&\propto\frac{1}{z-w}e^{i(\phi_1(w)-2\phi_2(w)+\phi_3(w))/3}+\ldots\end{split}\end{align} (up to non-singular terms and oscillations factors involving $e^{ik_f\mathsf{x}}$) using the correlation $\langle\phi_i(z)\phi_j(w)\rangle=-\delta_{ij}\log(z-w)+\ldots$ (c.f.~\eqref{E8bosonizedKMalgebra}). The exact constant phase in ``$\propto$" depends on the non-singular pieces in the above correlator that guarantee the fermion fields $d^j=e^{i\phi_j}$ mutually anti-commute. These phases are unimportant in this paper and will be omitted. The primary super-sector $\mathcal{E}$ transforms under the fundamental 3D irreducible representation of $SU(3)$, while its conjugate sector $\overline{\mathcal{E}}=\mathcal{E}^\dagger$ transforms under the conjugate representation. They both carry fractional spin (i.e., conformal scaling dimensions) $h_{\mathcal{E}}=h_{\overline{\mathcal{E}}}=1/3$. Fields within the same primary sector differ from each other by the local $SU(3)$ currents, which are integral combinations of electrons. On the other hand, fields belonging to different primary sectors are not related by any local integral product of electrons.

The particle-hole conjugate coupled-wire model $\mathcal{H}[E_6]$ can be constructed by reversing the role of $SU(3)$ and $E_6$ so that the former (latter) is gapped by an intra-wire (inter-wire) backscattering interaction. The $E_6$ quantum Hall state also has three degenerate ground states on a closed torus geometry. They correspond to the three anyon classes of bulk quasiparticle excitations and primary field super-selection sectors $1,\mathcal{E},\overline{\mathcal{E}}$ on the chiral edge CFT. Specifically, the primary super-sector $\mathcal{E}$ is spanned by 27 non-local fields \begin{align}\begin{split}\mathcal{E}&=\mathrm{span}\left\{e^{i2(\phi_1+\phi_2+\phi_3)/3}\right\}\\&\quad\cup\left\{e^{i(\phi_1+\phi_2+\phi_3)/6+\sum_{j=4}^8\varepsilon^j\phi_j/2}\right\}_{\varepsilon^{j=4,\ldots,8}=\pm1,\prod_{j=4}^8\varepsilon^j=-1}\\&\quad\cup\left\{e^{-i(\phi_1+\phi_2+\phi_3)/3\pm\phi_j}\right\}_{j=4,\ldots,8}.\end{split}\label{E6primary}\end{align} These fields transform irreducibly under the OPE of the $E_6$ current algebra (see \eqref{currentirrep}). The conjugate primary super-sector $\overline{\mathcal{E}}=\mathcal{E}^\dagger$ forms another irreducible representation. They both carry spin $h_{\mathcal{E}}=h_{\overline{\mathcal{E}}}=2/3$. The product of primary fields $\mathcal{E}_{SU(3)}\times\mathcal{E}_{E_6}$ and $\overline{\mathcal{E}}_{SU(3)}\times\overline{\mathcal{E}}_{E_6}$ generate the remaining 162 $E_8$ roots in the complement $\Delta_{E_8}\backslash(\Delta_{SU(3)}\cup\Delta_{E_6})$. Therefore, if both $SU(3)$ and $E_6$ were gapped by inter-wire backscattering interactions, their anyons would pairwise condense as local bosons and the resulting quantum Hall phase would be identical to the topologically trivial $E_8$ state.

We complete the construction of the $SU(3)$ and $E_6$ models by presenting their charge fractionalization and the momentum arrangement of electrons on each wire. 
For this, there are two issues to address in our coupled-wire construction. 
First, the $R$ matrix used in \eqref{diracRmatrix}---that enables the non-local Dirac fermion presentation of $E_8$ in \eqref{nonlocaldiracdef} by providing the transformation $\widetilde{\boldsymbol\Phi}=R\boldsymbol\phi$ between the Chevalley and Cartan-Weyl basis---has not been specified. Second, the root operators in \eqref{SU3roots} and \eqref{E6roots} carry oscillation factors $e^{i\alpha^jk_j\mathsf{x}}$. Like the $E_8$ model, the exact cancellation of these factors in the sine-Gordon interactions in \eqref{SU3E6sineGordon} requires a specific arrangement of electron Fermi momentum.

The construction of the $E_8$ quantum Hall state in \S\ref{sec:E8review} did not depend on the choice of the non-local Dirac fermions $d^j=e^{i\phi_j}$ presented in \eqref{nonlocaldiracdef}. This is because the $E_8$ current operators in \eqref{E8CartangeneratorsChevalley} and \eqref{E8rootsChevalley}, and in particular, the simple roots $e^{i\tilde\Phi_I}$ in \eqref{E8simpleroots} are fixed by their electronic origin. Here, however, the $SU(3)$ and $E_6$ models are constructed using a preferred set of simple roots in \eqref{SU3simpleroots} and \eqref{E6simpleroots} based on a particular choice of non-local Dirac fermions, or equivalently, their Cartan-Weyl bosonized variables $\phi_j$. While different choices are related by symmetries in the $E_8$ Weyl group \eqref{E8Weylgroup}, they lead to distinct bFQH states with inequivalent charge fractionalizations. This is because the fixed electric charges $\tilde{q}_I$ in \eqref{E8simlerootscharge} of the $E_8$ simple roots $e^{i\tilde\Phi_I}$ are not preserved by all $E_8$ Weyl symmetries. Given a specific choice of the non-local Dirac fermions $d_j$, their electric charges $q_j$ define a charge vector ${\bf q}=(q_1,\ldots,q_8)$. The collection of charge vectors for the various choices of $d_j$ was exhaustively presented above equation \eqref{E8fillingnumber}. Among them, they generate four inequivalent sets of $SU(3)$ and $E_6$ bFQH states with distinct filling numbers $\nu$ and quasiparticle charge assignments. 

In general, electric charge splits in the $SU(3)\times E_6\subseteq E_8$ decomposition. From the electric response on the edge, $\sigma=\nu e^2/h$, the filling numbers of the $SU(3)$ and $E_6$ bFQH states can be read off from the length of the charge vector when projecting it onto the $SU(3)$ and $E_6$ subspaces (c.f.~\eqref{E8fillingnumber} for the $E_8$ state): \begin{align}\begin{split}\nu_{SU(3)}&=\left|P_{SU(3)}{\bf q}\right|^2={\bf q}^{SU(3)}\cdot K_{SU(3)}^{-1}{\bf q}^{SU(3)},\\\nu_{E_6}&=\left|P_{E_6}{\bf q}\right|^2={\bf q}^{E_6}\cdot K_{E_6}^{-1}{\bf q}^{E_6}.\end{split}\label{SU3E6filling}\end{align} Here, $P_{SU(3)}$ and $P_{E_6}=\mathbb{1}_8-P_{SU(3)}$ are the projection matrices (each obeying $P^2=P$) onto subspaces spanned by the $SU(3)$ and $E_6$ roots, respectively: \begin{align}\begin{split}P_{SU(3)}&=A_{SU(3)}^TK^{-1}_{SU(3)}A_{SU(3)},\\P_{E_6}&=A_{E_6}^TK^{-1}_{E_6}A_{E_6},\label{PSU3PE6}\end{split}\end{align} where the rows of $A_{SU(3)}$ and $A_{E_6}$ are the the simple roots in the Euclidean 8-space chosen in \eqref{SU3simpleroots} and \eqref{E6simpleroots}. The electric charge of the simple roots of the two algebras are the entries of the charge vectors ${\bf q}^{SU(3)}=(q^{SU(3)}_{M=1,2})=A_{SU(3)}{\bf q}$ and ${\bf q}^{E_6}=(q^{E_6}_{N=1,\ldots,6})=A_{E_6}{\bf q}$. Since the $SU(3)$ and $E_6$ roots are orthogonal, $P_{SU(3)}{\bf q}\perp P_{E_6}{\bf q}$. Thus, the filling numbers in \eqref{SU3E6filling} obey the particle-hole conjugation relation: \begin{align}\nu_{SU(3)}+\nu_{E_6}=|{\bf q}|^2=\nu_{E_8}=16.\end{align}

Let's see specifically how this works for the charge vector ${\bf q}=(4,0,0,0,0,0,0,0)$ of a set of non-local Dirac fermions chosen using the $R$ matrix in \eqref{eq:newRmatrixwith4}. In this case, the bFQH models constructed have filling numbers $\nu_{SU(3)}=32/3$ and $\nu_{E_6}=16/3$. The charge vector ${\bf q}$ also dictates the charge assignments of a vertex field operator, $Q(e^{i{\bf m}\cdot\boldsymbol\phi})={\bf m}\cdot{\bf q}$. This decomposes the $SU(3)$ and $E_6$ WZW current algebras and their primary field super-selection sectors into subspaces with different electric charges. For example, with the same charge vector as above, the eight-dimensional $SU(3)$ WZW algebra---which is spanned by the two Cartan generators \eqref{SU3Cartangenerators} and the six roots \eqref{SU3roots}---splits into $8=4(0)\oplus2(+4)\oplus2(-4)$, a four-dimensional neutral subspace, a two-dimensional subspace of charge 2 currents and a two-dimensional subspace of charge $-2$ currents. The three dimensional $SU(3)$ primary sector $\mathcal{E}$ in \eqref{SU3primary} decomposes into $3=2(4/3)\oplus1(-8/3)$, a two-dimensional subspace of charge $4/3$ fields and a one-dimensional subspace of charge $-8/3$ fields. For the particle-hole conjugate $E_6$ state, its 78-dimensional WZW current algebra splits into $78=46(0)\oplus16(+2)\oplus16(-2)$, and its primary sector $\mathcal{E}$ in \eqref{E6primary} decomposes into $27=1(8/3)\oplus16(2/3)\oplus10(-4/3)$. 
As above, the numbers in each term $\ast\ (\star)$ specify the dimension $\ast$ of the subspace consisting of fields with charge $\star$. 

Under an $E_8$ Weyl symmetry, the set of reflected/rotated non-local Dirac fermions obtain a new charge vector. The $SU(3)$ and $E_6$ subalgebras are subsequently reflected/rotated and may lead to different electric responses from before. The four distinct classes of $SU(3)$ and $E_6$ bFQH states can be represented by the following four charge vectors: \begin{align}\begin{split}{\bf q}=\;&(0,0,0,0,0,0,0,4),\quad(4,0,0,0,0,0,0,0),\\&(2,0,0,0,2,2,2,0),\quad(2,-2,0,0,2,2,0,0).\end{split}\end{align} The bFQH coupled-wire models constructed have, respectively, the distinct filling numbers \begin{align}\nu_{SU(3)}=0,\;\frac{32}{3},\;\frac{8}{3},\;8,\quad\nu_{E_6}=16,\;\frac{16}{3},\;\frac{40}{3},\;8,\end{align} 
as summarized in Fig.\ref{fig:exceptionclassesstates} along with central charges of $SU(3)_1$ and $(E_6)_1$.
The charge assignments of the WZW current algebras and primary field super-selection sectors are summarized in Tables~\ref{tab:ExceptionalChargeAssignmentRoot} and \ref{tab:SU3E6SuperselectionsectorChargeAssgn}. The $E_8$ Weyl inversion that flips ${\bf q}\to-{\bf q}$ corresponds to the $\mathbb{Z}_2$ anyonic symmetry~\cite{khan2014,Teotwistdefectreview} (also referred to as outer automorphism) of the topological phases that relabels the anyon classes $\mathcal{E}\leftrightarrow\overline{\mathcal{E}}$. Since $\overline{\mathcal{E}}=\mathcal{E}^\dagger$, the two conjugate classes carry fields with conjugate electric charges. The charge assignment pattern of each primary field sector is invariant (changed) under the $\mathbb{Z}_2$ inversion flip for the bFQH states with integral (fractional) filling numbers. Therefore, the $\mathbb{Z}_2$ symmetry is preserved (resp.~broken) by the electric charge assignment. 
Since switching the anyon labels $\mathcal{E}\leftrightarrow\overline{\mathcal{E}}$ does not alter the fusion and statistics data, the bFQH states constructed from opposite charge vectors are indistinguishable in the absence of other physical symmetries.

Next, we address the momentum conservation in the coupled-wire model. The $SU(3)$ and $E_6$ root operators \eqref{SU3roots} and \eqref{E6roots} carry oscillation factors $e^{i\alpha^jk_j\mathsf{x}}$. The same goes for the integrated fermions $f_{n=1,2,3}$. These factors cancel and do not appear in the sine-Gordong interactions \eqref{SU3E6sineGordon} and \eqref{E8Hintra} when the bare Fermi momenta $k_{F,a}$ of the electron channels (see \eqref{electronmomentum}) take a set of specific values. 
A detailed discussion for how this cancellation can be arranged is presented in Appendix~\ref{app:momentum}. Given any fixed $R$ matrix \eqref{diracRmatrix}, the momentum conserving $SU(3)_1$ model is constructed with bare Fermi momenta, \begin{align}k_{F,a}=\frac{1}{2}\frac{eBd}{\hbar c}\sum_{j,j',J=1}^8\left(U^{++}_{Ja}+U^{+-}_{Ja}\right)\left(R^{-1}\right)^j_JP_{SU(3)}^{jj'}q_{j'},\label{FermimomentaSU3}\end{align} where only the first 8 rows of the $U^{+\pm}$ matrices in \eqref{Umatrix} are summed. (A general equation can be found in \eqref{appmomentumAbelian}.) The $E_6$ model is constructed with a different set of $k_{F,a}$, which are obtained by replacing the projection matrix $P_{SU(3)}\to P_{E_6}$ in \eqref{FermimomentaSU3}. 
Using these $k_{F,a}$, the filling number \eqref{fillingnumber} reproduces the value \eqref{SU3E6filling} predicted from the edge-state response.

\subsubsection{The Abelian \texorpdfstring{$\mathbb{Z}_2$}{Z2} \texorpdfstring{$SU(2)$}{SU(2)} and \texorpdfstring{$E_7$}{E7} states}\label{sec:SU2E7}

We now construct the $SU(2)_1$ and $(E_7)_1$ bFQH states. The construction method is identical to the $SU(3)_1$ and $(E_6)_1$ states presented above. Here, we highlight the essential features and results. First, these bFQH states rely on the $SU(2)\times E_7$ conformal bipartition of $E_8$. The $SU(2)_1$ and $(E_7)_1$ WZW subalgebras can be chosen by fixing two decoupled subsets of current operators inside $(E_8)_1$. Using the non-local Dirac fermions $d_j\sim e^{i\phi_j}$ (see \eqref{nonlocaldiracdef}), we fix the $SU(2)_1$ by choosing its generators \begin{align}\begin{split}\left[\mathtt{H}_{SU(2)}(\mathsf{x})\right]&=\frac{\partial_{\mathsf{x}}\phi_1-\partial_{\mathsf{x}}\phi_2}{\sqrt{2}},\\\left[\mathtt{E}_{SU(2)}(\mathsf{x})\right]^\pm&=e^{\pm i(\phi_1(\mathsf{x})-\phi_2(\mathsf{x})+(k_1-k_2)\mathsf{x})}.\end{split}\label{SU2currents}\end{align} The $SU(2)$ root system $\Delta_{SU(2)}$ contains the positive root $\boldsymbol\alpha_{SU(2)}={\bf e}_1-{\bf e}_2$ and the negative one $-\boldsymbol\alpha_{SU(2)}$. They correspond to the raising and lowering operators $[\mathtt{E}_{SU(2)}]^\pm=e^{\pm i\boldsymbol\alpha_{SU(2)}\cdot\boldsymbol\phi}$. The Cartan matrix of $SU(2)$ is $K_{SU(2)}=\boldsymbol\alpha_{SU(2)}\cdot\boldsymbol\alpha_{SU(2)}=2$. Since the $SU(2)_1$ WZW CFT is generated by a single bosonized variable, it is identical to $U(1)_2$. The $(E_7)_1$ theory is the subalgebra in $(E_8)_1$ that commutes with the $SU(2)_1$. The $E_7$ root system $\Delta_{E_7}$ consists of root vectors in $E_8$ that are orthogonal to $\boldsymbol\alpha_{SU(2)}$. It is the union of the following three sets: (i) the 2 vectors $\pm({\bf e}_1+{\bf e}_2)$, (ii) the 60 integral vectors of the form $(0,0,\alpha^3,\ldots,\alpha^8)$, where two of the  $\alpha^{3,\ldots,8}$ are $\pm1$ and the rest are 0, and (iii) the 64 half-integral vectors $(\varepsilon^1,\ldots,\varepsilon^8)/2$, where $\varepsilon^j=\pm1$, $\varepsilon^1=\varepsilon^2$ and $\prod_{j=1}^8\varepsilon^j=+1$. $\Delta_{E_7}$ can be generated by the set of simple root vectors that form the rows of the following matrix \begin{align}A_{E_7}=\left(\begin{smallmatrix}
&&1&-1&&&&\\
&&&1&-1&&&\\
&&&&1&-1&&\\
&&&&&1&-1&\\
&&&&&&1&1\\
-1/2&-1/2&-1/2&-1/2&-1/2&-1/2&-1/2&-1/2\\
&&&&&&1&-1
\end{smallmatrix}\right)\label{E7simpleroots}\end{align} and produce the Cartan matrix \begin{align}K_{E_7}=A_{E_7}A_{E_7}^T=\left(\begin{smallmatrix}
2&-1&&&&&\\
-1&2&-1&&&&\\
&-1&2&-1&&&\\
&&-1&2&-1&&-1\\
&&&-1&2&-1&\\
&&&&-1&2&\\
&&&-1&&&2
\end{smallmatrix}\right)\end{align} The $(E_7)_1$ WZW subalgebra is spanned by the $126$ root operators, \begin{align}[\mathtt{E}_{E_7}(\mathsf{x})]_{\boldsymbol\alpha_{E_7}}&=e^{i\alpha_{E_7}^j(\phi_j(\mathsf{x})+k_j\mathsf{x})},\quad\boldsymbol\alpha_{E_7}\in\Delta_{E_7}\label{E7roots}\end{align} and the 7 Cartan generators $\left[\mathtt{H}_{E_7}\right]_{b=1,\ldots,7}=\gamma^j_b\partial_{\mathsf{x}}\phi_j$, where $\boldsymbol\gamma_{b=1,\ldots,7}=(\gamma_b^1,\ldots,\gamma_b^8)$ is an orthonormal basis of the $E_7$ root subspace.

The coupled-wire Hamiltonians of the $SU(2)_1$ and $(E_7)_1$ bFQH states are constructed in a similar fasion as the $SU(3)_1$ and $(E_6)_1$ states. The model Hamiltonian \eqref{HA} contains the intra-wire backscattering $\mathcal{H}^f_{\mathrm{intra}}$ of the integrated fermions defined in \eqref{E8Hintra}, and the intra-wire and inter-wire backscattering of the two current algebras (c.f.~\eqref{SU3E6sineGordon} for the $SU(3)_1$ state). For the $SU(2)$ state, the inter-wire backscattering interactions involve the $SU(2)$ currents, whereas the intra-wire ones involve the $E_7$ currents. For the particle-hole conjugate $E_7$ state, the backscattering pattern of the two current algebras are switched. The resulting topological phases have a finite bulk excitation energy gap and chiral gapless boundary edges described by corresponding $SU(2)_1$ or $(E_7)_1$ CFTs. The thermal responses $\kappa_{xy}$ \eqref{electricthermalresponse} of the bFQH states are determined by the central charges $c_{SU(2)_1}=1$ and $c_{(E_7)_1}=7$, which are identical to the ranks of the simply-laced Lie algebras.

The two bFQH states have a $\mathbb{Z}_2=\left\{1,\mathcal{S}\right\}$ topological order and each supports a semion quasiparticle excitation $\mathcal{S}$. It obey the fusion rule $\mathcal{S}\times\mathcal{S}=1$ and therefore is its own anti-partner. ${\cal S}$ has spin $h=1/4$ in the $SU(2)_1$ state or $h=3/4$ in the $(E_7)_1$ state. The semion primary field super-selection sector in $SU(2)_1$ is spanned by the two non-local vertex fields $e^{\pm i\boldsymbol\alpha_{SU(2)}\cdot\boldsymbol\phi/2}$, which together rotate irreducibly under the $SU(2)_1$ algebra (see \eqref{currentirrep}). The semion primary super-sector in $(E_7)_1$ is spanned by 56 non-local fields \begin{align}\begin{split}\mathcal{S}&=\mathrm{span}\left\{e^{\pm i(\phi_1+\phi_2)/2\pm i\phi_j}\right\}_{j=3,\ldots,8}\\&\quad\quad\cup\left\{e^{i\sum_{j=3}^8\varepsilon^j\phi_j/2}\right\}_{\varepsilon^{j=3,\ldots,8}=\pm1,\prod_{j=3}^8\varepsilon^j=-1},\end{split}\end{align} which irreducibly represent $E_7$. Like in the previous $SU(3)\times E_6$ case, if both the $SU(2)$ and $E_7$ currents were gapped by inter-wire backscattering interactions, the $2\times56=112$ semion pairs from $SU(2)\times E_7$ would anyon condense~\cite{PhysRevB.79.045316, 2018ARCMP...9..307B} to form the remaining $E_8$ root currents outside of $SU(2)\times E_7$. The resulting state would have trivial bosonic topological order equivalent to $(E_8)_1$. We notice in passing that semion primary super-sectors $\mathcal{S}$ in $SU(2)_1$ as well as $(E_7)_1$ are closed under the $\mathbb{Z}_2$ involution symmetry $\phi_j\to-\phi_j$ in the Weyl group $\mathrm{Aut}(E_8)$. The involution acts differently on the $SU(3)_1$ and $(E_6)_1$ states where the non-trivial anyon classes are flipped,  $\mathcal{E}\leftrightarrow\overline{\mathcal{E}}$, under the symmetry.

The filling numbers and electric charge assignments of the bFQH states depend on the $R$ matrix that specifies the $E_8$ simple roots $e^{i\tilde\Phi_J}=e^{iR^j_J\phi_j}$ in Euclidean 8-space and defines the non-local Dirac fermions $d_j\sim e^{i\phi_j}$ (see \eqref{nonlocaldiracdef}). The allowed charge vectors ${\bf q}=(q_{j=1,\ldots,8})$ of the fermions were presented above \eqref{E8fillingnumber}. There are three inequivalent classes of the $SU(2)_1$ and $(E_7)_1$ states, where the filling numbers are \begin{align}\nu_{SU(2)}=0,\;2,\;8,\quad\nu_{E_7}=16-\nu_{SU(2)}.\end{align} The three classes can be respectively represented by three particular charge vectors \begin{align}{\bf q}&=4{\bf e}_8,\quad2{\bf e}_1+2{\bf e}_3+2{\bf e}_5+2{\bf e}_7,\quad4{\bf e}_1.\end{align} The electric charge assignments of the WZW currents and the primary fields are summarized in Tables~\ref{tab:ExceptionalChargeAssignmentRoot} and \ref{tab:SU2E7SuperselectionsectorChargeAssgn}.

Lastly, the coupled-wire models are exactly solvable when the bare electron momenta $k_{F,a}$ (see \eqref{electronmomentum}) take a set of specific values. These values are computed in \eqref{appmomentumAbelian} in Appendix~\ref{app:momentum} by applying the projection matrix $P_A=P_{SU(2)}$ or $P_{E_7}$ on the Euclidean 8-space. Small deviations away from these fine-tuned values are perturbations that should not alter the topological phases, assuming they are not strong enough the overcome the bulk excitation energy gap.

\subsubsection{The Abelian \texorpdfstring{$\mathbb{Z}_5$}{Z5} \texorpdfstring{$SU(5)$}{SU(5)} states}\label{sec:SU5}

Next we consider the Abelian $\mathbb{Z}_5$ $SU(5)$ bFQH states.
These bFQH states are based on the conformal embedding of $SU(5)^A\times SU(5)^B$ in $E_8$. Starting with the $A$ sector, the simply-laced Lie algebra is spanned by its $4$ Cartan generators and $20$ root operators
\begin{align}
\begin{split}
    \left[\mathtt{H}_{SU(5)^A}(\mathsf{x})\right]_{b=1,\ldots,4}
    &=\frac{1}{\sqrt{b(b+1)}}\left[
    \sum_{a=1}^b\partial_\mathsf{x}\phi_a-b\partial_\mathsf{x}\phi_{b+1}
    \right],\\
\left[\mathtt{E}_{SU(5)^A}(\mathsf{x})\right]_{\boldsymbol{\alpha}_{SU(5)}}
&=e^{i\alpha^j_{SU(5)}(\phi_j(\mathsf{x})+k_j\mathsf{x})}.
\end{split}\label{SU5Acurrents}
\end{align}
The $SU(5)$ root system $\Delta_{SU(5)^A}$ is composed of root vectors $\boldsymbol{\alpha}_{SU(5)}=
\pm({\bf e}_a-{\bf e}_b)$ where $1\leq a<b\leq5$.
The simple roots ${\bf e}_j-{\bf e}_{j+1}$, for $j=1,\ldots,4$, form the rows of the $4\times 8$ matrix $A_{SU(5)^A}$. The Cartan matrix of $SU(5)$ is the Gram matrix
\begin{align}
    K_{SU(5)}=A_{SU(5)^A}A^T_{SU(5)^A}=
    \left(\begin{smallmatrix}
    2&-1&&\\-1&2&-1&\\
    &-1&2&-1\\&&-1&2
    \end{smallmatrix}\right).
    \label{KSU5}
\end{align}

The $SU(5)^B$ algebra is the complement of $SU(5)^A$ in $E_8$. The Cartan generators $\left[\mathtt{H}_{SU(5)^B(\mathsf{x})}\right]_b=\gamma^j_b\partial_{\mathsf{x}}\phi_j$ in the $B$ sector can be chosen with orthonormal vectors $\boldsymbol{\gamma}_{b=1,2,3,4}$ perpendicular to the $SU(5)^A$ root system. For example, one can set $\boldsymbol\gamma_1={\bf e}_6$, $\boldsymbol\gamma_2={\bf e}_7$, $\boldsymbol\gamma_3={\bf e}_8$ and $\boldsymbol\gamma_4=\sum_{i=1}^5{\bf e}_i/\sqrt{5}$.
The $SU(5)^B$ root system $\Delta_{SU(5)^B}$ consists of the $E_8$ roots that are perpendicular to $\Delta_{SU(5)^A}$. $\Delta_{SU(5)^B}$ contains $\pm{\bf e}_a\pm{\bf e}_b$ for $6\leq a<b\leq8$ and $\varepsilon^a{\bf e_a}/2$ where $\varepsilon_a=\pm1$, $\prod_a\varepsilon^a=+1$ and $\varepsilon^1=\ldots=\varepsilon^5$. The simple roots of $SU(5)^B$ can be chosen to be the rows of 
\begin{align}
A_{SU(5)^B}=\left(\begin{smallmatrix}
&&&&&1&-1&\\&&&&&&1&-1\\
1/2&1/2&1/2&1/2&1/2&-1/2&-1/2&1/2\\-1/2&-1/2&-1/2&-1/2&-1/2&-1/2&-1/2&-1/2
\end{smallmatrix}\right),
\label{eq:SU5Bsimpleroots}
\end{align} so that their scalar product recovers $K_{SU(5)}=A_{SU(5)^B}A^T_{SU(5)^B}$ in \eqref{KSU5}. The mutual OPEs between current operators in the $A$ and $B$ sectors are non-singular while OPEs within the same sector obey the $SU(5)$ WZW current algebra at level 1 (c.f.~\eqref{WZWcurrentOPE}). The two sectors are decoupled. Their Sugawara energy-momentum tensors add up to $T_{SU(5)_1^A}+T_{SU(5)_1^B}=T_{(E_8)_1}$ and thus the central charges are $c_{SU(5)^A}=c_{SU(5)^B}=4$, which is half of that of $(E_8)_1$.

The $SU(5)$ bFQH states can be constructed by the coupled-wire Hamiltonians $\mathcal{H}[SU(5)]$ that contain the inter- and intra-wire current backscattering interactions $\mathcal{H}^{SU(5)^A}_{\mathrm{inter}}+\mathcal{H}^{SU(5)^B}_{\mathrm{intra}}$. The interactions are analogous to those in \eqref{SU3E6sineGordon}. They are non-competing sine-Gordon interactions that create a finite excitation energy gap in the bulk, but leave behind gapless chiral edge states on the boundaries described by the $SU(5)_1$ WZW CFT. The model conserves $\mathsf{x}$ momentum when the electron Fermi momenta $k_{F,a}$ obey \eqref{appmomentumAbelian} under the projection matrix $P_A=P_{SU(5)^A}$. 
Like the previous Abelian models, the $SU(5)$ bFQH state depends on the choice of the non-local Dirac fermions $d_j\sim e^{i\phi_j}$ and their electric charges $q_j$. These choices give rise to different $SU(5)_1$ bFQH states at various fillings with distinct charge assignments.

The topological phase carries $|K_{SU(5)}|=5$ primary field super-selection sectors. Each sector, labelled by $\mathcal{E}^{m=-2,-1,0,1,2}$, forms a $C^5_{|m|}=5!/(|m|!(5-|m|)!)$ dimensional irreducible representation of $SU(5)$. $\mathcal{E}^0$ is the trivial vacuum sector. In $SU(5)_1^A$, for $m>0$, \begin{align}
    \mathcal{E}^m =\mathrm{span}\left\{ e^{-i\sum_{i=1}^m\phi_{j_i}+im\sum_{l=1}^5\phi_l/5}\right\}_{1\leq j_1<\ldots<j_m\leq5}.
\end{align} For $m<0$, $\mathcal{E}^m=(\mathcal{E}^{-m})^\dagger$. The vertex fields in $\mathcal{E}^m$ all carry spins (scaling dimensions) $h_m=|m|(5-|m|)/10$. Primary fields in the same super-sector can be rotated irreducibly into one another under the $SU(5)$ current OPE (c.f.~\eqref{currentirrep}).
These bFQH states have the Abelian $\mathbb{Z}_5$ topological order. Their anyon classes follow the fusion rules $\mathcal{E}^m\times\mathcal{E}^{m'}=\mathcal{E}^{[m+m']}$ where $[m+m']=-2,-1,0,1,2$ and $[m+m']\equiv m+m'$ modulo 5. 

The possible Hall conductivities of the $SU(5)$ bosonic states can be exhausted by applying the projection operator $P_{SU(5)^A}=A_{SU(5)^A}K^{-1}_{SU(5)}A_{SU(5)^A}$ to the possible charge vectors ${\bf q}=(q_{i=1,\ldots,8})$ that were presented above \eqref{E8fillingnumber}. The resulting filling numbers of the $SU(5)_1$ states are
\begin{align}
\begin{split}
 \nu_{SU(5)_1}&=|P_{SU(5)^A}{\bf q}|^2
 =\frac{4}{5}\sum_{j=1}^5q_j^2-\frac{2}{5}\sum_{1\leq i<j\leq5}q_iq_j
 \\&=0,\frac{16}{5},\frac{24}{5},8,\frac{56}{5},\frac{64}{5},16.
\end{split}
\label{eq:SU5fillingPairs}
\end{align}
Pairs of bFQH states with fillings $\nu$ and $16-\nu$ are related by particle-hole conjugation. The $SU(5)_1$ state at filling $\nu=8$ is particle-hole symmetric. The electric charge assignments of the $SU(5)$ WZW currents and the primary fields are summarized in  
Table \ref{tab:SU5ChargeAssignmentRoot} and \ref{tab:SU5ChargeAssignmentPrimary1to2}. Since $\mathcal{E}^{-m}=(\mathcal{E}^m)^\dagger$, the primary fields in the conjugate sectors $\mathcal{E}^{1,2}$ and $\mathcal{E}^{-1,-2}$ carry opposite electric charges. At the same time, the $E_8$ Weyl inversion that flips $\phi_j\to-\phi_j$ corresponds to the $\mathbb{Z}_2$ anyonic symmetry (or outer automorphism)~\cite{khan2014,Teotwistdefectreview} of $SU(5)_1$ that switches the anyon classes $\mathcal{E}^m\leftrightarrow\mathcal{E}^{-m}$. Similar to the $SU(3)_1$ and $(E_6)_1$ states, the charge assignments of $\mathcal{E}^m$ and $\mathcal{E}^{-m}$, for $m\neq0$, are opposite (identical) when the filling number $\nu_{SU(5)}$ is fractional (resp.~integral). Thus, at fractional fillings, the $\mathbb{Z}_2$ anyonic symmetry is broken by the primary field charge assignments. Although there are opposite charge patterns in each of these fractional cases, the two bFQH states are physically indistinguishable and can be identified by relabeling the anyon classes $m\leftrightarrow-m$.

The fractional filling $\nu=64/5$ is special. There are multiplicities in the charge patterns of the WZW current operators as well as the primary fields that are {\em not} mutually related by the $\mathbb{Z}_2$ charge flip. (Similar multiplicities of charge assignments will also later appear in $SU(8)_1$ and $SO(N)_1$, for $N=9,\ldots,15$.) In case (i), the 24 WZW currents decompose into $24=16(0)\oplus4(+4)\oplus4(-4)$, where the numbers inside the parentheses indicate the electric charge. In case (ii), the charge decomposition is $24=10(0)\oplus4(+2)\oplus4(-2)\oplus3(+4)\oplus3(-4)$. The charge decompositions of the primary fields of cases (i) and (ii) are also $\mathbb{Z}_2$ inequivalent (see table~\ref{tab:SU5ChargeAssignmentPrimary1to2}). Despite the charge differences, these two cases belong in the same bFHQ phase. When juxtaposing states (i) and (ii), the shared 1D edge is gappable by charge-preserving higher-order interactions. Since the $SU(5)$ currents in (i) and (ii) have unequal charges, they cannot all backscatter on the shared edge without violating charge conservation. Alternatively, the shared edge can be gapped by the charge-conserving sine-Gordon interactions
\begin{align}
\begin{split}
    \mathcal{U}&=
    -u\sum_{\ell=1}^3\cos\left(\phi^R_\ell-\phi^R_{\ell+1}-\phi^L_\ell+\phi^L_{\ell+1}\right)\\
    &\;\;\;\;-u\cos\left(\sum_{\ell=1}^4\phi^R_\ell-4\phi^R_5+\sum_{\ell=1}^4\phi^L_\ell-4\phi^L_5\right).
\end{split}
\label{eq:SU5edgeUrecon}
\end{align}
Here, we assume the Dirac fermions $d_j^R\sim e^{i\phi_j^R}$ in case (i) carries charge $q_j=(0,0,0,-4,0,0,0,0)$, and the fermions $d_j^L\sim e^{i\phi_j^L}$ in case (ii) carries $q_j=(2,2,2,-2,0,0,0,0)$. The last sine-Gordon potential back-scatters a non-primitive boson with scaling dimension $h=10$. It becomes relevant in the RG sense given a sufficiently strong density-density interaction.

\subsubsection{The Abelian \texorpdfstring{$SO(2r)$}{SO(2r)} states and emergent Dirac fermions}\label{SOeven}

\begin{figure}[htbp]
\includegraphics[width=.47\textwidth]{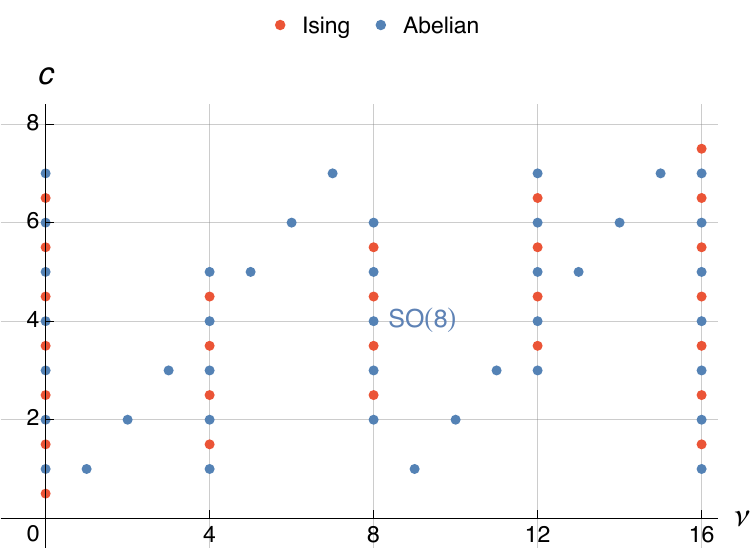}
\caption{Bosonic fractional quantum Hall (bFQH) states $SO(N)_1$. 
All $SO(2r)_1$ states are Abelian (blue dots); and all $SO(2r+1)_1$ states have non-Abelian Ising topological orders (red dots). The bFQH states are related under the particle-hole conjugation $(\nu,c)\leftrightarrow(16-\nu,8-c)$ about the PH symmetric $SO(8)_1$ at $(\nu,c)=(8,4)$.}
\label{fig:SOphasesfillings}
\end{figure}

We now construct the $SO(2r)_1$ bFQH states, for $r=1,\ldots,7$. The construction relies on the $SO(2r)\times SO(16-2r)$ conformal decomposition of $E_8$. Given a particular choice of non-local Dirac fermions $d_j\sim e^{i\phi_j}$, for $j=1,\ldots,8$, in each wire (see \eqref{nonlocaldiracdef}), the splitting is set by generating the $SO(2r)_1$ by $d_{l=1,\ldots,r}$ and $d_{l=1,\ldots,r}^\dagger$ and its particle-hole conjugate $SO(16-2r)_1$ by $d_{m=r+1,\ldots,8}$ and $d_{m=r+1,\ldots,8}^\dagger$. The WZW current operators of $SO(2r)_1$ are the fermion bilinears $d_ld_{l'}$, $d_ld_{l'}^\dagger$, $d_l^\dagger d_{l'}$ and $d_l^\dagger d_{l'}^\dagger$ for $1\leq l\leq l'\leq r$. They can be bosonized into $r$ Cartan generators and $2r(r-1)$ root operators: \begin{align}\begin{split}&\left[\mathtt{H}_{SO(2r)}(\mathsf{x})\right]_l=d_l(\mathsf{x})^\dagger d_l(\mathsf{x})=\partial_{\mathsf{x}}\phi_l(\mathsf{x}),\\&\left[\mathtt{E}_{SO(2r)}(\mathsf{x})\right]_{\boldsymbol\alpha_{SO(2r)}}=e^{i\alpha_{SO(2r)}^j\left(\phi_j(\mathsf{x})+k_j\mathsf{x}\right)}.\end{split}\label{SO2rcurrents}\end{align} The root system $\Delta_{SO(2r)}$ contains the integral length $\sqrt{2}$ vectors $\boldsymbol\alpha_{SO(2r)}=\pm{\bf e}_l\pm{\bf e}_{l'}$, where $1\leq l<l'\leq r$. They are integral combinations of a set of simple roots $\boldsymbol\alpha_1={\bf e}_1-{\bf e}_2,\ldots,\boldsymbol\alpha_{r-1}={\bf e}_{r-1}-{\bf e}_r$, and $\boldsymbol\alpha_r={\bf e}_{r-1}+{\bf e}_r$ that produce the Cartan matrix by scalar product: \begin{align}K^{SO(2r)}_{ll'}=\boldsymbol\alpha_l\cdot\boldsymbol\alpha_{l'}=\left(\begin{smallmatrix}2&-1&&&&&\\-1&2&&&&&\\&&\ddots&&&&\\&&&2&-1&&\\&&&-1&2&-1&-1\\&&&&-1&2&\\&&&&-1&&2\end{smallmatrix}\right).\label{SO2rCartanmatrix}\end{align} The conjugate $SO(16-2r)_1$ sector commutes with $SO(2r)_1$. The root system $\Delta_{SO(16-2r)}$ consists of $E_8$ root vectors that are perpendicular to $\Delta_{SO(2r)}$. In other words, $\boldsymbol\alpha_{SO(16-2r)}=\pm{\bf e}_m\pm{\bf e}_{m'}$, where $r+1\leq m<m'\leq 8$.

The coupled-wire model of the $SO(2r)_1$ state is constructed in a similar fashion as the previous Abelian quantum Hall states. The inter-wire and intra-wire current backscattering terms in the exactly solvable model Hamiltonian \eqref{HA} involve the $SO(2r)_1$ and $SO(16-2r)_1$ sector respectively. Generically, they are the two-fermion backscattering Gross-Neveu-type interactions:
\begin{widetext}
\begin{align}\mathcal{H}_{\mathrm{inter}}^{SO(2r)}&=u_{\mathrm{inter}}\sum_y\sum_{1\leq p<q\leq 2r}\psi^R_{yp}\psi^R_{yq}\psi^L_{y+1,p}\psi^L_{y+1,q},\quad\mbox{for $r=2,\ldots,7$}\label{SO2rHinter}\\
&=u_{\mathrm{inter}}\sum_y\left[\sum_{l=1}^r{d_{yl}^R}^\dagger d_{yl}^R{d_{y+1,l}^L}^\dagger d_{y+1,l}^L+\sum_{1\leq l<l'\leq r}\left(d_{yl}^Rd_{yl'}^R{d_{y+1,l}^L}^\dagger{d_{y+1,l'}^L}^\dagger+{d_{yl}^R}^\dagger d_{yl'}^Rd_{y+1,l}^L{d_{y+1,l'}^L}^\dagger+h.c.\right)\right]\nonumber\\
&=u_{\mathrm{inter}}\sum_y\left[\sum_{l=1}^r\left[\mathtt{H}_{SO(2r)}\right]^R_{yl}\left[\mathtt{H}_{SO(2r)}\right]^L_{y+1,l}-\sum_{\boldsymbol\alpha_{SO(2r)}}\cos\left(\boldsymbol\alpha_{SO(2r)}\cdot\boldsymbol\theta_{y+1/2}\right)\right],\nonumber
\end{align}
\begin{align}
\mathcal{H}_{\mathrm{intra}}^{SO(16-2r)}&=u_{\mathrm{intra}}\sum_y\sum_{2r+1\leq p<q\leq 16}\psi^R_{yp}\psi^R_{yq}\psi^L_{yp}\psi^L_{yq},\quad\mbox{for $r=1,\ldots,6$}\label{SO2rHintra}\\
&=u_{\mathrm{intra}}\sum_y\left[\sum_{m=r+1}^{16}{d_{ym}^R}^\dagger d_{ym}^R{d_{ym}^L}^\dagger d_{ym}^L+\sum_{r+1\leq m<m'\leq 16}\left(d_{ym}^Rd_{ym'}^R{d_{ym}^L}^\dagger{d_{ym'}^L}^\dagger+{d_{ym}^R}^\dagger d_{ym'}^Rd_{ym}^L{d_{ym'}^L}^\dagger+h.c.\right)\right]\nonumber\\
&=u_{\mathrm{intra}}\sum_y\left[\sum_{m=r+1}^{16}\left[\mathtt{H}_{SO(16-2r)}\right]^R_{ym}\left[\mathtt{H}_{SO(16-2r)}\right]^L_{ym}-\sum_{\boldsymbol\alpha_{SO(16-2r)}}\cos\left(\boldsymbol\alpha_{SO(16-2r)}\cdot\boldsymbol\theta_y\right)\right].\nonumber\end{align}
\end{widetext}
Here, $\psi$ are the real and imaginary Majorana components of the complex Dirac fermions $d_j=(\psi_{2j-1}+i\psi_{2j})/\sqrt{2}$;  $\boldsymbol\theta_{y+1/2}=\boldsymbol\phi^R_y-\boldsymbol\phi^L_{y+1}$; and $\boldsymbol\theta_y=\boldsymbol\phi^R_y-\boldsymbol\phi^L_y$. Similar to the $SU(3)_1$ and $(E_6)_1$ states previously presented in \eqref{SU3E6sineGordon}, the sine-Gordon interactions here are marginally relevant when $u>0$. They introduce a finite excitation energy gap in the bulk and leave behind the chiral gapless $SO(2r)_1$ WZW CFTs on boundary edges. The thermal response $\kappa_{xy}$ \eqref{electricthermalresponse} is determined by the central charge $c_{SO(2r)_1}=r$, which is the rank of the simply-laced $SO(2r)$ Lie algebra. The model Hamiltonian of the particle-hole conjugate $SO(16-2r)_1$ state can be constructed by interchanging the intra-wire and inter-wire gapping pattern between the $SO(2r)$ and $SO(16-2r)$ sectors. 

Cases involving $SO(2)_1$ when $r=1$ or $7$ are special. This is because the $SO(2)$ algebra is Abelian and the $SO(2)_1=U(1)_4$ WZW algebra is generated by a single current operator $\partial_{\mathsf{x}}\phi$. The current backscattering interaction $\partial_{\mathsf{x}}\phi^R\partial_{\mathsf{x}}\phi^L$ is a density interaction that does not open an energy gap by itself. Instead, the $SO(2)_1$ inter-wire or intra-wire gapping potential is \begin{align}\mathcal{U}^{SO(2)}&=u\partial_{\mathsf{x}}\phi^R_y\partial_{\mathsf{x}}\phi^L_{y'}-u'\cos\left(4\theta\right),\label{SO2H}\end{align} where $2\theta=\phi^R_y-\phi^L_{y'}$ (the factor of 2 is adopted here to match with the usual convention in Luttinger liquid theory), and $y'=y$ ($y'=y+1$) for an intra-wire (resp., inter-wire) interaction. The sine-Gordon potential here back-scatters the spin-2 local bosons $e^{i2\phi^\sigma}$ rather than spin-1 currents in \eqref{SO2rHinter} and \eqref{SO2rHintra}. It becomes relevant when the density interaction $u$ becomes ``repulsive" (i.e., negative) enough so that the Luttinger parameter $g=\sqrt{\frac{1+2\pi u/\tilde{v}}{1-2\pi u/\tilde{v}}}$ is smaller than $1/2$, where $\tilde{v}$ is the velocity of the fermions $d_j$ appeared in \eqref{E8CartanWeylLH}. Odd fermion backscattering terms, such as the single fermion backscattering $\cos(2\theta)\sim (d^R)^\dagger d^L+h.c.$, are non-local for both the inter- and intra-wire cases, and do not have an integral electron origin. Higher order even terms $\cos(4n\theta)$, for $n\geq2$, are irrelevant or less relevant. 

The $SO(2r)_1$ WZW CFT has four primary field super-selection sectors. The non-trivial ones consist of the spin $h=1/2$ fermion sector $f$ spanned by the Dirac fermions $d_{l=1,\ldots,r}$ and $d_{l=1,\ldots,r}^\dagger$, and the spin $h=r/8$ even ($+$) and odd ($-$) spinor sector $s_\pm$ spanned by the vertex operators $e^{i\varepsilon^l\phi_l/2}$, where $\varepsilon^{l=1,\ldots,r}=\pm1$ and $\prod_{l=1}^r\varepsilon^a=\pm1$. Fields within the same super-selection sector differ from each other by local bosons and can be rotated into each other by the WZW algebra (c.f.~\eqref{currentirrep}). For even $r$, they follow the $\mathbb{Z}_2\times\mathbb{Z}_2$ fusion rules, where $(s_\pm)^2=f^2=1$ and $s_\pm\times f=s_\mp$. For odd $r$, they follow the $\mathbb{Z}_4$ fusion rules, $s_\pm\times s_\mp=f^2=1$ and $s_\pm\times f=s_\mp$. Unlike the previously considered $SU(2)_1$, $SU(3)_1$ and $(E_{6,7,8})_1$ states, the non-local Dirac fermions $d_{a=1,\ldots,r}$ now emerge as quasiparticle excitations in $SO(2r)_1$ and exist in the form of deconfined anyons in the bulk and primary fields on the edge.

The $SO(2r)_1$ WZW CFT is generically symmetric under a twofold outer automorphism and the topological phase has the corresponding $\mathbb{Z}_2$ anyonic relabelling symmetry. The outer automorphism can be generated by the involution symmetry $w$ in $\mathrm{Aut}(E_8)$ that flips $\phi_r\to-\phi_r$ and $\phi_8\to-\phi_8$ while keeping the rest of the bosonized variables $\phi_{j\neq r,8}$ unchanged. The involution acts as a reflection on the $SO(2r)$ root system. But the mirror plane is not perpendicular to any $SO(2r)$ root and the reflection falls outside of the Weyl group of $SO(2r)$. The reflection flips the simple roots $\boldsymbol\alpha_{r-1}\leftrightarrow\boldsymbol\alpha_r$ and leaves $\boldsymbol\alpha_{l=1,\ldots,r-2}$ unchanged. It can be represented by the matrix $M=\mathbb{1}_{r-2}\oplus\sigma_x$, which commutes with the Cartan matrix \eqref{SO2rCartanmatrix}. The fermion primary sector $f$ of $SO(2r)_1$ is closed under the $\mathbb{Z}_2$ symmetry, but the even and odd spinor sectors are switched, $s_+\leftrightarrow s_-$ (c.f.~the switching action $\mathcal{E}\leftrightarrow\overline{\mathcal{E}}$ in $SU(3)_1$ and $(E_6)_1$). The $SO(8)_1$ theory is special and carries a triality $S_3$ symmetry. This is because, when $r=4$, in addition to the $w$ symmetry, the $SO(8)\times SO(8)$ root system is also preserved by another twofold symmetry $w'=(H_4\oplus H_4)/2$ in $\mathrm{Aut}(E_8)$, where $H_4$ is the Hadamard matrix \begin{align}H_4=\left(\begin{smallmatrix}1&1&1&1\\1&-1&1&-1\\1&1&-1&-1\\1&-1&-1&1\end{smallmatrix}\right).\label{hadamard}\end{align} Combining the two, $ww'$ is a threefold symmetry that rotates the primary field sectors $f\to s_-\to s_+\to f$. It is not a coincidence that all three non-trivial primary sectors in $SO(8)_1$ have identical spin $h=1/2$ and interchangeable fusion rules so that the theory is symmetric under the permutation group $S_3$ of three elements. 

The electric response of the $SO(2r)_1$ bFQH states depend on the choice of the eight non-local Dirac fermions in \eqref{nonlocaldiracdef}, $d_j\sim e^{i\phi_j}$, which are specified by the $R$ matrix that corresponds to the $E_8$ simple roots $e^{i\tilde\Phi_J}=e^{iR^j_J\phi_j}$. The allowed charge vectors ${\bf q}=(q_{j=1,\ldots,8})$ of the fermions were presented above \eqref{E8fillingnumber}. The filling numbers of the $SO(2r)_1$ state and its particle-hole conjugate $SO(16-2r)_1$ are \begin{align}\nu_{SO(2r)_1}=\sum_{j=1}^rq_j^2,\quad\nu_{SO(16-2r)_1}=\sum_{j=r+1}^8q_j^2,\end{align} which add up to $\nu_{E_8}=16$. 
Fig.\ref{fig:SOphasesfillings}
summarizes the possible filling numbers and central charges of the Abelian $SO(2r)_1$ and $SO(16-2r)_1$ bFQH states.
The electric charge assignments of the WZW currents and the primary fields are summarized in Tables \ref{tab:SOevenChargeAssignmentRoot} and \ref{tab:SOevenChargeAssignmentPrimary}. The $SO(2)_1$ algebra does not contain any roots. Instead of listing the trivial electric charge of its only current operator $\partial_{\mathsf{x}}\phi_1$, the $SO(2)$ column in Table \ref{tab:SOevenChargeAssignmentRoot} counts the electric charges $\pm q_1$ of the smallest local boson $e^{\pm i2\phi_1}$.

We highlight two observations in our results. First, there are two distinct states at filling fraction $\nu=16$ for $SO(2r)_1=SO(10)_1$, $SO(12)_1$, or $SO(14)_1$, where the WZW current operators either (i) carry electric charges $0,\pm4$ or (ii) carry charges $0,\pm2,\pm4$ (see Table \ref{tab:SOevenChargeAssignmentRoot}). 
The coupled-wire models of case (i) are constructed with the charge vector ${\bf q}=(q_{j=1,\ldots,8})$ of non-local Dirac fermions that contains one and only one non-zero entry $q_j=4$ and is within the $SO(2r)$ subspace. Models in case (ii) are constructed with ${\bf q}$ that contains four non-zero entries $q_j=\pm2$ in the $SO(2r)$ subspace. The two cases also have unequal charge assignments for their primary fields. For case (i), the fermion sector $f$ contains fields with charges $0,\pm4$ and the spinor sectors $s_\pm$ carry charges $\pm2$. For case (ii), $f$ has charges $0,\pm2$ and $s_\pm$ have charges $0,\pm2,\pm4$ (see Table \ref{tab:SOevenChargeAssignmentPrimary}).
Despite the distinct charge assignments, the two cases belong in the same bFQH phase at $\nu=16$. 
When juxtaposing the two states, the shared edge is gappable. 
Because the WZW currents in cases (i) and (ii) have different charge assignments, current backscattering interactions (that include all currents) on the shared edge generally violate charge conservation. 
To gap the shared edge in a way that preserves charge conservation, an alternative set of interactions needs to be considered: There exist sine-Gordon potentials that backscatter higher-spin bosons with the same charges between the (i) and (ii) boundaries. 
To demonstrate this, we consider $SO(2r)_1=SO(10)_1$ where the shared edge carries 5 counter propagating pairs of non-local Dirac fermions $d^\sigma_{j=1,\ldots,5}\sim e^{i\phi^\sigma_j}$. The fermion charges are $q^R_{j=1,\ldots,5}=(0,0,0,0,4)$ and $q^L_{j=1,\ldots,5}=(2,2,2,2,0)$. The shared edge can be gapped by the charge-preserving, local sine-Gordon potentials: \begin{align}\begin{split}\mathcal{U}&=-u\sum_{l=1}^3\cos\left(\phi^R_l-\phi^R_{l+1}-\phi^L_l+\phi^L_{l+1}\right)\\&\;\;-u\cos\left(2\phi^L_5+\sum_{l=1}^4\phi^R_l\right)-u\cos\left(2\phi^R_5-\sum_{l=1}^4\phi^L_l\right).\end{split}\label{SO101214sharededge}\end{align} 
${\cal U}$ breaks the $SO(10)$ symmetry. 
The last two terms in ${\cal U}$ backscatter spin-2 bosons; these terms are relevant for sufficiently strong density-density interactions. 
Therefore, the two states (i) and (ii) are only distinguishable when the $SO(10)$ symmetry is preserved.

Second, in Table \ref{tab:SOevenChargeAssignmentPrimary}, we observe that a $SO(2r)_1$ state may carry multiple distinct charge assignment patterns for its anyons or edge primary fields at the same filling number. For example, for $SO(4)_1$ at filling $\nu=2$, the charges of the even spinors $q(s_+)=\pm1$ while the odd ones are neutral $q(s_-)=0$ in one state, but the charge pattern is reversed with $q(s_+)=0$ and $q(s_-)=\pm1$ in another state. For $SO(6)_1$ at filling $\nu=3$, the charges of the spinors are $q(s_\pm)=\mp3/2,\pm1/2$ for one state, but are flipped to $q(s_\pm)=\pm3/2,\mp1/2$ for another. This degeneracy stems from the $\mathbb{Z}_2$ outer automorphism (anyonic symmetry) of $SO(2r)_1$ that switches between the $s_+\leftrightarrow s_-$ anyon classes. Each state in the above examples illustrates a weak $\mathbb{Z}_2$ symmetry breaking by electric charge in the sense that the charge assignments of two anyon classes related by the $\mathbb{Z}_2$ symmetry are not identical (even up to charge 2 local bosons). In addition, the spinors $s_\pm$ are self-conjugate $(s_\pm)^2=1$ when $r$ is even, and are anti-partners of each other $s_+\times s_-=1$ when $r$ is odd. This requires the charges of each of $s_\pm$ to be closed under $q\to-q$ when $r$ is even or requires $s_\pm$ to have opposite charges when $r$ is odd. For $SO(8)_1$, the $S_3$ symmetry that permutes the fermions $f,s_+,s_-$ gives rise to the three charge patterns for some filling numbers. For example in any given $SO(8)_1$ state at $\nu=4$, one fermion class carries odd charges while the other two carry even charges. The $S_3$ symmetry is weakly broken down by the charge assignments into $\mathbb{Z}_2$, which switches the two even-charged fermion classes. There are filling numbers where the charge pattern is non-degenerate and the anyonic symmetry of the state is preserved by electric charge. In particular, the particle-hole symmetric $SO(8)_1$ state at filling $\nu=8$ fully preserves its $S_3$ triality symmetry where the three fermion classes $f,s_+,s_-$ have identical charges.

Lastly, we address the momentum conservation of the backscattering interactions in the coupled-wire models. They conserve $\mathsf{x}$-momentum when the bare electron momenta $k_{F,a}$ (see \eqref{electronmomentum}) take a set of specific values. These values are computed in \eqref{appmomentumAbelian} in Appendix~\ref{app:momentum} by applying the projection matrix $P_A=\mathbb{1}_r\oplus\mathbb{0}_{8-r}$ for $SO(2r)_1$ or $P_B=\mathbb{0}_r\oplus\mathbb{1}_{8-r}$ for the particle-hole conjugate $SO(16-2r)$ state. 
This allows the $L$ and $R$ non-local Dirac fermions to have the same $\mathsf{x}$-momentum if they are pairwise back-scattered within a wire, or different momenta that are commensurate with the magnetic field, $k^R_j-k^L_j=\frac{eBd}{\hbar c}q_j$, if they are pairwise back-scattered between wires.

\subsubsection{The Abelian orbifold \texorpdfstring{$U(1)_8$}{U(1)} and \texorpdfstring{$SU(8)_1$}{SU(8)} states}\label{sec:abelianorbifolds}
\begin{figure}[htbp]
\includegraphics[width=.47\textwidth]{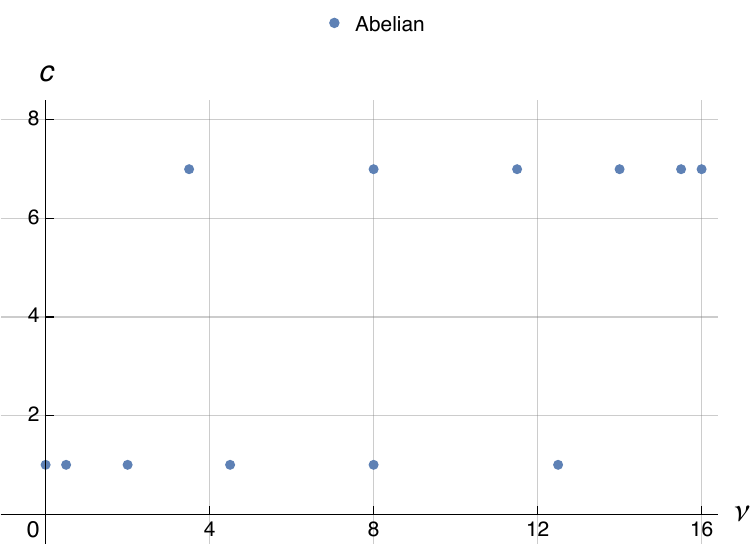}
\caption{$SU(8)_1$ bFQH states with central charge $c=7$ and $U(1)_8$ with central charge $c=1$. 
}\label{fig:su8u1phases}
\end{figure}

We demonstrate the concept of symmetry ``gauging"~\cite{TeoHughesFradkin15,BarkeshliBondersonChengWang14} by constructing the Abelian orbifold bFQH states \begin{align}U(1)_8=\frac{SU(2)_1}{\mathbb{Z}_2},\quad SU(8)_1=\frac{(E_7)_1}{\mathbb{Z}_2}.\label{U1SU8orbifolds}\end{align} 
These are the simplest examples originating from the $\mathbb{Z}_2$ symmetry that flips the signs of the eight non-local Dirac fermions, $d_j\to-d_j$. It is an internal symmetry because all local operators, including the WZW currents in $E_8$ and any of its subalgebras, are unchanged under the symmetry. In \S\ref{sec:SU2E7}, we constructed the $SU(2)_1$ and $(E_7)_1$ bFQH states. The corresponding WZW theories contain $U(1)_8$ and $SU(8)_1$ as subalgebras. Together, $U(1)_8\times SU(8)_1$ conformally splits $(E_8)_1$. The two sectors can be gapped within a wire or in between wires by sine-Gordon potentials or current backscattering interactions. However, the coupled-wire models constructed this way do not carry the $U(1)_8$ and $SU(8)_1$ topological orders. This is because both subalgebras carry spin-1 bosonic primary field sectors that are local integral combinations of electrons. These spin-1 local fields ``anyon condense"~\cite{PhysRevB.79.045316, 2018ARCMP...9..307B} and then extend the WZW algebras from $U(1)_8$ to $SU(2)_1$ and from $SU(8)_1$ to $(E_7)_1$. Therefore, the bFQH states still carry the $SU(2)_1$ and $(E_7)_1$ topological order. 

Here, we will construct $U(1)_8$ and $SU(8)_1$ bFQH states by designing a conformal embedding $U(1)_8\times SU(8)_1\subseteq(E_8)_1$ that cannot be extended to $SU(2)_1\times(E_7)_1$ without violating electron locality. The spin-1 bosonic primary fields will be non-local and therefore will not be able to ``anyon condense." These bosons will be odd under the $\mathbb{Z}_2$ symmetry and will be identified as $\mathbb{Z}_2$ gauge charges. Deconfined gauge fluxes will emerge in the topological phases as anyons that exhibit $\pi$ monodromy with the gauge charges.

We begin by defining a mixed set of non-local Dirac fermions. Starting with $d_j\sim e^{i\phi_j}$, $j=1,\ldots,8$, in \eqref{nonlocaldiracdef}, we consider the following basis transformation of the Cartan-Weyl bosonized variables 
\begin{align}\tilde\phi_{2l-1}=\phi_{2l-1},\quad\tilde\phi_{2l}=\frac{1}{2}\sum_{l'=1}^4\left(H_4\right)_l^{l'}\phi_{2l'}\label{mixedphidef}\end{align} where $l=1,2,3,4$ and $H_4$ is the Hadarmard matrix \eqref{hadamard}. The new bosonized variables $\tilde\phi$ are still described by the same free theory \eqref{E8CartanWeylLH} as the old ones, and obey the same equal-time commutation relations $\left[\tilde\phi_j^\sigma(\mathsf{x}),\partial\tilde\phi_{j'}^{\sigma'}(\mathsf{x})\right]=2\pi i\sigma\delta^{\sigma\sigma'}\delta_{jj'}\delta(\mathsf{x}-\mathsf{x}')$. However, the basis transformation is {\em not} a Weyl group symmetry in $\mathrm{Aut}(E_8)$. Not all $E_8$ root vectors $\boldsymbol\alpha$ in $\Delta_{E_8}$ correspond to local vertex operators $e^{i\alpha^j\tilde\phi_j}$. In particular, the $\mathbb{Z}_2$ gauge transformation $\phi_j\to\phi_j+\pi$ (i.e., $d_j\to-d_j$), for all $j=1,\ldots,8$, now only alters the signs of the new fermions $\tilde{d}_j=e^{i\tilde\phi_j}\to(-1)^j\tilde{d}_j$ with odd indices because \begin{align}\mathbb{Z}_2:\tilde\phi_j\to\tilde\phi_j+\pi g_j,\label{Z2gaugesymm0}\end{align} for $g_{j=1,\ldots,8}=(1,2,1,0,1,0,1,0)$. 
Operators such as $\tilde{d}_1\tilde{d}_2\sim e^{i(\tilde\phi_1+\tilde\phi_2)}$ are no longer local because they are odd under $\mathbb{Z}_2$. A root operator $e^{i\alpha^j\tilde\phi_j}$ is a local integral electronic combination only when $\boldsymbol\alpha$ has even product with ${\bf g}$, i.e., $\alpha^jg_j\equiv0$ mod 2. There are 112 such even root vectors, and they form a root system for $SO(16)$. We caution that the $SO(16)_1$ algebra generated by these local roots is a sub-algebra sitting inside the $(E_8)_1$. One of the spinor sectors of the $SO(16)_1$ contains local fields that extend $SO(16)_1$ back to the original $(E_8)_1$. Consequently, a coupled-wire model that back-scatters the $SO(16)_1$ between wires would still belong in the $(E_8)_1$ phase with trivial topological order. On the other hand, if the $SO(16)_1$ is split by some conformal embedding $\mathcal{G}_A\times\mathcal{G}_B\subseteq SO(16)_1$, which we are going to perform in this subsection, the coupled-wire models of $\mathcal{G}_A$ and $\mathcal{G}_B$ may exhibit an orbifold structure in which a deconfined $\mathbb{Z}_2$ gauge theory emerges.

First, we consider the $SU(2)\times E_7$ algebra by replacing $\phi$ in \S\ref{sec:SU2E7} by the mixed variables $\tilde\phi$ defined in \eqref{mixedphidef}. Since $\phi$ and $\tilde\phi$ are described by the same Lagrangian density, the $SU(2)$ and $E_7$ currents still obey the same OPEs. However, only the root operators that are invariant under $\mathbb{Z}_2$ are local. The $\mathbb{Z}_2$ even current operators span the $U(1)_8$ and $SU(8)_1$ subalgebras of $SU(2)_1$ and $(E_7)_1$, respectively. To see this, we begin with the substitution $\phi\to\tilde\phi$ in the $SU(2)_1$ currents in \eqref{SU2currents}. While the Cartan generator $\mathtt{H}_0=\partial_{\mathsf{x}}(\tilde\phi_1-\tilde\phi_2)/\sqrt{2}$ is even under \eqref{Z2gaugesymm0}, the raising and lowering operators, \begin{align}\mathtt{E}^\pm=e^{\pm i(\tilde\phi_1-\tilde\phi_2)}=e^{\pm i[\phi_1-(\phi_2+\phi_4+\phi_6+\phi_8)/2]},\end{align} are $\mathbb{Z}_2$ odd and no longer local. Instead, the primitive local bosons are $e^{\pm2i(\tilde\phi_1-\tilde\phi_2)}$, which have spin 4. It generates the $U(1)_8$ CFT, which we will later identify as the $SU(2)_1/\mathbb{Z}_2$ orbifold. 

Performing the same substitution $\phi\to\tilde\phi$ for the $(E_7)_1$ currents in \eqref{E7roots}, the $\mathbb{Z}_2$ symmetric root operators $e^{i\alpha^j\tilde\phi_j}$ are associated with the even root vectors $\boldsymbol\alpha_{SU(8)}$ that form the root system, \begin{align}\Delta_{SU(8)}=\left\{\boldsymbol\alpha\in\Delta_{E_7}:\alpha^jg_j\equiv0\mbox{ mod 2}\right\}.\end{align} A set of simple root vectors $\boldsymbol\alpha_{l=1,\ldots,7}$ can be chosen to be the rows of the rectangular matrix, \begin{align}A_{SU(8)}=\left(\begin{smallmatrix}&&&&1&&1&\\&&1&&-1&&&\\&&&&1&&-1&\\-1/2&-1/2&-1/2&1/2&-1/2&1/2&1/2&1/2\\&&&-1&&-1&&\\&&&&&1&&-1\\&&&1&&-1&&\end{smallmatrix}\right).\label{ASU8}\end{align} The scalar products, \begin{align}K^{SU(8)}_{ll'}=\boldsymbol\alpha_l\cdot\boldsymbol\alpha_{l'}=2\delta_{ll'}-\delta_{l,l'+1}-\delta_{l,l'-1},\end{align} are the entries of the Cartan matrix of $SU(8)$. The local roots $e^{i\alpha^j_{SU(8)}\tilde\phi_j}$ and the Cartan generators $\mathtt{H}_1=\partial_{\mathsf{x}}(\tilde\phi_1+\tilde\phi_2)/\sqrt{2}$, $\mathtt{H}_2=\partial_{\mathsf{x}}\tilde\phi_3$, $\ldots$, $\mathtt{H}_7=\partial_{\mathsf{x}}\tilde\phi_8$ span the $SU(8)_1$ WZW algebra. We will identify this CFT with the $(E_7)_1/\mathbb{Z}_2$ orbifold after constructing the coupled-wire models.

Like all models in this paper, the Hamiltonian begins with $\mathcal{H}_0+\mathcal{H}^f_{\mathrm{intra}}$ that reduces the electron wires to counter-propagating pairs of $E_8$ CFTs (see \eqref{H0}, \eqref{E8Hintra} and \eqref{HA}). Ignoring the gapped integrated fermions $f_{n=1,2,3}$, the free theory is described by \eqref{E8CartanWeylLH}. Since the basis transformation \eqref{mixedphidef} is orthogonal, the form of the free Lagrangian density \eqref{E8CartanWeylLH} is unchanged under the substitution $\phi\to\tilde\phi$. Using the splitting $U(1)_8\times SU(8)_1\subseteq(E_8)_1$, the two decoupled sectors can be gapped by backscattering interactions in different sectors. To construct the $U(1)_8$ state, the interactions are \begin{align}\begin{split}\mathcal{H}^{SU(8)_1}_{\mathrm{intra}}&=u_{\mathrm{intra}}\left[\sum_{l=1}^7\mathtt{H}_{l,y}^R\mathtt{H}_{l,y}^L-\sum_{\boldsymbol\alpha_{SU(8)}}\cos\left(\alpha_{SU(8)}^j\theta_{y,j}\right)\right],\\\mathcal{H}^{U(1)_8}_{\mathrm{inter}}&=u_{\mathrm{inter}}\mathtt{H}_{0,y}^R\mathtt{H}_{0,y+1}^L-u'_{\mathrm{inter}}\cos\left(2\theta_{y+1/2}^{U(1)_8}\right),\end{split}\end{align} where $\theta_{y+1/2}^{U(1)_8}=\tilde\phi^R_{y,1}-\tilde\phi^R_{y,2}-\tilde\phi^L_{y+1,1}+\tilde\phi^L_{y+1,2}$ and $\theta_{y,j}=\tilde\phi_{y,j}^R-\tilde\phi_{y,j}^L$. The intra-wire $SU(8)_1$ current backscattering interaction is marginally relevant when $u_{\mathrm{intra}}>0$. The sine-Gordon potential simultaneously pins the ground state expectation values $\langle\boldsymbol\alpha_{SU(8)}\cdot\boldsymbol\theta_y\rangle$ and gaps the $SU(8)_1$ sector on all wires. The inter-wire $U(1)_8$ sine-Gordon interaction back-scatters the spin-4 local boson $e^{\pm 2i(\tilde\phi_1-\tilde\phi_2)}$. Because of its higher spin $h>1$, this interaction is only relevant when the density interaction $u_{\mathrm{inter}}$ is negative enough so that the Luttinger parameter $g=\sqrt{\frac{1+2\pi u/\tilde{v}}{1-2\pi u/\tilde{v}}}$ is smaller than $1/4$. Under this condition, the inter-wire sine-Gordon interaction pins $\langle\theta^{U(1)_8}_{y+1/2}\rangle$ and gaps all $U(1)_8$ sector modes except the right and left moving ones on the top and bottom boundary edges. The particle-hole conjugate $SU(8)_1$ state can be constructed by a coupled-wire model that exchanges the inter-wire and intra-wire backscattering roles of $U(1)_8$ and $SU(8)_1$. 

Next, we justify the orbifold identifications \eqref{U1SU8orbifolds}. From \S\ref{sec:E8review}, we see that electron locality implies the large gauge invariance is $\phi_j\equiv\phi_j+2\pi r_j$, where ${\bf r}=(r_1,\ldots,r_8)$ lives inside the lattice $\mathcal{R}=\mathrm{span}_{\mathbb{Z}}(\Delta_{E_8})$ generated by the $E_8$ root vectors in Euclidean 8-space. This allows the bosonized variable $\phi_1-\phi_2$ that generates $SU(2)_1$ (see \eqref{SU2currents} in \S\ref{sec:SU2E7}) to be shifted by any integer multiple of $2\pi$. The substitution $\phi\to\tilde\phi$ in \eqref{mixedphidef} imposes the additional identification: \begin{align}\tilde\phi_1-\tilde\phi_2\equiv\tilde\phi_1-\tilde\phi_2+\pi.\label{Z2actionSU2}\end{align} Hence, the $U(1)_8$ bosonized variable $\tilde\phi_1-\tilde\phi_2$ is compactified on the circle $\mathbb{R}/\pi\mathbb{Z}$, which has half the circumference of the closed circle $\mathbb{R}/2\pi\mathbb{Z}$ where the $SU(2)_1$ variable $\phi_1-\phi_2$ lives. A similar distinction of compactifications applies to $(E_7)_1$ and $SU(8)_1$. The bosonized variables $(A_{E_7})^j_l\phi_j$ that generate the simple roots of $E_7$ in \eqref{E7simpleroots}, for $l=1,\ldots,7$, are left unchanged under the large gauge transformation $(A_{E_7})^j_l\phi_j\equiv(A_{E_7})^j_l\phi_j+2\pi n_l$, for any integers $n_l$. However, with the substitution $\phi\to\tilde\phi$, the large gauge transformation rules are modified to include \begin{align}(A_{E_7})^j_l\tilde\phi_j\equiv(A_{E_7})^j_l\tilde\phi_j+\pi,\label{Z2actionE7}\end{align} for all $l$, in addition to the ones above. Thus, while the $(E_7)_1$ bosonized variables $(A_{E_7})^j_l\phi_j$ are compactified on the torus $\mathbb{R}^7/2\pi\mathbb{Z}^7$, the new ones $(A_{E_7})^j_l\tilde\phi_j$ that generate $SU(8)_1$ live on the torus $\mathbb{R}^7/2\pi\mathrm{BCC}$ of half the size. Here, $\mathrm{BCC}$ is the 7-dimensional lattice containing vectors with all integral or all half-integral entries. The new compactifications \eqref{Z2actionSU2} and \eqref{Z2actionE7} are associated with the orbifold CFTs $SU(2)_1/\mathbb{Z}_2$ and $(E_7)_1/\mathbb{Z}_2$, where $\mathbb{Z}_2$ is the quotient groups $\mathbb{Z}/2\mathbb{Z}$ and $\mathrm{BCC}/\mathbb{Z}^7$ that differentiate the new large gauge transformations from the old ones.

The orbifolding of the CFTs corresponds to the gauging of anyon structures. Both the $U(1)_8$ and the $SU(8)_1$ topological states support eight Abelian anyon classes $\mathcal{E}^m$, for $m=-3,\ldots,4$. Their total quantum dimension is $\sqrt{8}$, which is larger than that of $SU(2)_1$ and $(E_7)_1$ by a factor of 2, the order of the gauge group $\mathbb{Z}_2$. For $U(1)_8$, the anyons correspond to the the fractional vertex operators $\mathcal{E}^m=e^{im(\tilde\phi_1-\tilde\phi_2)/4}$ of the edge CFT. They carry spins $h_m=m^2/16$. The spin-1 vertex $\mathcal{E}^4$ is exactly the raising operator $\mathtt{E}^+$ of $SU(2)_1$; it carries unit $\mathbb{Z}_2$ gauge charge. For $SU(8)_1$, the anyons correspond to 8 primary field super-selection sectors on the edge. Each super-sector is spanned by a collection of vertex fields that irreducibly ``rotate" under the $SU(8)_1$ current OPE (c.f.~\eqref{currentirrep}). They have spins $h_m=m(8-m)/16$. The exact forms of the primary fields can be deduced using \eqref{appSU8primary} in Appendix~\ref{topologicaldataSU8}. In particular, the super-sector $\mathcal{E}^4$ has 70 spin-1 primary fields that are odd under $\mathbb{Z}_2$. They are the $\mathbb{Z}_2$ gauge charges and are non-local boson fields that would extend $SU(8)_1$ to $(E_7)_1$ had they been integral. The anyon classes in both $U(1)_8$ and $SU(8)_1$ follow the $\mathbb{Z}_8$ fusion rules \begin{align}\mathcal{E}^m\times\mathcal{E}^{m'}=\mathcal{E}^{[m+m']},\label{Z8fusion}\end{align} where $[n]$ puts $n$ back in the range $-3,\ldots,4$ by subtracting or adding an integer multiple of 8. The anyons $\mathcal{E}^m$ with odd index $m=\pm1,\pm3$ all carry a $\mathbb{Z}_2$ flux because of their $\pi$ monodromy with the $\mathbb{Z}_2$ charge $\mathcal{E}^4$. The remaining anyons $\mathcal{E}^{\pm2}$ are semions with spin $1/4$ for $U(1)_8$ or $3/4$ for $SU(8)_1$. They have trivial monodromy with the $\mathbb{Z}_2$ charge and therefore have a trivial flux component. They are associated with the semions $\mathcal{S}$ in the un-gauged theories $SU(2)_1$ and $(E_7)_1$. Each $\mathbb{Z}_2$ gauge flux must square under fusion into a semion instead of the trivial vacuum class. This property reflects the ``non-symmorphic" nature of the quantum $\mathbb{Z}_2$ symmetry~\cite{TeoHughesFradkin15} and is captured as a non-trivial element in the group cohomology $H^2(\mathbb{Z}_2,\mathcal{A})=\mathbb{Z}_2$, where $\mathcal{A}=\mathbb{Z}_2=\{1,\mathcal{S}\}$ is the fusion group of Abelian anyons in $SU(2)_1$ and $(E_7)_1$.

Now we present the electric responses of the $U(1)_8$ and $SU(8)_1$ bFQH states. The coupled-wire model depends on the particular choice of the eight non-local Dirac fermions in \eqref{nonlocaldiracdef}, $d_j\sim e^{i\phi_j}$, which are related to the local $E_8$ simple roots by $e^{i\tilde\Phi_J}=e^{iR^j_J\phi_j}$ for some $R$ matrix. The allowed charge vectors ${\bf q}=(q_{j=1,\ldots,8})$ of the fermions $d_j$ were presented above \eqref{E8fillingnumber}. Since the $U(1)_8$ is in the $\phi_1-(\phi_2+\phi_4+\phi_6+\phi_8)/2$ direction, its filling number can be deduced by the length square of the projection the ${\bf q}$ vector along this direction. The filling number of $SU(8)_1$ is the length square of the orthogonal projection.
We have:
\begin{align}\begin{split}\nu_{U(1)_8}({\bf q})&=\frac{1}{8}(2q_1-q_2-q_4-q_6-q_8)^2,\\\nu_{SU(8)_1}({\bf q})&=16-\nu_{U(1)_8}({\bf q}).\end{split}\end{align} 
Fig.\ref{fig:su8u1phases}
summarizes the possible filling numbers and central charges of the Abelian orbifold $U(1)_8$ and $SU(8)_1$ states.
The electric charge assignments of the $SU(8)_1$ WZW currents are summarized in Table~\ref{tab:AclassChargeAssgn2}. Like the $SO(2)_1$ algebra in the previous subsection, the $U(1)_8$ algebra also does not contain any roots, and  the $U(1)_8$ column in the same table counts the electric charges $\pm(2q_1-q_2-q_4-q_6-q_8)$ of the smallest local boson $e^{\pm i(2\phi_1-\phi_2-\phi_4-\phi_6-\phi_8)}$. The electric charges carried by the primary fields $\mathcal{E}^m$ of $U(1)_8$ and $SU(8)_1$ are summarized in Table~\ref{tab:SU8ChargeAssignmentPrimary1to2}. In particular, we notice that the charge and anyon data for the $U(1)_8$ bFQH state at $\nu=1/2$ agrees with the strongly paired state that theoretically may occupy the half-filled Landau level.

We observe that there are multiplicities in the charge assignments. First, primary fields in the conjugate sectors $\mathcal{E}^m$ and $\mathcal{E}^{-m}$ carry opposite charges. Therefore, unless the set of primary fields' charges within a super-selection sector $\mathcal{E}^m$ is closed under $q\to-q$, fields in the conjugate sector $\mathcal{E}^{-m}$ must carry unequal charges. This imbalance occurs for all primary sectors in $U(1)_8$ in all filling numbers and in most cases in $SU(8)_1$. Thus, the electric charge assignment weakly breaks the conjugation symmetry $\mathcal{E}^m\leftrightarrow\mathcal{E}^{-m}$. Conversely, the conjugation $\phi_j\to-\phi_j$ inverts the charge for each individual primary sector and leads to the double degeneracy of charge assignments in all filling numbers in Table~\ref{tab:SU8ChargeAssignmentPrimary1to2}. The same phenomenon was observed for the previous $SU(3)_1$, $(E_6)_1$, and $SO(2r)_1$ states.

Second, an additional multiplicity arises for $SU(8)_1$ at filling numbers $14$ and $31/2$. Table~\ref{tab:AclassChargeAssgn2} shows two distinct charge patterns for the $SU(8)_1$ WZW currents at each of these fillings. The two states also have unequal charge assignments for their primary fields that cannot be attributed to the conjugation symmetry. Similar degeneracy was seen for $SO(10)_1$, $SO(12)_1$ and $SO(14)_1$ at filling 16. Like the previous cases, the two $SU(8)_1$ states still belong in the same bFQH phase. When juxtaposing the two states, the shared boundary edge is gappable by interactions that backscatter higher spin bosons and break the $SU(8)$ symmetry. The details can be found in \eqref{SU8sharededge} in Appendix~\ref{topologicaldataSU8}.

\subsection{Non-Abelian states}\label{nonabeliansection}
 
We now construct non-Abelian bosonic fractional quantum Hall (bFQH) states that partially fill the $E_8$ state. The construction is similar to the previous examples (c.f.~\eqref{HA}) and relies on the bipartite conformal embeddings $\mathcal{G}_A\times\mathcal{G}_B\subseteq E_8$ (see \eqref{E8embeddings}). Here, the topological states $\mathcal{G}_{A/B}$ support non-Abelian quasiparticle excitations. They exhibit multi-channel fusion rules and carry non-unit quantum dimensions, $d>1$. The anyon braiding operations do not all mutually commute. We focus on three classes of non-Abelian bFQH states: Ising, Fibonacci, and metaplectic topological orders. The chiral CFTs on the boundary edges of all but one of these states are affine WZW simple Lie algebras $\mathcal{G}$ at level 1 that are {\em not} simply-laced. The exceptional case is the metaplectic orbifold state $SO(3)_2=SU(2)_4=SU(3)_1/\mathbb{Z}_2$. Each long root current operator in $\mathcal{G}$ is an $E_8$ root, but each short root in $\mathcal{G}$ is a linear combination of multiple $E_8$ roots. Consequently, the inter/intra-wire current backscattering interactions \eqref{HinterA} and \eqref{HintraB} consist of competing sine-Gordon potentials that collectively gap degrees of freedom carrying fractional central charges $c$. 

\subsubsection{The \texorpdfstring{$SO(2r+1)$}{SO(2r+1)} Ising states and emergent Majorana fermions}\label{SOodd}

We now construct the $SO(2r+1)_1$ bFQH states, for $r=0,1,\ldots,7$. They all carry an Ising-like topological order~\cite{NayakWilczek96,kitaev2006anyons}. In the generic range for $r=1,\ldots,6$, the bFQH state has an edge-state theory described by the $SO(2r+1)$ WZW theory at level 1. 
$SO(15)_1$ and its particle-hole conjugate the Ising CFT, which we denote by $SO(1)_1$, require a special treatment (because the Ising CFT is not a WZW theory) and will therefore be presented last.

The generic construction for $r=1,\ldots,6$ relies on the $SO(2r+1)\times SO(15-2r)$ conformal bipartition of $E_8$. A particular decomposition is chosen by fixing a set of non-local Dirac fermions $d_{j=1,\ldots,8}\sim e^{i\phi_j}$ (see \eqref{nonlocaldiracdef}) that represent the $E_8$ WZW CFT in each chiral sector $\sigma=R,L$ on each wire. Each Dirac fermion can be formally split into real and imaginary Majorana components, $d_j=(\psi_{2j-1}+i\psi_{2j})/\sqrt{2}$. The $SO(2r+1)_1$ sector is generated by the first $2r+1$ Majorana fermions $\psi_{p=1,\ldots,2r+1}$. The (real) WZW algebra is spanned by the current operators $J_{pq}=i\psi_p\psi_q$,  
for $1\leq p<q\leq2r+1$. Similarly, the $SO(15-2r)_1$ sector is generated by the remaining fermions $\psi_{p=2r+2,\ldots,16}$.

The (complexified) $SO(2r+1)_1$ algebra can be obtained by first bosonizing the $SO(2r)_1$ subalgebra generated by $\psi_{p=1,\ldots,2r}$. The Cartan generators and root operators of $SO(2r)_1$ were presented in \eqref{SO2rcurrents}.
These operators coincide with the Cartan generators and long root operators of $SO(2r+1)_1$.
Because the $SO(2r+1)_1$ algebra extends $SO(2r)_1$, we must also include the following current operators: \begin{align}\left[\mathtt{E}_{SO(2r+1)}(\mathsf{x})\right]_{\pm{\bf e}_j}=i\psi_{2r+1}e^{\pm i\left(\phi_j(\mathsf{x})+k_j\mathsf{x}\right)},\label{eq:SOoddshortroots}\end{align} which pair the Majorana fermion $\psi_{2r+1}$ with one of the Dirac fermions $d_{j=1,\ldots,r}$ or $d^\dagger_{j=1,\ldots,r}$ in $SO(2r)_1$. Each of the additional currents in \eqref{eq:SOoddshortroots}, referred to as a short root operator, is a linear combination of two $E_8$ root operators: \begin{align}\begin{split}&\left[\mathtt{E}_{SO(2r+1)}(\mathsf{x})\right]_{\pm{\bf e}_j}\\&e^{i\left(\pm\phi_j(\mathsf{x})+\phi_{2r+1}(\mathsf{x})\right)\pm ik_j\mathsf{x}}+e^{i\left(\pm\phi_j(\mathsf{x})-\phi_{2r+1}(\mathsf{x})\right)\pm ik_j\mathsf{x}},\end{split}\end{align} because $\psi_{2r+1}=(d_{r+1}+d_{r+1}^\dagger)/\sqrt{2}$. In order for the short root current backscattering to preserve charge and momentum conservation, the two $E_8$ roots must carry identical charge and momentum; this ensures the linear combination transforms homogeneously. This requires $d_{r+1}\sim e^{i\phi_{r+1}}$ to be electrically neutral and have trivial momentum: \begin{align}q_{r+1}=k_{r+1}=0.\label{SOoddchargemomentumrequirement}\end{align} 
The conjugate $SO(15-2r)_1$ algebra can be organized in a similar manner. It contains the bosonized $SO(14-2r)_1$ subalgebra, which is generated by $\psi_{p=2r+3,\ldots,16}$ and is associated with the long root vectors $\boldsymbol\alpha_{SO(14-2r)}=\pm{\bf e}_j\pm{\bf e}_{j'}$, for $r+2\leq j<j'\leq8$. $SO(15-2r)_1$ extends $SO(14-2r)_1$ by including the short roots \begin{align}\left[\mathtt{E}_{SO(15-2r)}(\mathsf{x})\right]_{\pm{\bf e}_j}=i\psi_{2r+2}e^{\pm i\left(\phi_j(\mathsf{x})+k_j\mathsf{x}\right)},\end{align} for $j=r+2,\ldots,8$. 

The coupled-wire model of the $SO(2r+1)_1$ bFQH state follows the recipe given in \eqref{HA}. Each 11-channel electron wire is turned into the bosonic $E_8$ CFT by the intra-wire backscattering interaction $\mathcal{H}^f_{\mathrm{intra}}$ from \eqref{E8Hintra} that gaps all odd fermion excitations. The $(E_8)_1$ WZW algebra is split into $SO(2r+1)\times SO(15-2r)$, and the two decoupled sectors are gapped by inter-wire and intra-wire current backscattering interactions \eqref{HinterA} and \eqref{HintraB}. Similar to the $SO(2r)$ theory (c.f.~\eqref{SO2rHinter} and \eqref{SO2rHintra}), these current backscattering terms are quartic in the fermions, as in the Gross-Neveu model~\cite{GrossNeveu1974,PhysRevB.94.165142},
\begin{widetext}
\begin{align}\begin{split}
\mathcal{H}^{SO(2r+1)}_{\rm inter}&=u_{\rm inter}\sum_y\sum_{1\leq p<q\leq2r+1}\psi^R_{yp}\psi^R_{yq}\psi^L_{y+1,p}\psi^L_{y+1,q},\quad\mbox{for $r=1,\ldots,6$}\\
&=u_{\rm inter}\sum_y\Bigg[\sum_{l=1}^r \left[\mathtt{H}_{SO(2r)}\right]^R_{y,l}\left[\mathtt{H}_{SO(2r)}\right]^L_{y+1,l}-\sum_{\boldsymbol\alpha_{SO(2r)}}\cos\left(\boldsymbol\alpha_{SO(2r)}\cdot\boldsymbol\theta_{y+1/2}\right)\\&
\;\;\;\;-2i\psi^R_{y,2r+1}\psi^L_{y+1,2r+1}\sum_{j=1}^r\cos\left({\bf e}_j\cdot\boldsymbol\theta_{y+1/2}\right)\Bigg],\end{split}\label{SOoddHinter}\\
\begin{split}\mathcal{H}^{SO(15-2r)}_{\rm intra}
&=u_{\rm intra}\sum_y\sum_{2r+2\leq p<q\leq16}\psi^R_{yp}\psi^R_{yq}\psi^L_{yp}\psi^L_{yq},\quad\mbox{for $r=1,\ldots,6$}\\
&=u_{\rm intra}\sum_y\Bigg[\sum_{m=r+2}^8\left[\mathtt{H}_{SO(14-2r)}\right]^R_{y,m}\left[\mathtt{H}_{SO(14-2r)}\right]^L_{y,m}-\sum_{\boldsymbol\alpha_{SO(14-2r)}}\cos\left(\boldsymbol\alpha_{SO(14-2r)}\cdot
\boldsymbol\theta_y\right)\\
&\;\;\;\;-2i\psi^R_{y,2r+2}\psi^L_{y,2r+2}\sum_{j=r+2}^8\cos\left({\bf e}_j\cdot\boldsymbol\theta_y\right)\Bigg].\end{split}\label{SOoddHintra}
\end{align}
\end{widetext}
The sine-Gordon vector variables have entries $\theta_{y+1/2,j}=\phi^R_{y,j}-\phi^L_{y+1,j}$ and $\theta_{y,j}=\phi^R_{y,j}-\phi^L_{y,j}$, for $j=1,\ldots,8$. 
There is no sum over long roots $\boldsymbol\alpha_{SO(2r)} =\pm{\bf e}_j\pm{\bf e}_{j'}$ ($\boldsymbol\alpha_{SO(14-2r)}$) when $r=1$ ($r=6$) because $SO(2)$ does not have a root system. 
The second line of \eqref{SOoddHinter} is identical to the inter-wire interactions $\mathcal{H}^{SO(2r)}_{\rm inter}$ encountered in \eqref{SO2rHinter}. They are marginally relevant when $u_{\rm inter}>0$ and simultaneously pin the ground-state expectation values $\left\langle\theta_{y+1/2,j}(\mathsf{x})\right\rangle=\pi n_j$, for $j=1,\ldots,r$, where $n_{j=1,\ldots,r}$ are either all even integers or all odd integers. Terms in the last line of \eqref{SOoddHinter} back-scatter the short roots. 
At low energies, they effectively become the single-fermion backscattering interaction, \begin{align}&-2u_{\rm inter}\sum_yi\psi^R_{y,2r+1}\psi^L_{y+1,2r+1}\sum_{j=1}^r\left\langle\cos\left(\theta_{y+1/2,j}\right)\right\rangle\nonumber\\&=-2ru_{\rm inter}\sum_y(-1)^{n_{y+1/2,j}}i\psi^R_{y,2r+1}\psi^L_{y+1,2r+1},\label{Majoranamass}\end{align} which gaps the remaining Majorana fermions, and pins $\left\langle i\psi^R_{y,2r+1}(\mathsf{x})\psi^L_{y+1,2r+1}(\mathsf{x})\right\rangle\sim(-1)^{n_{y+1/2,j}}$. The intra-wire interactions \eqref{SOoddHintra} similarly gaps all degrees of freedom in $SO(15-2r)_1$. Together, $\mathcal{H}^{SO(2r+1)}_{\rm inter}$ and $\mathcal{H}^{SO(15-2r)}_{\rm intra}$ produce a state with a finite bulk excitation energy gap and leave behind the gapless chiral $SO(2r+1)_1$ WZW CFT on the edge.

The $SO(2r+1)_1$ topological phase has an Ising topological order that corresponds to the $SO(2r+1)_1$ WZW CFT on any boundary edge. There are three super-selection sectors of primary fields (anyons) $1$, $f$, and $\sigma$. The fermion sector $f$ has spin $h=1/2$ and is spanned by the $2r+1$ Majorana fermions $\psi_{p=1,\ldots,r+1}$. The Ising twist field (also known as Ising anyon) $\sigma$ has spin $h=(2r+1)/16$. The Ising sector consists of spinor fields \begin{align}
 \sigma={\rm span}
 \left\{
 \sigma_{2r+1}\exp\left(\frac{i}{2}\sum_{j=1}^r\varepsilon^j\phi_j\right):\varepsilon_{j=1,\ldots,r}=\pm1\right\}.\label{eq:primarysigmaSOodd}
\end{align}
Here, the vertex field $e^{i\varepsilon^j\phi_j/2}$ is the spinor field $s_\pm$ of $SO(2r)_1$. There is a $\pi$-monodromy between this field and any of the first $2r$ Majorana fermions $\psi_{p=1,\ldots,2r}$. It corresponds to a $\pi$-kink in $\langle\theta_j(\mathsf{x})\rangle$ where the ground state expectation value jumps by an odd integer multiple of $\pi$ from one side to another. The Majorana mass $2r(-1)^{n_j}u_{\mathrm{inter}}$ of $\psi_{2r+1}$ in \eqref{Majoranamass} therefore changes sign at the kink, and traps a Majorana zero mode at the domain wall. This corresponds to the Ising twist field $\sigma_{2r+1}$ that has a $\pi$-monodromy with $\psi_{2r+1}$. Since the fermion pairs $\psi_p\psi_{2r+1}$ are local on any given wire in each chiral sector, the Ising twist field $\sigma_{2r+1}$ and the spinor field $s_\pm=e^{i\varepsilon^j\phi_j/2}$ are confined together. $\sigma_{2r+1}$ has spin $1/16$ while $s_\pm$ has spin $r/8$. They add up to the total spin $h_{\boldsymbol\sigma}=(2r+1)/16$. The three anyon sectors have the fusion rules:
\begin{align}
\begin{split}
f\times f=1,\quad f\times\sigma=\sigma,\quad \sigma\times \sigma=1+f.
\end{split}\label{Isingfusion}
\end{align}

Next, we address the charge and momentum conservation of the $SO(2r+1)_1$ coupled-wire model. The model relies on a particular set of non-local Dirac fermions $d_j\sim e^{i\phi_j}$ (see \eqref{nonlocaldiracdef}) that are defined using a $R$ matrix that specifies the $E_8$ simple roots $e^{i\tilde\Phi_J}=e^{iR^j_J\phi_j}$ in Euclidean 8-space. Unlike the previous Abelian models, in order for $d_{r+1}$ to split into Majorana components $\psi_{2r+1}$ and $\psi_{2r+2}$ that back-scatter in different direction, the $R$-matrix must be chosen so that $d_{r+1}$ is electrically neutral, i.e.~$q_{r+1}=(R^{-1})_{r+1}^J\tilde{q}_J=0$ where the simple root electric charges $\tilde{q}_J$ were presented in \eqref{E8simlerootscharge}. The model conserves momentum when the electron Fermi momenta $k_{F,a}$ are set by \eqref{appmomentumAbelian} under the projection $P_A^{jj'}=1$ if $1\leq j=j'\leq r$ and 0 otherwise. This ensures (i) $d^R_{y,j}$ and $d^L_{y+1,j}$ have identical momenta for $j=1,\ldots,r$, (ii) $d^R_{y,j}$ and $d^L_{y,j}$ have identical momenta for $j=r+2,\ldots,8$, and (iii) $d^\sigma_{y,r+1}$ has vanishing momentum. Since $\psi_{2r+1}$ is electrically neutral, $d_{j=1,\ldots,r}$ are responsible for the charge response of the chiral $SO(2r+1)_1$ WZW CFT on the boundary edge. Therefore, the filling number is \begin{align}\nu_{SO(2r+1)}=\sum_{j=1}^rq_j^2.\end{align} Moreover, as $q_{r+1}=0$ and $\nu_{E_8}=\sum_{j=1}^8q_j^2=16$, the particle-hole conjugate $SO(15-2r)$ carries the conjugate filling number $\nu_{SO(15-2r)}=16-\nu_{SO(2r+1)}$.  
Fig.~\ref{fig:SOphasesfillings} summarizes the possible filling numbers and central charges $c_{SO(2r+1)}=(2r+1)/2$ of the various $SO(2r+1)_1$ bFQH states.

The allowed electric charges $q_j$ of the non-local Dirac fermions were presented above \eqref{E8fillingnumber}. With the additional charge neutral condition $q_{r+1}=0$, the electric charge of the WZW currents and primary fields of $SO(2r+1)_1$ can be read-off from their $SO(2r)$ components. The charge assignments are summarized in Tables~\ref{tab:SOoddChargeAssignmentRoot} and \ref{tab:SOoddChargeAssignmentPrimary}. Similar to the Abelian $SO(10)_1$, $SO(12)_1$, and $SO(14)_1$, we notice that there are two distinct charge patterns at filling $\nu=16$ for each of $SO(9)_1$, $SO(11)_1$, $SO(13)_1$ and $SO(15)_1$. The two cases correspond to the two charge vectors ${\bf q}=(4,0,0,0,0,0,0,0)$ and $(2s_1,2s_2,2s_3,2s_4,0,0,0,0)$, for $s_j=\pm1$, of the non-local Dirac fermions. When the two states are put side by side so that they share a boundary, the distinct WZW current charge patterns forbid all currents to be back-scattered from one state to another without violating charge conservation. On the other hand, there may exist alternative gapping interactions on the shared edge that involve the backscattering of higher-spin bosons. For example, this has been seen in \eqref{SO101214sharededge} for the two $SO(10)_1$ states at filling 16. 

We now present such an alternative gapping interaction on the shared edge of the two $SO(11)_1$ states at filling 16:
\begin{align}\mathcal{U}'=\mathcal{U}+iu\psi^R_{11}\psi^L_{11}\cos\left(\phi^R_3+\phi^R_4+\phi^R_5-\phi^L_3-\phi^L_4+\phi^L_5\right),\label{SO11sharededge}\end{align} where the potential $\mathcal{U}$ was defined in \eqref{SO101214sharededge}. $\mathcal{U}$ gaps the $SO(10)_1$ sub-sector consisting of the Dirac fermions $d^{L/R}_{j=1,\ldots,5}=e^{i\phi^{L/R}_{j=1,\ldots,5}}$. The last term in \eqref{SO11sharededge} is an integral electronic combination because it is proportional to the 4-body backscattering $(\psi^R_{11}d^R_3d^R_4d^R_5)(\psi^L_{11}{d^R_3}^\dagger {d^R_4}^\dagger d^R_5)$ and any fermion pair from the same edge is local. The sine-Gordon variable combination $\langle\phi^R_3+\phi^R_4+\phi^R_5-\phi^L_3-\phi^L_4+\phi^L_5\rangle$ takes a finite ground state expectation value in $\pi\mathbb{Z}$ because it can be expressed as a half-integral linear combination of the five sine-Gordon variables in potential $\mathcal{U}$ defined in \eqref{SO101214sharededge}. Therefore, at low energies, the last term of \eqref{SO11sharededge} is effectively the Majorana fermion backscattering $\pm i\psi^R_{11}\psi^L_{11}$, which gaps the remaining Majorana sector in $SO(11)_1$. 

Similar to \eqref{SO11sharededge}, alternative gapping potentials on the shared edges of the two $\nu=16$ $SO(13)_1$ states and the two $\nu=16$ $SO(15)_1$ states exist. The two $SO(9)_1$ states at filling 16 are special. The $SO(9)_1$ shared edge with distinct charge patterns between the left and right WZW currents cannot be gapped while preserving charge conservation. There is no even-body fermion backscattering involving $\psi^R_9$ and $\psi^L_9$ that preserves charge. However, the shared edge is gappable under an edge reconstruction that extends $SO(9)_1$ and includes additional counter-propagating fermion channels without changing the bulk topological order. For example, the $R$ edge, consisting of $\psi^R_{p=1,\ldots,9}$, can be extended to $\psi^R_1,\ldots,\psi^R_{10},\psi^L_{10}$, where any fermion pair within the set is local. The shared edge can then be gapped after extending both the $L$ and $R$ edges by a potential similar to \eqref{SO11sharededge}. To summarize, the two distinct charge patterns at filling 16 for each of the $SO(9)_1$, $SO(11)_1$, $SO(13)_1$ and $SO(15)_1$ state belong in the same bFQH phase. They are distinguishable only when the $SO(2r+1)$ symmetry is preserved.

\begin{figure}[htbp]
(a)\includegraphics[width=.20\textwidth]{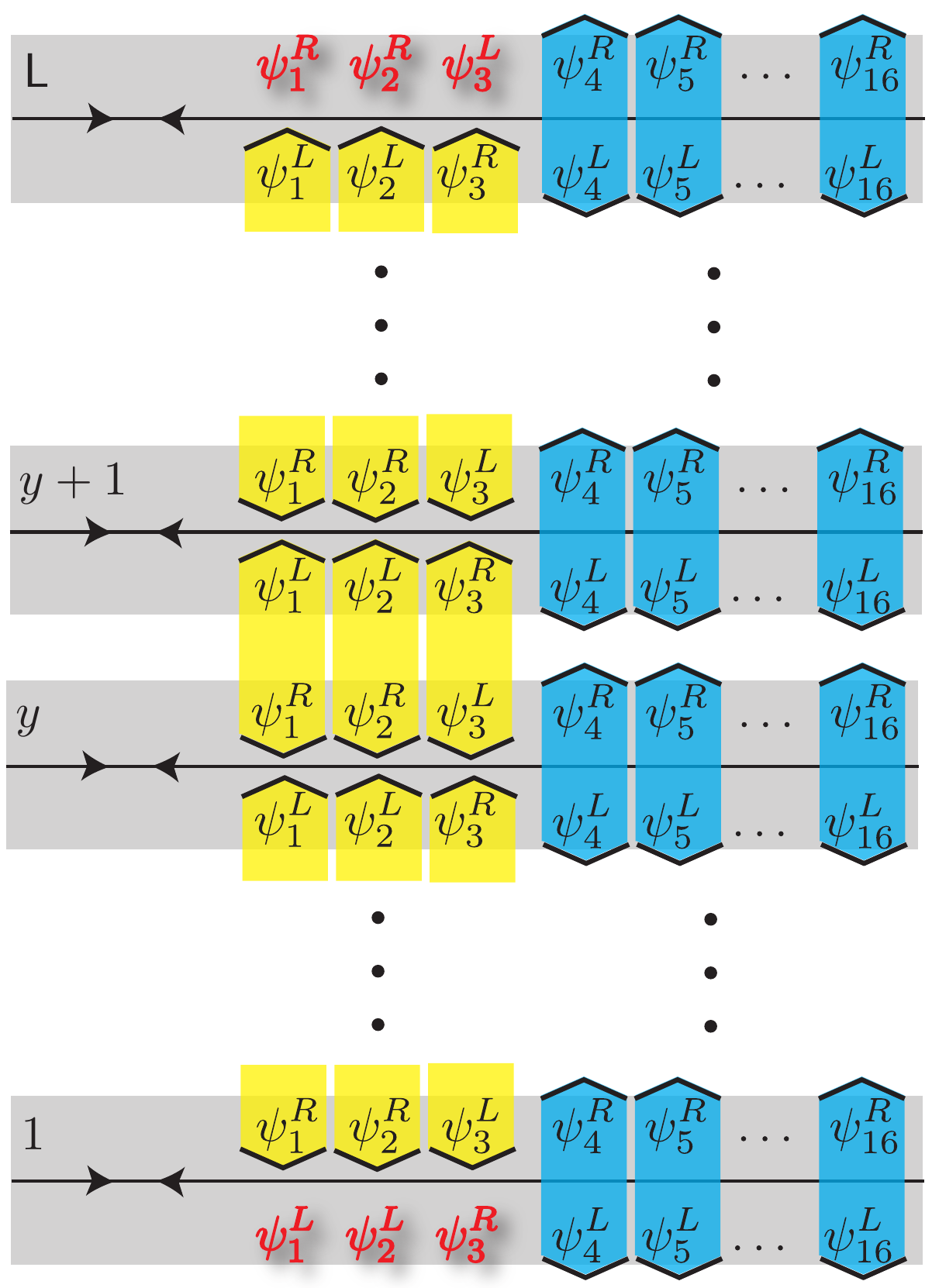}
(b)\includegraphics[width=.22\textwidth]{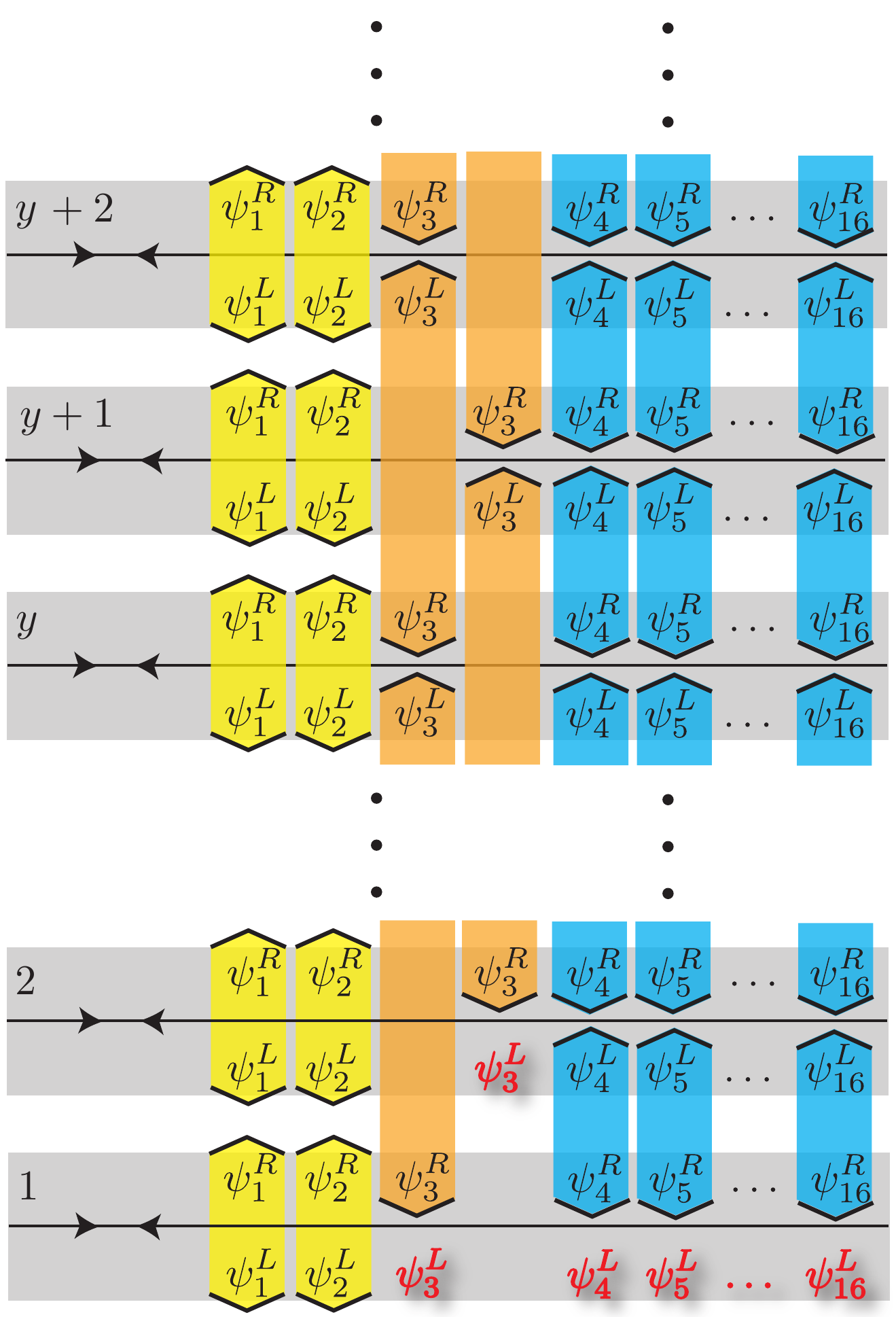}
\caption{Fermion bilinears take finite ground state expectation values $\langle\psi^R_{yp}\psi^L_{y'p}\rangle$ in the (a) $SO(1)_1$ and (b) $SO(15)_1$ coupled-wire models. The chiral Majorana fermions in red near the boundary edges remain gapless.}\label{fig:so1so15scatter}
\end{figure}

Lastly, we address the coupled-wire construction of the $SO(15)_1$ and $SO(1)_1$ states. Recall that $SO(1)_1$ refers to the $c=1/2$ Ising minimal model CFT $\mathcal{M}(4,3)$ \cite{francesco2012conformal}, which consists of a single Majorana fermion.
This CFT is {\em not} a WZW theory. 
Although $SO(1)_1 \times SO(15)_1$ still conformally embeds into $(E_8)_1$, in the sense that the energy-momentum tensor decomposes $T_{E_8}=T_{SO(15)}+T_{SO(1)}$, a coupled-wire model cannot be constructed based on the splitting. This is because the Ising sector cannot be gapped by backscattering local currents. Single Majorana fermion backscattering $\psi^L\psi^R$ within the same wire or between wires is {\em not} local because the Majorana fermion is not an integral combination of electrons. Here, we present an alternative construction for the $SO(15)_1$ bFQH state at filling 16 and the electrically neutral $SO(1)_1=\mathrm{Ising}$ states. They support anyon quasiparticle excitations $f$ and $\sigma$ in the bulk that follow the fusion rules \eqref{Isingfusion}. They have spins $h_f=1/2$ and $h_\sigma=15/16$ or $1/16$ for $SO(15)_1$ or $SO(1)_1$ respectively.

The coupled-wire models are based on the same 11-channel electron wire arrays, as before. The intra-wire backscattering $\mathcal{H}_{\mathrm{intra}}^f$ in \eqref{E8Hintra} gaps the three integrated fermions and leave behind the $E_8$ WZW CFTs on each wire. To construct the $SO(1)_1$ state, the model Hamiltonian $\mathcal{H}[SO(1)_1]=\mathcal{H}_0+\mathcal{H}_{\mathrm{intra}}^f+\mathcal{H}_{\mathrm{inter}}+\mathcal{H}_{\mathrm{intra}}$ in \eqref{HA} consists of the following inter-wire and intra-wire backscattering interactions.
\begin{align}
\begin{split}
    \mathcal{H}_{\rm intra}
    &=-u_{\rm intra}\sum_{y}\sum_{4\leq p<q\leq 16}
\psi^R_{y,p}\psi^R_{y,q}\psi^L_{y,p}\psi^L_{y,q},\\
\mathcal{H}_{\rm inter}&=-u_{\rm inter}
\sum_{y} \Big[ 
\psi^R_{y,1}\psi^R_{y,2}\psi^L_{y+1,1}\psi^{L}_{y+1,2}\\
&+\lambda_0^2
\psi^R_{y,1}\psi^R_{y+1,3}
\psi^L_{y+1,1}\psi^{L}_{y,3}
\psi^R_{y,4}\psi^L_{y,4}\psi^R_{y+1,4}\psi^L_{y+1,4}
\\
&+\lambda_0^2
\psi^R_{y,2}\psi^R_{y+1,3}
\psi^L_{y+1,2}\psi^{L}_{y,3}
\psi^R_{y,4}\psi^L_{y,4}
\psi^R_{y+1,4}\psi^L_{y+1,4}
\Big].
\label{eq:SO1Interactions}
\end{split}
\end{align}
The model preserves charge and momentum conservation when (i) the $R$-matrix is chosen so that the Dirac fermion $d_2=(\psi_3+i\psi_4)/\sqrt{2}$ has zero electric charge, $q_2=(R^{-1})^J_2\tilde{q}_J=0$, and (ii) the electron Fermi momenta $k_{F,a}$ are set by \eqref{appmomentumAbelian} under the projection $P_A^{jj'}=1$ if $j=j'=1$ and 0 otherwise. $\mathcal{H}_{\mathrm{intra}}$ contains local $SO(13)_1$ current backscattering terms that collectively gap the fermions $\psi_{4,\ldots,16}$ within a wire. At low-energy, fermion pairs $i\psi^R_{y,p}\psi^L_{y,p}$, for $p=4,\ldots,16$, can be replaced by their ground state (mean-field) expectation values $\langle i\psi^R_{y,p}\psi^L_{y,p}\rangle=(-1)^{s_y}/\lambda$. Here, the sign is uniform and independent from $p$, and $\lambda$ is some finite positive scalar with units of length. 

The inter-wire interactions $\mathcal{H}_{\mathrm{inter}}$ are a combination of products of local fermion pairs $\psi^\sigma_{y,r}\psi^\sigma_{y,p}$. The two 4-body terms in \eqref{eq:SO1Interactions} consist of products of the form: \begin{align}(\psi^R_{y,r}\psi^R_{y,4})(\psi^R_{y+1,3}\psi^R_{y+1,4})
(\psi^L_{y+1,r}\psi^L_{y+1,4})(\psi^{L}_{y,3}\psi^L_{y,4}),\end{align} for $r=1,2$, where the two fermions in each parentheses have the same wire and chiral labels and thus the pair is a local integral combination of electrons. The scalar $\lambda_0$ has units of length so that the two 4-body terms share the same dimension as the two-body one. Although the 4-body terms are by themselves irrelevant, they become marginally relevant when $u_{\mathrm{inter}}>0$ in the presence of $\mathcal{H}_{\mathrm{intra}}$. This is because, at low-energy, the fermion bilinears $\psi^R_4\psi^L_4$ take finite expectation values, and in the mean-field approximation, $\mathcal{H}_{\rm inter}$ becomes the two-body interaction,
\begin{align}\begin{split}
    \mathcal{H}^{\rm mf}_{\rm inter}&=-u_{\rm inter}
\sum_{y=1}^{\mathsf{L}-1} \Big[ 
\psi^R_{y,1}\psi^R_{y,2}\psi^L_{y+1,1}\psi^{L}_{y+1,2}\\
&+
\lambda_0^2\psi^R_{y,1}\psi^R_{y+1,3}\psi^L_{y+1,1}\psi^{L}_{y,3}
\langle \psi^R_{y,4}\psi^L_{y,4} \rangle
\langle \psi^R_{y+1,4}\psi^L_{y+1,4} \rangle\\
&+
\lambda_0^2\psi^R_{y,2}\psi^R_{y+1,3}\psi^L_{y+1,2}\psi^{L}_{y,3} 
\langle \psi^R_{y,4}\psi^L_{y,4} \rangle
\langle \psi^R_{y+1,4}\psi^L_{y+1,4} \rangle\Big]\\
&=-u_{\rm inter}
\sum_{y=1}^{\mathsf{L}-1} \Big[ 
\psi^R_{y,1}\psi^R_{y,2}\psi^L_{y+1,1}\psi^{L}_{y+1,2}\\
&\;\;\;\;+(-1)^{1+s_y+s_{y+1}}\frac{\lambda_0^2}{\lambda^2}\psi^R_{y,1}\psi^R_{y+1,3}\psi^L_{y+1,1}\psi^{L}_{y,3}\\
&\;\;\;\;+(-1)^{1+s_y+s_{y+1}}\frac{\lambda_0^2}{\lambda^2}\psi^R_{y,2}\psi^R_{y+1,3}\psi^L_{y+1,2}\psi^{L}_{y,3}\Big].
\end{split}
\label{eq:SO1intermf}
\end{align}
It is crucial to recognize that the last two terms in $\mathcal{H}^{\rm mf}_{\rm inter}$ are not local because the four fermions in each term all have different wire and chirality indices. Therefore, $\mathcal{H}^{\rm mf}_{\rm inter}$ cannot be directly used in the electron-based coupled-wire model, and instead has to originate from the full many-body interactions in \eqref{eq:SO1Interactions}. The signs $(-1)^{1+s_y+s_{y+1}}$ depend on the ground states of $\mathcal{H}_{\mathrm{intra}}$. When $(-1)^{1+s_y+s_{y+1}}\lambda_0^2/\lambda^2=1$, the mean-field Hamiltonian is identical to the current backscattering $\mathcal{H}^{\rm mf}_{\rm inter}=u_{\mathrm{inter}}{\bf J}_{SO(3)_1}^R\cdot{\bf J}_{SO(3)_1}^L$, where the $R$-chiral ($L$-chiral) $SO(3)_1$ currents are generated by fermion pairs within the chiral set $\psi^R_{y,1},\psi^R_{y,2},\psi^R_{y+1,3}$ (resp.~$\psi^L_{y+1,1},\psi^L_{y+1,2},\psi^L_{y,3}$). It gaps the three counter-propagating pairs of Majorana fermions. In general, by bosonizing $(\psi^R_{y,1}+i\psi^R_{y,2})/\sqrt{2}\sim e^{i\phi^R_{y,1}}$ and $(\psi^L_{y+1,1}+i\psi^R_{y+1,2})/\sqrt{2}\sim e^{i\phi^L_{y+1,1}}$, the mean-field Hamiltonian \eqref{eq:SO1intermf} can be expressed as \begin{align}\begin{split}\mathcal{H}^{\rm mf}_{\rm inter}&=u_{\mathrm{inter}}\sum_y\big[\partial_{\mathsf{x}}\phi^R_{y,1}\partial_{\mathsf{x}}\phi^L_{y+1,1}\\&\quad\quad-(-1)^{1+s_y+s_{y+1}}\frac{\lambda_0^2}{\lambda^2}i\psi^R_{y+1,3}\psi^L_{y,3}\cos\theta_{y+1/2,1}\big],\end{split}\label{eq:SO1intermf2}\end{align} where $\theta_{y+1/2,1}=\phi^R_{y,1}-\phi^L_{y+1,1}$. Similar to \eqref{Majoranamass}, in order to minimize the energy \eqref{eq:SO1intermf2}, the sine-Gordon angle parameter and Majorana mass must have correlated ground state expectation values:
 \begin{align}\begin{split}&\langle\theta_{y+1/2,1}(\mathsf{x})\rangle=\pi m_{y+1/2},\\&\left\langle i\psi^R_{y+1,3}(\mathsf{x})\psi^L_{y,3}(\mathsf{x})\right\rangle\sim(-1)^{m_{y+1/2}},\end{split}\end{align} where $m_{y+1/2}$ is an integer. The sine-Gordon term fixes the same sign $(-1)^{m_{y+1/2}}$ for $\langle i\psi^R_{y,1}\psi^L_{y,1}\rangle$ and $\langle i\psi^R_{y,2}\psi^L_{y,2}\rangle$.

We now see that $\mathcal{H}_{\rm inter}+\mathcal{H}_{\rm intra}$ gaps all bulk Majorana fermions in a closed torus geometry. On the other hand, in an open geometry with boundary edges at $y=1$ and $y=\mathsf{L}$, the sum of wires in $\mathcal{H}_{\mathrm{inter}}$ only runs over $y=1,\ldots,\mathsf{L}-1$. There remains gapless Majorana fermions $\psi^L_{y=1,1}$, $\psi^L_{y=1,2}$, $\psi^R_{y=1,3}$ on the bottom edge and $\psi^R_{y=\mathsf{L},1}$, $\psi^R_{y=\mathsf{L},2}$, $\psi^L_{y=\mathsf{L},3}$ on the top edge as indicated in Fig.~\ref{fig:so1so15scatter}(a). At neutral filling $\nu=0$ when the Dirac fermion $d^\sigma_{y,1}=(\psi^\sigma_{y,1}+i\psi^\sigma_{y,2})/\sqrt{2}$ is electrically neutral, the edge states are unstable to the local edge perturbation, \begin{align}\begin{split}
   \mathcal{U}_{\rm edges}&=-u\psi^R_{1,3}\psi^R_{1,4}\psi^L_{1,2}\psi^L_{1,4}-u\psi^R_{\mathsf{L},3}\psi^R_{\mathsf{L},4}\psi^L_{\mathsf{L},2}\psi^L_{\mathsf{L},4}\\
   &\rightarrow u\psi^R_{1,3}\psi^L_{1,2} \langle \psi^R_{1,4}\psi^L_{1,4} \rangle+u\psi^R_{\mathsf{L},3}\psi^L_{\mathsf{L},2} \langle \psi^R_{\mathsf{L},4}\psi^L_{\mathsf{L},4} \rangle.
\end{split}\label{SO1edgereconstruction}\end{align} The pair $\langle\psi^R_{y,4}\psi^L_{y,4}\rangle$ takes finite ground state expectation values from $\mathcal{H}_{\mathrm{intra}}$. $\mathcal{U}_{\mathrm{edges}}$ introduces a Majorana mass for $\psi^\sigma_2$ and $\psi^{-\sigma}_3$ on the top and bottom edges. This leaves behind the single chiral Ising CFT $SO(1)_1$ on the boundaries, generated by the lone $\psi^L_{y=1,1}$ on the bottom edge and $\psi^R_{y=\mathsf{L},1}$ on the top edge. 

On the other hand, at non-zero fillings $\nu=1,4,9,16$ when $d^\sigma_{y,1}$ carries electric charge $q_1=\pm1,\pm2,\pm3,\pm4$ respectively, \eqref{SO1edgereconstruction} does not preserve charge conservation. 
The edge CFT $U(1)_4\times\overline{\mathrm{Ising}}$---consisting of the charge carrying forward moving $d_1=(\psi_1+i\psi_2)/\sqrt{2}\sim e^{i\phi_1}$ and neutral backward moving $\bar\psi_3$---cannot be reduced. The fermions  $\psi_{1,2}$ and $\bar\psi_3$ differs from each other by local bosons (up to the ground state expectation value $\langle\psi^R_4\psi^L_4\rangle$) and belong to the same fermion anyon class $f$. The spinor fields $e^{\pm i\phi_1/2}$ (spin $1/8$) and the Ising twist field $\bar\sigma_3$ (spin $-1/16$) of $\bar\psi_3$ are confined together and can only appear simultaneously. The combinations belong in the Ising anyon class $\sigma$, which has spin $h_\sigma=1/8-1/16=1/16$. This recovers the $SO(1)_1$ topological order. Unlike the electrically neutral case, the super-selection sectors $f$ and $\sigma$ now each contain a collection of fields with mixed chirality and electric charges, $Q_f=\{\pm q_1,0\}$ and $Q_\sigma=\{\pm q_f/2\}$. These charged $SO(1)_1$ bFQH states are not included in Figs.~\ref{fig:fillingcentralchargeplot}, \ref{fig:SOphasesfillings} and Tables~\ref{tab:SOoddChargeAssignmentRoot}, \ref{tab:SOoddChargeAssignmentPrimary}.

Moving on to the construction of the $SO(15)_1$ bFQH state, we apply the particle-hole conjugation on the $SO(1)_1$ state by shifting the wire and chiral labels in the Hamiltonian \eqref{eq:SO1Interactions} from $\psi^L_{y,p}\to\psi^R_{y,p}$ and $\psi^R_{y,p}\to\psi^R_{y+1,p}$. The model Hamiltonian $\mathcal{H}[SO(15)_1]=\mathcal{H}_0+\mathcal{H}_{\mathrm{intra}}^f+\mathcal{H}_{\mathrm{inter}}+\mathcal{H}_{\mathrm{intra}}$ consists of the following shifted inter-wire and intra-wire backscattering terms (see also Fig.~\ref{fig:so1so15scatter}(b)): \begin{align}\begin{split}
    &\mathcal{H}_{\rm intra}+\mathcal{H}_{\rm inter}\\
    &=-u\sum_{y}\sum_{4\leq p<q\leq 16}
\psi^L_{y+1,p}\psi^L_{y+1,q}\psi^R_{y,p}\psi^R_{y,q}\\
&\;\;\;-u\sum_{y} \Big[ 
\psi^L_{y,1}\psi^L_{y,2}\psi^R_{y,1}\psi^R_{y,2}\\
&\;\;\;+\lambda_0^2
\psi^L_{y+1,1}\psi^L_{y+2,3}
\psi^R_{y+1,1}\psi^R_{y,3}
\psi^L_{y+1,4}\psi^R_{y,4}\psi^L_{y+2,4}\psi^R_{y+1,4}
\\
&\;\;\;+\lambda_0^2
\psi^L_{y+1,2}\psi^L_{y+2,3}
\psi^R_{y+1,2}\psi^R_{y,3}
\psi^L_{y+1,4}\psi^R_{y,4}
\psi^L_{y+2,4}\psi^R_{y+1,4}
\Big].
\label{eq:SO15Interactions}
\end{split}\end{align}
The model preserves charge and momentum conservation when (i) the $R$-matrix is chosen so that the Dirac fermion $d_2=(\psi_3+i\psi_4)/\sqrt{2}$ has zero electric charge, $q_2=(R^{-1})^J_2\tilde{q}_J=0$, and (ii) the electron Fermi momenta $k_{F,a}$ are set by \eqref{appmomentumAbelian} under the projection $P_A^{jj'}=1$ if $3\leq j=j'\leq8$ and 0 otherwise. Using a similar mean-field analysis as in \eqref{eq:SO1intermf}, the model gaps all fermions in the bulk but leaves behind gapless Majorana fermions on the boundary. Since the model involves second nearest wire interactions, the Majorana fermions that remain gapless near the bottom edge are $\psi_{y=1,p=1,\ldots,16}^L$, $\psi^R_{y=1,p=1,2}$, and $\psi^L_{y=2,p=3}$. Two pairs of Majoranan fermions can further be gapped by the edge potential at $y=1$: \begin{align}\mathcal{U}_{\mathrm{edge}}=u\partial_{\mathsf{x}}\phi^R_1\partial_{\mathsf{x}}\phi^L_1-u'\cos\left[2\left(\phi^R_1-\phi^L_1\right)\right],\end{align} for a sufficiently strong ``repulsive" interaction $u<-\frac{3\tilde{v}}{10\pi}$, so that the sine-Gordon potential is relevant. There remains 15 chiral Majoranan fermions $\psi^L_{y=1,p=3,\ldots,16}$ and $\psi^L_{y=2,p=3}$ (see Fig.~\ref{fig:so1so15scatter}(b)). They generate the chiral $SO(15)_1$ WZW CFT on the edge. All 15 fermion fields belong to the same anyon class $f$. For instance, $\psi^L_{y=1,p}$ and $\psi^L_{y=2,3}$ differ from each other by local bosons and ground state expectation values $\langle\psi^L_{y=2,4}\psi^R_{y=1,4}\rangle$. The Ising twist fields of each of the 15 fermions must appear simultaneously due to electron locality. Therefore, the bFQH state carries the $SO(15)_1$ topological order. Although the possible filling numbers of the $SO(15)_1$ bFQH models are $\nu=16,15,12,7,0$, depending on the electric charge $q_1=0,1,2,3,4$ of the Dirac fermion $d_1=(\psi_1+i\psi_2)/\sqrt{2}$, we only include the filling 16 case in Figs.~\ref{fig:fillingcentralchargeplot}, \ref{fig:SOphasesfillings} and Tables~\ref{tab:SOoddChargeAssignmentRoot}, \ref{tab:SOoddChargeAssignmentPrimary}. This is because it is the only bFQH state that is particle-hole conjugate to the neutral $\nu = 0$ $SO(1)_1$ Ising state.

\subsubsection{The \texorpdfstring{$G_2$}{G2} and \texorpdfstring{$F_4$}{F4} Fibonacci states}

We now construct bosonic fractional quantum Hall (bFQH) states with Fibonacci topological order. These phases support Fibonacci anyon excitations~\cite{ReadRezayi,SlingerlandBais01}. The gapless edge modes are described by either the $G_2$ or $F_4$ WZW CFT at level 1. These states were first constructed by the coupled-wire approach in Ref.~\onlinecite{PhysRevB.100.085116} at filling $\nu=8$. We begin by briefly reviewing this construction. Then, we present the following newly-discovered properties of these (and related) Fibonacci states: (i) the explicit charge assignments of the $G_2$ and $F_4$ currents and Fibonacci primary fields, (ii) the new Fibonacci states at filling $\nu=0$ and 16, and (iii) the realization of bulk Fibonacci anyons using an open string of local boson operators.

The coupled-wire models begin with the embedding of the $(G_2)_1$ and $(F_4)_1$ WZW subalgebras in $(E_8)_1$. With the help of the 8 non-local fermions $d_j\sim e^{i\phi_j}$ that represent the $E_8$ (see \eqref{nonlocaldiracdef}), the embedding makes use of the ``parafermion" decomposition of the current algebra~\cite{GEPNER198710}, \begin{align}(G_2)_1\simeq SU(3)_1\times\mathbb{Z}_3,\label{G2SU3Z3}\end{align} where ``$\mathbb{Z}_3$" stands for the $\mathbb{Z}_3$ parafermion CFT~\cite{ZamolodchikovFateev85} with central charge $c=4/5$. 
$G_2$ has $2$ Cartan generators, $6$ long roots, and $6$ short roots.
The Cartan generators and long roots of $G_2$ coincide with those of $SU(3)_1$, presented in \eqref{SU3roots} and \eqref{SU3Cartangenerators} in \S\ref{sec:SU3E6}. The $(G_2)_1$ WZW algebra extends $SU(3)_1$ by the $\mathbb{Z}_3$ parafermion CFT, which is the coset $SU(2)_3/U(1)_6$ (or equivalently, $(G_2)_1/SU(3)_1$). 
The $\mathbb{Z}_3$ parafermion primary field $\Psi$ and its hermitian conjugate $\Psi^\dagger$ carry spin $h=2/3$ and obey the 3-nilpotent fusion rule.
The latter means that $\Psi^3=(\Psi^\dagger)^3=\Psi\times\Psi^\dagger=1$ belong in the trivial primary sector of local fields. The parafermion field can be chosen to be the linear combination of vertex operators \begin{align}
    \Psi=\frac{1}{\sqrt{3}}\left[-e^{-i2(\phi_1+\phi_2+\phi_3)/3}
    +2e^{i(\phi_1+\phi_2+\phi_3)/3}\sin\phi_4\right]
    \label{eq:Z3parafermion}
\end{align} for each wire $y$ and chiral sector $\sigma=R,L$. The coefficients of the linear combination are chosen so that, according to the bosonized variables correlations \eqref{TOcorrelDirac}, the parafermion fields obey the OPE \begin{align}\Psi(z)\Psi^\dagger(w)&=\frac{1}{(z-w)^{4/3}}\left[1+\frac{5}{3}(z-w)^2T_{\mathbb{Z}_3}(w)\right]+\ldots,\nonumber\\\Psi(z)\Psi(w)&=\frac{2/\sqrt{3}}{(z-w)^{2/3}}\Psi^\dagger(w)+\ldots,\end{align} where $T_{\mathbb{Z}_3}(z)$ is the energy-momentum tensor of the $\mathbb{Z}_3$ parafermion CFT. The 6 short roots of $(G_2)_1$ are the decoupled products $\mathcal{E}^\dagger\Psi$ and $\mathcal{E}\Psi^\dagger$, where $\mathcal{E}$ is the $SU(3)_1$ primary field triplet defined in \eqref{SU3primary}. We choose the positive short roots to be
\begin{align}
\begin{split}
[\mathtt{E}_{G_2}]_1 &=
e^{-i(\phi_1+\phi_2-2\phi_3)/3}\Psi,\\
[\mathtt{E}_{G_2}]_2 &=
e^{-i(\phi_1-2\phi_2+\phi_3)/3}\Psi,\\
[\mathtt{E}_{G_2}]_3 &=
e^{i(-2\phi_1+\phi_2+\phi_3)/3}\Psi^\dagger.
\end{split}\label{eq:G2shorts}
\end{align}
The negative short roots are the Hermitian conjugates of the positive ones. These operators have spin $h=h_{\mathcal{E}}+h_\Psi=1/3+2/3=1$, appropriate for a boson. Moreover, each of them is a linear combination of three $E_8$ roots and therefore is an even integral combination of local electrons. The locality of the $G_2$ short roots means the bosonic pairs $\mathcal{E}\Psi^\dagger$ and $\mathcal{E}^\dagger\Psi$ in the tensor product \eqref{G2SU3Z3} are ``anyon condensed"~\cite{PhysRevB.79.045316, 2018ARCMP...9..307B}. Together with the $SU(3)_1$ currents, they span the closed $G_2$ WZW algebra at level 1. In the coupled-wire model, in order for the short root current backscattering to conserve charge and momentum, all three vertex operator components of each short root must have identical charge and $\mathsf{x}$-momentum. This requires $d_1d_2d_3\sim e^{i(\phi_1+\phi_2+\phi_3)}$ and $d_4\sim e^{i\phi_4}$ to both have zero charge and vanishing momentum.

We define the $(F_4)_1$ WZW subalgebra as the complement subalgebra of $(G_2)_1$ in $(E_8)_1$. 
The $(F_4)_1$ WZW algebra has 4 Cartan generators, 24 long roots, and 24 short roots. It is an extension of $SO(9)_1$, \begin{align}(F_4)_1\simeq\mathcal{M}(5,4)\times SO(9)_1,\label{F4SO9M54}\end{align} where $\mathcal{M}(5,4)$ is the tricritical Ising CFT~\cite{francesco2012conformal} with central charge $c=7/10$.
Here, we have taken $SO(9)_1$ to be generated by $\psi_8,\ldots,\psi_{16}$. $(F_4)_1$ contains the simply-laced algebra $SO(8)_1$, which is generated by $d_{j=5,\ldots,8}=(\psi_{2j-1}+i\psi_{2j})/\sqrt{2}\sim e^{i\phi_{5,\ldots,8}}$. The 4 Cartan generators of $(F_4)_1$ are those of $SO(8)_1$: $[\mathtt{H}_{F_4}]_l=\partial_{\mathsf{x}}\phi_{4+l}$, for $l=1,\ldots,4$. The $24$ long roots of $(F_4)_1$ are the $SO(8)_1$ roots:
\begin{align}
[\mathtt{E}_{F_4}(\mathsf{x})]_{\boldsymbol{\alpha}_{SO(8)}}=e^{i\boldsymbol{\alpha}_{SO(8)}\cdot(\boldsymbol{\phi}(\mathsf{x})+\boldsymbol{k}\mathsf{x})},
\end{align}
where the long root vectors are $\boldsymbol{\alpha}_{SO(8)}=\{\pm({\bf e}_i\pm{\bf e}_j): 5\leq i<j\leq8\}$ and $\boldsymbol\phi=(\phi_5,\ldots,\phi_8)$. The $24$ short roots pair the $8$ vector fermions $e^{\pm i\phi_{5,\ldots,8}}$ and $16$ spinors $e^{i\boldsymbol\varepsilon\cdot\boldsymbol\phi/2}$ of $SO(8)_1$ with Majorana fermions $\psi_8$ and $\psi_\pm$, respectively~\cite{PhysRevB.100.085116}: 
\begin{align}
    [\mathtt{E}_{F_4}(\mathsf{x})]_{\pm j}&=\psi_8(\mathsf{x})e^{\pm i(\phi_{4+j}(\mathsf{x})+k_{4+j}\mathsf{x})}, \quad j=1,\ldots,4,\nonumber\\
    [\mathtt{E}_{F_4}(\mathsf{x})]_{\boldsymbol\varepsilon_{\pm}}&=\psi_{\pm}(\mathsf{x})e^{i\varepsilon^j_\pm(\phi_{4+j}(\mathsf{x})+k_{4+j}\mathsf{x})/2},
\label{F4shortroots}\end{align}
where $\varepsilon^j_\pm=1$ or $-1$ and $\prod_{j=1}^4\varepsilon^j_\pm=\pm1$. The 8 short roots in the first line of \eqref{F4shortroots} are the short roots of $SO(9)_1$ (c.f.~\eqref{eq:SOoddshortroots} in section~\ref{SOodd}). The remaining 16 short roots in the second line extend $SO(9)_1$ to $(F_4)_1$. The Majorana fermions $\psi_\pm$ are \begin{align}\psi_\pm=\frac{1}{\sqrt{2}}\omega_\pm e^{i(\phi_1+\phi_2+\phi_3\pm\phi_4)/2}+h.c.,\end{align} where $\omega_+=e^{i3\pi/8}$ and $\omega_-=e^{-i\pi/8}$. The $F_4$ short roots are linear combinations of $E_8$ roots. In the coupled-wire model, the current backscattering conserves charge and momentum when all vertex components in a given short root have identical charge and momentum. Like the $G_2$ case, this condition is satisfied when $d_1d_2d_3$ and $d_4$ both have vanishing charge and momentum.

The fermions $\psi_\pm$ are clearly decoupled from $SO(8)_1$ because they involve disjoint sets of bosonized variables. Moreover, $\psi_\pm$ are also decoupled from $SU(3)_1$, which is generated by the traceless combinations $\phi_1-\phi_2$, $\phi_2-\phi_3$ and $\phi_3-\phi_1$ that have non-singular correlations with $\phi_1+\phi_2+\phi_3$. The phases $\omega_\pm$ are chosen so that the OPE $\Psi(z)\psi_\pm(w)$ and $\Psi^\dagger(z)\psi_\pm(w)$ with the $\mathbb{Z}_3$ parafermion $\Psi$ in \eqref{eq:Z3parafermion} are non-singular according to the bosonized variables correlations \eqref{TOcorrelDirac}. This ensures the $(G_2)_1$ and $(F_4)_1$ currents have non-singular mutual OPE. It is essential to recognize that, although $\psi_8,\psi_\pm$ each has self-fermionic statistics and obeys the 2-nilpotent fusion rules, $\psi_8^2=\psi_\pm^2=1$, the three fermions have mutual {\em semionic} statistics due to the cross fusion rule, $\psi_8\times\psi_\pm=\psi_\mp$, that is due to the OPE, $\psi_8(z)\psi_\pm(w)\sim(z-w)^{-1/2}\psi_\mp(w)+\ldots$. 
The short roots $[\mathtt{E}_{F_4}]_{\boldsymbol\varepsilon_{\pm}}$ in \eqref{F4shortroots} can be decomposed in terms of primary fields in $\mathcal{M}(5,4)$ and $SO(9)_1$: \begin{align}[\mathtt{E}_{F_4}]_{\boldsymbol\varepsilon_{\pm}}\sim\left[\frac{7}{16}\right]\otimes\left[\frac{9}{16}\right]_{\boldsymbol\varepsilon_{\pm}}.\label{F4shortrootdecomposition}\end{align} $[7/16]$ is the primary field with spin $h=7/16$ at position $(r,s)=(2,1)$ on the conformal grid of the tricritical Ising CFT $\mathcal{M}(5,4)$~\cite{francesco2012conformal}. $[9/16]_{\boldsymbol\varepsilon_{\pm}}$ is the spinor field $\sigma_{\boldsymbol\varepsilon_\pm}=\sigma_8e^{i\varepsilon_\pm^j\phi_{4+j}/2}$ with spin $h=9/16$ in $SO(9)_1$, where $\sigma_8$ is the Ising twist field associated to $\psi_8$. The locality of $[\mathtt{E}_{F_4}]_{\boldsymbol\varepsilon_{\pm}}$ means the bosonic pairs \eqref{F4shortrootdecomposition} in the tensor product \eqref{F4SO9M54} are ``anyon condensed"~\cite{PhysRevB.79.045316, 2018ARCMP...9..307B}.

Having defined the WZW algebra embedding $(G_2)_1\times(F_4)_1\subseteq(E_8)_1$, we are ready to review the coupled-wire construction of the Fibonacci bFQH states~\cite{PhysRevB.100.085116}. Following \eqref{HA}, the electron-based models begin with the intra-wire backscattering $\mathcal{H}^f_{\mathrm{intra}}$ of the integrated fermions (see \eqref{E8Hintra}). This gaps all local odd fermion degrees of freedom and leaves behind the counter-propagating $(E_8)_1$ bosons on each wire. The Fibonacci states are constructed by back-scattering the $(G_2)_1$ currents and $(F_4)_1$ currents in complementary ways. 

For example, the $(G_2)_1$ model consists of the inter-wire and intra-wire interactions $\mathcal{H}^{G_2}_{\mathrm{inter}}+\mathcal{H}^{F_4}_{\mathrm{intra}}$:
\begin{align}
\mathcal{H}^{G_2}_{\rm inter}&=u_{\rm inter}\sum_y\boldsymbol{J}^R_{y,G_2}\cdot\boldsymbol{J}^L_{y+1,G_2}\label{HG2inter}\\
&=\mathcal{H}^{SU(3)}_{\rm inter}+u_{\rm inter}\sum_y\left(\mathcal{O}^{SU(3)}_{y+1/2}{\Psi^R_y}^\dagger\Psi^L_{y+1}+h.c.\right),\nonumber\\
\mathcal{H}^{F_4}_{\rm intra}&=u_{\rm intra}\sum_y\boldsymbol{J}^R_{y,F_4}\cdot\boldsymbol{J}^L_{y,F_4}\label{HF4intra}\\
&=\mathcal{H}^{SO(9)}_{\rm intra}+u_{\rm intra}\sum_y\sum_{\boldsymbol\varepsilon_\pm}{\left[\mathtt{E}_{F_4}\right]^R_{y,\boldsymbol\varepsilon_\pm}}^\dagger\left[\mathtt{E}_{F_4}\right]^L_{y,\boldsymbol\varepsilon_\pm},\nonumber
\end{align}
where $\mathcal{H}^{SU(3)}_{\rm inter}$ and $\mathcal{H}^{SU(3)}_{\rm inter}$ were defined in \eqref{SU3E6sineGordon} and \eqref{SOoddHintra} respectively. The operator $\mathcal{O}^{SU(3)}_{y+1/2}$ is the combination of the sine-Gordon vertex operators, \[e^{i(\theta_1+\theta_2-2\theta_3)/3}+e^{i(\theta_1-2\theta_2+\theta_3)/3}+e^{i(-2\theta_1+\theta_2+\theta_3)/3},\] where $\theta_j=\phi^R_{y,j}-\phi^L_{y+1,j}$. It is the singlet non-chiral product $\mathcal{E}^R_y\cdot{\mathcal{E}^L_{y+1/2}}^\dagger$ of $SU(3)_1$ primary fields in \eqref{SU3primary}; it is pinned at a finite ground state expectation value under $\mathcal{H}^{SU(3)}_{\rm inter}$. At low energies, the short root part of $\mathcal{H}^{G_2}_{\rm inter}$ is approximated by the $\mathbb{Z}_3$ parafermion backscattering terms, \begin{align}\mathcal{H}^{G_2}_{\rm inter}\to u_{\rm inter}\sum_y\left\langle\mathcal{O}^{SU(3)}_{y+1/2}\right\rangle{\Psi^R_y}^\dagger\Psi^L_{y+1}+h.c.\label{Z3parafermionbackscattering}\end{align}
This gaps the counter-propagating $\mathbb{Z}_3$ parafermion sectors~\cite{Fateev91}.

Next, we focus on the $F_4$ sector. The Gross-Neveu interaction $\mathcal{H}^{SO(9)}_{\mathrm{intra}}$ in \eqref{HF4intra} gaps the counter-propagating $SO(9)_1$ CFT on each wire (c.f.~\eqref{SOoddHintra} in \S\ref{SOodd}). On any given wire $y$, the non-chiral singlet combination, \begin{align}\sum_{\boldsymbol\varepsilon_\pm}\sigma^R_{\boldsymbol\varepsilon_\pm}\sigma^L_{-\boldsymbol\varepsilon_\pm}=\sum_{\boldsymbol\varepsilon_\pm}\sigma_8^R\sigma^L_8e^{i\varepsilon^j_\pm(\phi^R_{4+j}-\phi^L_{4+j})},\end{align} takes a finite ground state expectation value~\cite{PhysRevB.94.165142}. Therefore, using the identification \eqref{F4shortrootdecomposition}, the backscattering terms for the remaining $F_4$ short roots $[\mathtt{E}_{F_4}]_{\boldsymbol\varepsilon_\pm}$ are approximated at low energy by \begin{align}\mathcal{H}^{F_4}_{\mathrm{intra}}\to u_{\mathrm{intra}}\sum_y\left[\sum_{\boldsymbol\varepsilon_\pm}\left\langle\sigma^R_{y,\boldsymbol\varepsilon_\pm}\sigma^L_{y,-\boldsymbol\varepsilon_\pm}\right\rangle\right]\sigma'_y,\label{M54gapping}\end{align} where $\sigma'=\Phi_{\frac{7}{16},\frac{7}{16}}$ is the diagonal non-chiral product $[7/16]^R[7/16]^L$. 
The field $\sigma'$ is identical to the (subleading) magnetization operator in the tricritical Ising model.
Perturbation by this term is known to deform the model away from its critical point into a gapped phase~\cite{LASSIG1991591}.

$\mathcal{H}^{G_2}_{\mathrm{inter}}+\mathcal{H}^{F_4}_{\mathrm{intra}}$ together introduce a finite excitation energy gap in the bulk of the coupled-wire model. In an open cylinder geometry, chiral $G_2$ WZW CFTs at level 1 are left behind on the two boundaries. The particle-hole conjugate $F_4$ bFQH state can be constructed by switching the inter-wire and intra-wire backscattering patterns between $G_2$ and $F_4$. Charge conservation requires $d_1d_2d_3$ and $d_4$ to be electrically neutral. Hence, the charge vector ${\bf q}=(q_1,\ldots,q_8)$ of the 8 non-local Dirac fermions (see above \eqref{E8fillingnumber}) must be further restricted by $q_1+q_2+q_3=q_4=0$. This forces $(q_1,q_2,q_3)=(0,0,0)$ or $(\pm2,\mp2,0)$. The filling numbers of the Fibonacci bFQH states are \begin{align}\begin{split}\nu_{G_2}&=\sum_{j=1}^3q_j^2=0\;\mbox{or}\;8,\\\nu_{F_4}&=\sum_{j=5}^8q_j^2=16-\nu_{G_2}.\end{split}\end{align} The quantized thermal Hall conductances are determined by the chiral central charges of the level 1 WZW algebras $c_{G_2}=14/5$ and $c_{F_4}=26/5$. The $G_2$ coupled-wire model conserves momentum along the $\mathsf{x}$ direction when the Fermi momenta of the electron channels obey \eqref{appmomentumAbelian} under the projection matrix $P_A=P_{SU(3)_1}$ (see \eqref{PSU3PE6}). The same goes for the $F_4$ model by substituting $P_A=P_{SO(8)_1}=\mathbb{0}_4\oplus\mathbb{1}_4$ in \eqref{appmomentumAbelian}.

Both $G_2$ and $F_4$ states have Fibonacci topological order: These states support Fibonacci anyons that follow the fusion rule $\tau\times\tau=1+\tau$ and carry the golden ratio quantum dimension $d_\tau=(1+\sqrt{5})/2$. The Fibonacci primary fields of the $(G_2)_1$ and $(F_4)_1$ WZW CFTs can be found in Ref.~\onlinecite{PhysRevB.100.085116}. 
They can be summarized as follows. 
The $G_2$ Fibonacci super-selection sector is the 7-dimensional irreducible representation of the $G_2$ algebra spanned by \begin{align}
\begin{split}
[\tau]&=
(1\otimes\tau)\oplus(\mathcal{E}\otimes\tau\Psi^\dagger)\oplus(\mathcal{E}^\dagger\otimes\tau\Psi)\\
&=
{\rm span}
\left\{\begin{array}{@{} c @{}}
\tau, e^{i(\phi_1+\phi_2-2\phi_3)/3}\tau\Psi^\dagger,\\
e^{i(\phi_2+\phi_3-2\phi_1)/3}\tau\Psi^\dagger, e^{i(\phi_3+\phi_1-2\phi_2)/3}\tau\Psi^\dagger,\\
e^{-i(\phi_1+\phi_2-2\phi_3)/3}\tau\Psi,
e^{-i(\phi_2+\phi_3-2\phi_1)/3}\tau\Psi,\\
e^{-i(\phi_3+\phi_1-2\phi_2)/3}\tau\Psi
\end{array}\right\}.
\end{split}\label{eq:G2supersector}
\end{align}
Here, the tensor product splits each field in the $\mathbb{Z}_3$ parafermion CFT sector and $SU(3)_1$ sector (see \eqref{G2SU3Z3}). $\tau$ is the spin $2/5$ primary field in the $\mathbb{Z}_3$ parafermion CFT. $\tau\Psi$ and $\tau\Psi^\dagger$ are the two primary fields with spin $1/15$. The vertex operators are $SU(3)_1$ primary fields in $\mathcal{E}$ and $\mathcal{E}^\dagger$ (see \eqref{SU3primary}). The $F_4$ Fibonacci super-selection sector is the 26-dimensional irreducible representation of $F_4$ algebra spanned by
\begin{align}
\begin{split}
[\bar\tau] &=
\left(\Phi_{(1,3)}\otimes1\right)\oplus\left(\Phi_{(1,2)}\otimes f\right)\oplus\left(\Phi_{(2,2)}\otimes\sigma\right),
\label{eq:F4supersector}
\end{split}
\end{align}
where the fields are split in the tricritical Ising sector and $SO(9)_1$ sector (see \eqref{F4SO9M54}).  Here, $\Phi_{(r,s)}$ are the primary fields located at the $(r,s)$ entry in the conformal grid of the CFT minimal model $\mathcal{M}(5,4)$. $\Phi_{(1,3)}=\bar\tau$, $\Phi_{(1,2)}$ and $\Phi_{(2,2)}$ have spins $3/5$, $1/10$ and $3/80$, respectively. $f$ is the $SO(9)_1$ vector spanned by $\psi_8$ and $e^{\pm i\phi_{j=5,\ldots,8}}$. $\sigma$ is the $SO(9)_1$ spinor spanned by $\sigma_8e^{i\varepsilon^j\phi_{4+j}}$, for $\varepsilon^{j=1,\ldots,4}=\pm$. 
The electric charges of the fields in \eqref{eq:G2supersector} and \eqref{eq:F4supersector} are completely determined by their vertex component. The charge assignments of the $G_2$ and $F_4$ currents and their Fibonacci primary fields are summarized in Tables \ref{tab:ExceptionalChargeAssignmentRoot} and \ref{tab:G2F4SuperselectionsectorChargeAssgn}.

We conclude this section by showing Fibonacci anyon pairs can be created in the bulk by applying an open string $\mathcal{S}$ of electron operators on the ground state. This operator string makes use of the $E_8$ currents that lie outside of the $G_2\times F_4$ subalgebra. The $E_8$ algebra has dimension 248, while $G_2\times F_4$ has dimension $14+52=66$. The complement $(G_2\times F_4)^\perp$ is spanned by $182=7\times26$ current fields that can be decomposed into the Fibonacci pair $[\tau][\bar\tau]$ from \eqref{eq:G2supersector} and \eqref{eq:F4supersector}. The operator string is a product $\mathcal{S}=\prod_{y=y_1}^{y_2}\mathcal{O}^R_y(\mathsf{x}_0)\mathcal{O}^L_y(\mathsf{x}_0)$, where $\mathcal{O}$ is a current operator in $(G_2\times F_4)^\perp$. For example, we take $\mathcal{O}=\partial_{\mathsf{x}}(\phi_1+\phi_2+\phi_3)$. This operator is a $SU(3)_1\times SO(9)_1$ singlet and has trivial OPE with the $SU(3)_1$ and $SO(9)_1$ currents. Therefore, it splits into the product $\mathcal{O}\sim\tau\bar\tau$, where $\tau$ is the primary field with spin $2/5$ in the $\mathbb{Z}_3$ parafermion CFT and $\bar\tau=\Phi_{(1,3)}$ is the primary field with spin $3/5$ in the tricritical Ising CFT. The string operator becomes \begin{align}\mathcal{S}\sim\prod_{y=y_1}^{y_2}\tau^R_y(\mathsf{x}_0)\bar\tau_y^R(\mathsf{x}_0)\tau^L_y(\mathsf{x}_0)\bar\tau_y^L(\mathsf{x}_0).\label{Fibonaccistring}\end{align} 

To demonstrate this, we will now apply this operator to the ground state of the $G_2$ model. (The $F_4$ model has a similar structure.) The inter-wire $\mathbb{Z}_3$ parafermion backscattering interaction \eqref{Z3parafermionbackscattering} introduces the finite ground state expectation value $\langle\tau^R_y(\mathsf{x}_0)\tau^L_{y+1}(\mathsf{x}_0)\rangle$. At the same time, the intra-wire $\sigma'=\Phi_{\frac{7}{16},\frac{7}{16}}$ interaction in \eqref{M54gapping} pins the ground state expectation value $\langle\bar\tau_y^R(\mathsf{x}_0)\bar\tau_y^L(\mathsf{x}_0)\rangle$. To see this, we use the effective Landau-Ginzburg description of minimal CFT models~\cite{francesco2012conformal}: \begin{align}\mathcal{L}_{2D}=\frac{1}{2}\left(\nabla\mathsf\Phi\right)^2+V(\mathsf\Phi).\end{align} The relevant minimal models here are the tricritical Ising model $\mathcal{M}(5,4)$ with \begin{align}V(\mathsf\Phi)=\mathsf\Phi^6-\lambda_3\mathsf\Phi^3,\label{M54LGpotential}\end{align} and the 3-state Potts model $\mathcal{M}(6,5)$ with \begin{align}V(\mathsf\Phi)=\mathsf\Phi^8-\lambda_6\mathsf\Phi^6.\label{M65LGpotential}\end{align} When $\lambda_{3,6}=0$, the models are at their conformal critical points. The $\lambda_3$ term in \eqref{M54LGpotential} deforms the tricritical Ising model by the subleading magnetization operator $\mathsf\Phi^3\sim\sigma'=\Phi_{\frac{7}{16},\frac{7}{16}}$ that appears in \eqref{M54gapping}. 
(Here, ``subleading" means that $\sigma'$ is the next-most-relevant, in the renormalization group sense, magnetization operator of the tricritical Ising model \cite{francesco2012conformal}. A similar definition of ``subleading" applies below.)
For nonzero $\lambda_{3} >0$, the Landau-Ginzburg variable takes a nonzero ground state expectation value, and consequently, so too does the (tricritical Ising) subleading thermal operator $\langle\mathsf\Phi^4\rangle\sim\langle\varepsilon'\rangle=\langle\Phi_{\frac{3}{5},\frac{3}{5}}\rangle=\langle\bar\tau^R\bar\tau^L\rangle$. The $\lambda^6$ term in \eqref{M65LGpotential} corresponds to the $\mathbb{Z}_3$ parafermion backscattering $\mathsf\Phi^6\sim\Phi_{\frac{2}{3},\frac{2}{3}}={\Psi^R}^\dagger\Psi^L$ in \eqref{Z3parafermionbackscattering}. 
It pins the Landau-Ginzburg field and introduces the finite ground state expectation value for the (3-state Potts) thermal operator $\langle\mathsf\Phi^4\rangle\sim\langle\varepsilon\rangle=\langle\Phi_{\frac{2}{5},\frac{2}{5}}\rangle=\langle\tau^R\tau^L\rangle$.

By replacing $\tau^R_y\tau^L_{y+1}$ and $\bar\tau^R_y\bar\tau^L_y$ by scalar ground state expectation values, the operator string $\mathcal{S}$ in \eqref{Fibonaccistring} at low energy becomes \begin{align}\mathcal{S}&\to\tau^R_{y_2}(\mathsf{x}_0)\tau^L_{y_1}(\mathsf{x}_0)\\&\quad\quad\times\prod_{y=y_1}^{y_2-1}\left\langle\tau_y^R(\mathsf{x}_0)\tau_{y+1}^L(\mathsf{x}_0)\right\rangle\prod_{y=y_1}^{y_2}\left\langle\bar\tau_y^R(\mathsf{x}_0)\bar\tau_y^L(\mathsf{x}_0)\right\rangle.\nonumber\end{align} The Fibonacci field operators $\tau^R_{y_2}(\mathsf{x}_0)\tau^L_{y_1}(\mathsf{x}_0)$ create a pair of gapped excitations, one between $y_2$ and $y_2+1$ and another between $y_1-1$ and $y_1$.

\subsubsection{The  \texorpdfstring{$SU(2)_4$}{SU(2)4} and \texorpdfstring{$Sp(8)_1$}{Sp(8)} metaplectic states}\label{sec:Sp8}

We construct the non-Abelian orbifold bFQH states \begin{align}SU(2)_4=\frac{SU(3)_1}{\mathbb{Z}_2},\quad Sp(8)_1=\frac{(E_6)_1}{\mathbb{Z}_2}\label{SU2Sp8Z2}\end{align} by gauging the $\mathbb{Z}_2$ outer automorphism symmetry of $SU(3)_1$ and $(E_6)_1$. 
The concepts of orbifolding and gauging were introduced in the contexts of CFTs~\cite{DixonHarveyVafaWitten85I,DixonHarveyVafaWitten85II,Ginsparg88,DijkgraafVerlindeVerlinde88,MooreSeiberg89zoo,ChenRoyTeoRyu17} and topological field theories~\cite{BarkeshliBondersonChengWang14,TeoHughesFradkin15}.
In \S\ref{sec:abelianorbifolds}, we discussed the Abelian orbifold bFHQ states, $U(1)_8=SU(2)_1/\mathbb{Z}_2$ and $SU(8)_1=(E_7)_1/\mathbb{Z}_2$, where the $\mathbb{Z}_2$ symmetry is an inner automorphism that does not alter the anyon classes of $SU(2)_1$ and $(E_7)_1$. 
In contrast, the $\mathbb{Z}_2$ symmetry in $SU(3)_1$ and $(E_6)_1$ in \eqref{SU2Sp8Z2} is an outer automorphism~\cite{KhanTeoVishveshwara15, Teotwistdefectreview} that conjugates the anyon classes $\mathcal{E}\leftrightarrow\mathcal{E}^\dagger$. 
When such a symmetry is gauged, the topological phase is promoted to a non-Abelian orbifold phase, referred to as a twist liquid~\cite{TeoHughesFradkin15}. 
The edge-state theory of a twist liquid is an orbifold CFT. 
Ref.~\onlinecite{TeoHu2021} considered the coupled-wire construction of non-Abelian dihedral twist liquids, such as the $SU(2)_4=SO(3)_2=SU(3)_1/\mathbb{Z}_2$ states.
Here, we study bFQH symplectic $Sp(8)_1$ and $SU(2)_4$ states that arise from the $E_8$ state.

The origin of the $\mathbb{Z}_2$ symmetry is the internal gauge symmetry in the non-local Dirac fermion presentation \eqref{nonlocaldiracdef} of the $E_8$ state. The symmetry, for any give wire $y$ and chiral sector $\sigma=R,L$, flips the signs of all eight non-local Dirac fermions, $d_j\to-d_j$, through a shift of the bosonized variables $\phi_j\to\phi_j+\pi$. 
All local operators must be even under this $\mathbb{Z}_2$ symmetry. 
This includes, in particular, all $E_8$ current operators. 
In \eqref{mixedphidef} (see section~\ref{sec:abelianorbifolds}), we defined a new set of bosonized variables $\tilde\phi_{2l-1}=\phi_{2l-1}$, $\tilde\phi_{2l}=\left(H_4\right)_l^{l'}\phi_{2l'}/2$, where $H_4$ is the Hadamard matrix \eqref{hadamard}. These bosons were associated with the non-local Dirac fermions $\tilde{d}_j=(\tilde\psi_{2j-1}+i\tilde\psi_{2j})/\sqrt{2}\sim e^{i\tilde\phi_j}$ that transformed according to the $\mathbb{Z}_2$ symmetry as $\tilde{d}_j\to(-1)^j\tilde{d}_j$. Here, in addition, we assume that fermions $\tilde{d}_{1,2,3,4}$ carry zero electric charge and zero $\mathsf{x}$-momentum. 
This allows us to introduce a different basis of fermions: \begin{align}\begin{split}&\bar{d}_1=\frac{\tilde\psi_3+i\tilde\psi_1}{\sqrt{2}},\quad\bar{d}_2=\frac{\tilde\psi_4+i\tilde\psi_2}{\sqrt{2}},\\&\bar{d}_3=\frac{\tilde\psi_7+i\tilde\psi_5}{\sqrt{2}},\quad\bar{d}_4=\frac{\tilde\psi_8+i\tilde\psi_6}{\sqrt{2}},\\&\bar{d}_5=\tilde{d}_5,\quad\bar{d}_6=\tilde{d}_7,\quad\bar{d}_7=\tilde{d}_6,\quad\bar{d}_8=\tilde{d}_8.\end{split}\label{dbar}\end{align} 
These fermions are bosonized as $\bar{d}_j\sim e^{i\bar\phi_j}$. The local $\mathbb{Z}_2$ symmetry acts as follows: \begin{align}\mathbb{Z}_2:\quad\begin{split}&\bar{d}_{1,2,3,4}\to\bar{d}^\dagger_{1,2,3,4},\quad\bar\phi_{1,2,3,4}\to-\bar\phi_{1,2,3,4},\\&\bar{d}_{5,6}\to-\bar{d}_{5,6},\quad\bar\phi_{5,6}\to\bar\phi_{5,6}+\pi,\\&\bar{d}_{7,8}\to\bar{d}_{7,8},\quad\bar\phi_{7,8}\to\bar\phi_{7,8}.\end{split}\label{Z2gaugesymm1}\end{align} 

We now revisit the $SU(3)_1$ and $(E_6)_1$ WZW algebras (studied in \S\ref{sec:SU3E6}) to show how the $SU(2)_4$ and $Sp(8)_1$ symmetries arise. 
These algebras are generated by current operators $\partial_{\mathsf{x}}\bar\phi_j$ and $e^{i\alpha^j\bar\phi_j}$, obtained from those in \S\ref{sec:SU3E6} by replacing the bosonized variables $\phi_j\to\bar\phi_j$. 
Not all current operators are local because they may not be even under the internal $\mathbb{Z}_2$ symmetry \eqref{Z2gaugesymm1}. For example, the Cartan generators $\partial\bar\phi_{1,2,3,4}$ are now odd under $\mathbb{Z}_2$ and therefore are {\em not} local integral combinations of electrons. The local $(E_8)_1$ currents constructed by electrons are still the original ones associated with the old bosonized variables $\phi_j$. 
The local sub-algebras of $SU(3)_1$ and $(E_6)_1$---consisting of current operators that are even under the internal $\mathbb{Z}_2$ symmetry and lie inside the local $(E_8)_1$ algebra---are $SU(2)_4$ and $Sp(8)_1$, respectively. 
To see this, we start with the $SU(3)_1$ algebra, which is generated by $\mathtt{H}_1=\partial_{\mathsf{x}}(\bar\phi_1-\bar\phi_2)/\sqrt{2}$, $\mathtt{H}_2=\partial_{\mathsf{x}}(\bar\phi_1+\bar\phi_2-2\bar\phi_3)/\sqrt{6}$, and the roots $e^{\pm i(\bar\phi_a-\bar\phi_b)}$ for $1\leq a<b\leq3$. Its $\mathbb{Z}_2$ even sub-algebra is spanned by $J_1=2\sqrt{2}\cos\left(\bar\phi_2-\bar\phi_3\right)$, $J_2=2\sqrt{2}\cos\left(\bar\phi_3-\bar\phi_1\right)$, and $J_3=2\sqrt{2}\cos\left(\bar\phi_1-\bar\phi_2\right)$. (The presentation of $J_{1,2,3}$ using the $\tilde\phi$ variables can be found in \eqref{app:SU24current} in Appendix~\ref{app:Sp8}.) $J_{1,2,3}$ obey the $SU(2)_4$ current OPE, \begin{align}J_a(z)J_b(w)=\frac{4\delta_{ab}}{(z-w)^2}+\frac{i\sqrt{2}\epsilon_{abc}}{z-w}J_c(w)+\ldots,\end{align} where $\epsilon_{abc}$ is the Levi-Civita symbol for $a,b,c=1,2,3$, and the coefficient of the most singular term specifies the level $k=4$ of the $SU(2)$ WZW algebra.

Next, we consider the $(E_6)_1$ algebra. Among its Cartan generators $\mathtt{H}_b=\gamma_b^j\partial_{\mathsf{x}}\bar\phi_j$, only $\partial_{\mathsf{x}}\bar\phi_{5,6,7,8}$ are even under the $\mathbb{Z}_2$ symmetry \eqref{Z2gaugesymm1}. In addition, there are 32 root operator combinations of $E_6$ that are even under the $\mathbb{Z}_2$. They are \begin{align}\begin{split}&e^{\pm i(\bar\phi_5+\bar\phi_6)},e^{\pm i(\bar\phi_5-\bar\phi_6)},e^{\pm i(\bar\phi_7+\bar\phi_8)},e^{\pm i(\bar\phi_7-\bar\phi_8)},\\&\chi_0e^{\pm i\bar\phi_7},\chi_0e^{\pm i\bar\phi_8},\eta_0e^{\pm i\bar\phi_5},\eta_0e^{\pm i\bar\phi_6},\\&\eta_\pm e^{i\sum_{j=5}^8\varepsilon_\pm^j\bar\phi_j/2},\quad\mbox{if }\varepsilon_\pm^5=\varepsilon_\pm^6,\\&\chi_\pm e^{i\sum_{j=5}^8\varepsilon_\pm^j\bar\phi_j/2},\quad\mbox{if }\varepsilon_\pm^5=-\varepsilon_\pm^6,
%\eta_\pm\exp\left(\frac{i}{2}\sum_{j=5}^8\varepsilon_\pm^j\bar\phi_j\right)
\end{split}\label{Sp8roots1}\end{align} where $\varepsilon_\pm^{j=5,6,7,8}=\pm1$ with $\prod_{j=5}^8\varepsilon_\pm^j=\pm1$.
The Majorana fermions are \begin{align}\begin{split}&\chi_0=\sqrt{2}\cos\bar\phi_4,\quad\eta_0=\sqrt{2}\sin\bar\phi_4,\\&\chi_\pm=\sqrt{2}\cos\left[\left(\bar\phi_1+\bar\phi_2+\bar\phi_3\pm\bar\phi_4\right)/2\right],\\&\eta_\pm=\sqrt{2}\sin\left[\left(\bar\phi_1+\bar\phi_2+\bar\phi_3\pm\bar\phi_4\right)/2\right].\end{split}\label{Sp8chieta}\end{align} (The presentations of the Majorana fermions $\chi_{0,\pm}$ and $\eta_{0,\pm}$ in terms of the $\tilde\phi$ variables can be found in \eqref{app:Sp8chieta} in Appendix~\ref{app:Sp8}.)
While $\chi_{0,\pm}$ and $\eta_{0,\pm}$ have neutral charge and zero momentum, $\bar{d}_{5,6,7,8}\sim e^{i\bar\phi_{5,6,7,8}}$ can carry charge and momentum, in general. 
Thus, $\bar\phi_j$ in the vertex components in \eqref{Sp8roots1} are really abbreviations for $\bar\phi_j(\mathsf{x})+\bar{k}_j\mathsf{x}$ for $j=5,6,7,8$. The root operators \eqref{Sp8roots1} can be put in a more familiar form using the basis transformation, \begin{align}\begin{pmatrix}X_1\\X_2\\X_3\\X_4\end{pmatrix}=\frac{1}{2}\begin{pmatrix}1&1&0&0\\1&-1&0&0\\0&0&1&1\\0&0&1&-1\end{pmatrix}\begin{pmatrix}\bar\phi_5\\\bar\phi_6\\\bar\phi_7\\\bar\phi_8\end{pmatrix},\end{align} after which these operators become \begin{align}\begin{split}&e^{\pm 2iX_1},e^{\pm 2iX_2},e^{\pm 2iX_3},e^{\pm 2iX_4},\\&\chi_0e^{\pm i(X_3\pm X_4)},\eta_0e^{\pm i(X_1\pm X_2)},\\&\eta_+e^{\pm i(X_1\pm X_3)},\chi_+e^{\pm i(X_2\pm X_4)},\\&\eta_-e^{\pm i(X_1\pm X_4)},\chi_-e^{\pm i(X_2\pm X_3)}.\end{split}\label{Sp8roots2}\end{align} The $\mathbb{Z}_2$ symmetry acts as \begin{align}\mathbb{Z}_2:\quad\begin{split}&X_1\to X_1+\pi,\quad X_{2,3,4}\to X_{2,3,4},\\&\chi_\pm\to\chi_\pm,\quad\eta_\pm\to-\eta_\pm.\end{split}\label{Z2gaugesymm2}\end{align} The vertex components $e^{i\alpha^lX_l}$ are associated with the root vectors $\boldsymbol\alpha=(\alpha^1,\alpha^2,\alpha^3,\alpha^4)$ of $C_4=Sp(8)$. The operators in the first line of \eqref{Sp8roots2} corresponds to the 8 long roots, $\boldsymbol\alpha=\pm2{\bf e}_{1,2,3,4}$ of length 2. 
The remaining operators in \eqref{Sp8roots2} correspond to the 24 short roots, $\boldsymbol\alpha=\pm{\bf e}_l\pm{\bf e}_{l'}$ with length $\sqrt{2}$, for $1\leq l<l'\leq4$. 
Together with the 4 Cartan generators, $\mathtt{H}_l=\sqrt{2}\partial_{\mathsf{x}}X_l$, these operators span the $Sp(8)$ WZW current algebra at level 1.

The coupled-wire models of the $SU(2)_4$ and $Sp(8)_1$ states follow the general strategy in \eqref{HA}. 
The models begin with the intra-wire backscattering $\mathcal{H}^f_{\mathrm{intra}}$ of the integrated fermions (see \eqref{E8Hintra}). The remaining counter-propagating modes of the $(E_8)_1$ CFT on each wire are gapped by backscattering the $SU(2)_4$ and $Sp(8)_1$ currents in opposite directions. 
For example, for the $Sp(8)_1$ bFQH model, the backscattering interactions are the sum of 
\begin{widetext}
\begin{align}\begin{split}\mathcal{H}^{SU(2)_4}_{\mathrm{intra}}
&=-u_{\mathrm{intra}}\sum_y\sum_{1\leq a<b\leq3}\cos\left(\bar\phi^R_{ya}-\bar\phi^R_{yb}\right)\cos\left(\bar\phi^L_{ya}-\bar\phi^L_{yb}\right)\\&=-\frac{u_{\mathrm{intra}}}{2}\sum_y\sum_{1\leq a<b\leq3}\left(\cos\theta^y_{ab}+\cos\varphi^y_{ab}\right),\end{split}\label{HSU24intra}
\end{align}
\begin{align}
\begin{split}\mathcal{H}^{Sp(8)_1}_{\mathrm{inter}}&=u_{\mathrm{inter}}\sum_y\Bigg\{\sum_{l=1}^4\mathtt{H}^R_{y,l}\mathtt{H}^L_{y+1,l}-2\sum_{l=1}^4\cos\left(2\Xi^{y+1/2}_l\right)\\
&\quad\quad\quad-2i\eta^R_{y,0}\eta^L_{y+1,0}\sum_{s=\pm}\cos\left(\Xi^{y+1/2}_1+s\Xi^{y+1/2}_2\right)
-2i\chi^R_{y,0}\chi^L_{y+1,0}\sum_{s=\pm}\cos\left(\Xi^{y+1/2}_3+s\Xi^{y+1/2}_4\right)\\
&\quad\quad\quad-2i\eta^R_{y,+}\eta^L_{y+1,+}\sum_{s=\pm}\cos\left(\Xi^{y+1/2}_1+s\Xi^{y+1/2}_3\right)-2i\chi^R_{y,+}\chi^L_{y+1,+}\sum_{s=\pm}\cos\left(\Xi^{y+1/2}_2+s\Xi^{y+1/2}_4\right)\\
&\quad\quad\quad-2i\eta^R_{y,-}\eta^L_{y+1,-}\sum_{s=\pm}\cos\left(\Xi^{y+1/2}_1+s\Xi^{y+1/2}_4\right)-2i\chi^R_{y,-}\chi^L_{y+1,-}\sum_{s=\pm}\cos\left(\Xi^{y+1/2}_2+s\Xi^{y+1/2}_3\right)\Bigg\},
\end{split}\label{HSp8inter}\end{align}
\end{widetext}
where $\theta^y_{ab}=\bar\phi^R_{ya}-\bar\phi^R_{yb}-\bar\phi^L_{ya}+\bar\phi^L_{yb}$, $\varphi^y_{ab}=\bar\phi^R_{ya}-\bar\phi^R_{yb}+\bar\phi^L_{ya}-\bar\phi^L_{yb}$, and $\Xi^{y+1/2}_l=X^R_{y,l}-X^L_{y+1,l}$.

The gapping of the $SU(2)_4=SO(3)_2$ sector by current backscattering was shown in Ref.~\onlinecite{TeoHu2021}. The interaction \eqref{HSU24intra} fixes the finite ground state expectation values, \begin{align}s_y=\left\langle\sin\left(\bar\phi^R_{ya}-\bar\phi^R_{yb}\right)\sin\left(\bar\phi^L_{ya}-\bar\phi^L_{yb}\right)\right\rangle,\end{align} for $1\leq a<b\leq3$, where $S_{ab}=\sin\left(\bar\phi_a-\bar\phi_b\right)$ are non-local spin-1 primary fields that are odd under the $\mathbb{Z}_2$ symmetry and thus carry a non-trivial $\mathbb{Z}_2$ gauge charge. Applying the $\mathbb{Z}_2$ symmetry to one of the two chiral sectors produces a new ground state and flips the sign of $s_y$. Depending on the sign of the expectation value $s_y$, either $\theta^y_{ab}$ or $\varphi^y_{ab}$ is pinned. \begin{align}\begin{split}&s_y>0:\quad\left\langle\cos\theta^y_{ab}\right\rangle>0,\quad\left\langle\cos\varphi^y_{ab}\right\rangle=0,\\&s_y<0:\quad\left\langle\cos\theta^y_{ab}\right\rangle=0,\quad\left\langle\cos\varphi^y_{ab}\right\rangle>0.\end{split}\end{align}

We now explain the gapping of the $Sp(8)_1$ sector. The first line of \eqref{HSp8inter} introduces the finite ground state expectation values $\langle\Xi^{y+1/2}_l\rangle=\pi n^{y+1/2}_l$, where $n^{y+1/2}_l$ is an integer. Subsequently, the second line of \eqref{HSp8inter} is approximately proportional to $i\left(\chi^R_{y,0}\chi^L_{y+1,0}+s_{y+1/2}\eta^R_{y,0}\eta^L_{y+1,0}\right)=2\cos\left(\bar\phi^R_{y,4}-s_{y+1/2}\bar\phi^L_{y+1,4}\right)$, where the sign is \begin{align}s_{y+1/2}=(-1)^{n^{y+1/2}_1+n^{y+1/2}_2+n^{y+1/2}_3+n^{y+1/2}_4}=\pm 1.\end{align} 
The terms in the second line therefore pin either $\theta^{y+1/2}_4=\bar\phi^R_{y,4}-\bar\phi^L_{y+1,4}$ or $\varphi^{y+1/2}_4=\bar\phi^R_{y,4}+\bar\phi^L_{y+1,4}$. 
The $\mathbb{Z}_2$ symmetry, say in the $R$ sector on wire $y$, shifts $X^R_{y,1}\to X^R_{y,1}+\pi$, while leaving $X^R_{y,l=2,3,4}$ unchanged. Therefore, it produces a new ground state, changes $n^{y+1/2}_1\to n^{y+1/2}_1+1$, and switches the sign $s_{y+1/2}$. Similarly, depending on the sign $s_{y+1/2}$, the last two lines of \eqref{HSp8inter} at low energies become \begin{align}\begin{split}&i\left(\chi^R_{y,\pm}\chi^L_{y+1,\pm}+s_{y+1/2}\eta^R_{y,\pm}\eta^L_{y+1,\pm}\right)\\&\to2\cos\left(\theta^{y+1/2}_{123}\pm\left\langle\theta^{y+1/2}_4\right\rangle/2\right)\\&\;\;\;\mbox{ or }2\cos\left(\varphi^{y+1/2}_{123}\pm\left\langle\varphi^{y+1/2}_4\right\rangle/2\right),\end{split}\end{align} 
where $\theta^{y+1/2}_{123}=\sum_{a=1}^3\left(\bar\phi^R_{y,a}-\bar\phi^L_{y+1,a}\right)/2$ and $\varphi^{y+1/2}_{123}=\sum_{a=1}^3\left(\bar\phi^R_{y,a}+\bar\phi^L_{y+1,a}\right)/2$.

The coupled-wire model \eqref{HSU24intra} and \eqref{HSp8inter} describes a bFHQ state with $Sp(8)_1$ topological order. Recall that the model construction assumes $\tilde{d}_{5,6,7,8}$ are the only Dirac fermions that carry electric charge. They are responsible for the charge response of the entire $E_8$ sector. Since $\tilde{d}_{5,6,7,8}$ are all within the $Sp(8)_1$ sector and are decoupled from the $SU(2)_4$ sector, the $Sp(8)_1$ bFQH state must have the same filling number as the full $E_8$ state, $\nu_{Sp(8)_1}=16$. The central charge of the $Sp(8)_1$ CFT is identical to its parents state $(E_6)_1$. 
Therefore, $c_{Sp(8)_1}=6$. The intra-wire and inter-wire current backscattering interactions \eqref{HSU24intra} and \eqref{HSp8inter} preserve charge and momentum conservation when the Dirac fermions $\tilde{d}_{1,2,3,4}$ have vanishing electric charge and $\mathsf{x}$-momentum. These assumptions are realized when the model is constructed according to the following rules. (1) Recall that the 8 electronic $E_8$ simple roots are related to the Dirac fermions $d_j$ by a $R$-matrix (see \eqref{nonlocaldiracdef} and \eqref{diracRmatrix}). The $R$-matrix here needs to be chosen so that after the Hadamard transformation $d\to\tilde{d}$, $\tilde{d}_{1,2,3,4}$ have zero charge. Equivalently, the charges $q_j$ of the original fermions $d_j$ are restricted by $q_{1,3}=0$ and $\left(H_4\right)_{l=1,2}^{l'}q_{2l'}=0$. For example, one can choose a suitable $R$ matrix with ${\bf q}=(0,2,0,-2,0,-2,0,2)$. (2) The fermions $\tilde{d}_{1,2,3,4}$ have zero momentum when the Fermi momentum $k_{F,a}$ of the electron wire channels are chosen according to \eqref{appmomentumAbelian}, where the matrix $P_A=M(\mathbb{0}_4\oplus\mathbb{1}_4)M$ projects out $\tilde\phi_{1,2,3,4}$ and $\tilde\phi_j=M_j^{j'}\phi_{j'}$ is the basis transformation defined in \eqref{mixedphidef}. 

The coupled-wire model for the particle-hole conjugate state with $SU(2)_4$ topological order can be constructed by reversing the intra-wire and inter-wire backscattering patterns of \eqref{HSU24intra} and \eqref{HSp8inter}. It has trivial filling $\nu_{SU(2)_4}=0$ because the $SU(2)_4$ sector is electrically neutral. The chiral central charge is $c_{SU(2)_4}=3$ and is the same as its parent CFT $SU(3)_1$. 
For the $SU(2)_4$ state, the projection matrix in \eqref{appmomentumAbelian} needs to be trivial, $P_A=0$, in order for the model to conserve momentum.

We now discuss the primary fields and anyon excitations of the $Sp(8)_1$ and $SU(2)_4$ theory. We first present the primary field content of $SU(2)_4=SU(3)_1/\mathbb{Z}_2$ state. 
The 5 primary field super-selection sectors $[j]$ can be labeled by the $SU(2)$ ``spins" $j=0,1/2,1,3/2,2$. Each primary sector $[j]$ is spanned by $2j+1$ primary fields that irreducibly rotate among each other under the $SU(2)_4$ current algebra (c.f.~\eqref{currentirrep}). The conformal scaling dimensions (or spins) of the primary fields are $h_j=j(j+1)/6$, where the denominator $6=k+h_{SU(2)}$ is set by the dual Coxeter number $h_{SU(2)}=2$ and level $k=4$ of the WZW algebra. We begin with the free field representation of the bosonic Abelian primary fields in $[2]$, which is identified as the pure $\mathbb{Z}_2$ gauge charge sector in the orbifold theory. $[2]$ is the complement of $SU(2)_4$ in $SU(3)_1$: \begin{align}\begin{split}[2]&=\mathrm{span}\left\{\frac{\partial\bar\phi_1-\partial\bar\phi_2}{\sqrt{2}},\frac{\partial\bar\phi_1+\partial\bar\phi_2-2\partial\bar\phi_3}{\sqrt{6}}\right\}\\&\quad\quad\quad\quad\cup\left\{\sin\left(\bar\phi_a-\bar\phi_b\right):1\leq a<b\leq3\right\},\end{split}\end{align} which consists of the current operators that are odd under the internal $\mathbb{Z}_2$ symmetry \eqref{Z2gaugesymm1}. These fields obey the fusion rule $[2]\times[2]=[0]$ because any even combination is even under $\mathbb{Z}_2$ and is local. 

The $SU(3)_1$ primary sectors $\mathcal{E}$ and $\mathcal{E}^\dagger$ (by replacing $\phi\to\bar\phi$ in \eqref{SU3primary}) now becomes the primary sector $[1]$. Since $\mathcal{E}\leftrightarrow\mathcal{E}^\dagger$ are flipped under the internal $\mathbb{Z}_2$ symmetry, they are now the same primary sector in $SU(3)_1/\mathbb{Z}_2$. They obey the fusion rules \begin{align}[2]\times[1]=[1],\quad [1]\times[1]=[0]+[1]+[2].\end{align} The former originates from the $SU(3)_1$ current OPE of $\mathcal{E}$ and $\mathcal{E}^\dagger$. The latter holds because (i) operator products from $\mathcal{E}\times\mathcal{E}^\dagger$ produce both local bosons, such as $SU(2)_4$ currents, as well as $\mathbb{Z}_2$ charges in $[2]$; and (ii) pair products in $\mathcal{E}\times\mathcal{E}$ give fields in $\mathcal{E}^\dagger$ that now also belong in $[1]$. $[1/2]$ and $[3/2]$ are twist fields (also known as $\mathbb{Z}_2$ fluxes). They have spin $1/8$ and $5/8$ respectively, and follow the fusion rules \begin{align}\begin{split}&[1/2]\times[1/2]=[0]+[1],\\&[1]\times[1/2]=[1/2]+[3/2],\\&[2]\times[1/2]=[3/2].\end{split}\end{align} The twist fields in $SU(2)_4$ do not have free field representations. 
However, we will see below that they can be paired with twist fields in $Sp(8)_1$ to form local $E_8$ currents. Since $SU(2)_4$ is electrically neutral, so are all of its primary fields.

We now present the primary fields of $Sp(8)_1=(E_6)_1/\mathbb{Z}_2$. Particle-hole symmetry dictates that $Sp(8)_1$ and $SU(2)_4$ have the same number of primary field super-selection sectors, identical fusion rules, and conjugate spins. We label the 5 primary sectors of $Sp(8)_1$ by $1,\sigma,\mathsf{E},\tau,S$, which are the particle-hole conjugates of $[0],[1/2],[1],[3/2],[2]$, respectively. Each sector is spanned by $\#$ primary fields that rotate irreducibly under the $Sp(8)_1$ current algebra (c.f.~\eqref{currentirrep}). The spin $h$, quantum dimension $d$, and the number of fields $\#$ in each primary sector are listed in Table~\ref{tab:Sp8}. The fusion rules mirror $SU(2)_4$: \begin{align}\begin{split}&S^2=1,\quad S\times\sigma=\tau,\quad S\times\mathsf{E}=\mathsf{E},\\&\mathsf{E}\times\mathsf{E}=1+\mathsf{E}+S,\quad\mathsf{E}\times\sigma=\sigma+\tau,\\&\sigma\times\sigma=1+\mathsf{E}.\end{split}\end{align}

\begin{table}[htbp]
\centering
\begin{tabular}{c|cccc}
&$\sigma$&$\mathsf{E}$&$\tau$&$S$\\\hline
$h$&$3/8$&$2/3$&$7/8$&$1$\\
$d$&$\sqrt{3}$&$2$&$\sqrt{3}$&$1$\\
$\#$&$8$&$27$&$48$&$42$
\end{tabular}
\caption{The spin $h$, quantum dimension $d$, and the number of fields $\#$ of each non-trivial primary sector of $Sp(8)_1$.}\label{tab:Sp8}
\end{table}

We begin with the free field representation of the $\mathbb{Z}_2$ charges in sector $S$, which is the complement of $Sp(8)_1$ in $(E_6)_1$. $S$ is spanned by the 42 fields: \begin{align}\begin{split}&\frac{\partial\bar\phi_1+\partial\bar\phi_2+\partial\bar\phi_3}{\sqrt{3}},\partial\bar\phi_4,\\&e^{\pm i(\bar\phi_5+\bar\phi_7)},e^{\pm i(\bar\phi_5-\bar\phi_7)},e^{\pm i(\bar\phi_5+\bar\phi_8)},e^{\pm i(\bar\phi_5-\bar\phi_8)},\\&e^{\pm i(\bar\phi_6+\bar\phi_7)},e^{\pm i(\bar\phi_6-\bar\phi_7)},e^{\pm i(\bar\phi_6+\bar\phi_8)},e^{\pm i(\bar\phi_6-\bar\phi_8)},\\&\chi_0e^{\pm i\bar\phi_5},\chi_0e^{\pm i\bar\phi_6},\eta_0e^{\pm i\bar\phi_7},\eta_0e^{\pm i\bar\phi_8},\\&\chi_\pm e^{i\sum_{j=5}^8\varepsilon_\pm^j\bar\phi_j/2},\quad\mbox{if }\varepsilon_\pm^5=\varepsilon_\pm^6,\\&\eta_\pm e^{i\sum_{j=5}^8\varepsilon_\pm^j\bar\phi_j/2},\quad\mbox{if }\varepsilon_\pm^5=-\varepsilon_\pm^6.\end{split}\end{align} The electric charges of these operators can be read off from that of the Dirac fermions $\bar{d}_{5,6,7,8}=e^{i\bar\phi_{5,6,7,8}}$: one possibility is that all these fermions carry $\pm2$ charge; another possibility is to have one carry charge $\pm4$ with the remaining fermions neutral. 
Next, we move on to the $\mathsf{E}$ sector. It originates from the $(E_6)_1$ primary field sectors $\mathcal{E}$ and $\mathcal{E}^\dagger$ by replacing $\phi\to\bar\phi$ in \eqref{E6primary}. Like the $SU(2)_4$ primaries, these fields are flipped and conjugated $\mathcal{E}\leftrightarrow\mathcal{E}^\dagger$ under the internal $\mathbb{Z}_2$ symmetry, and therefore both $\mathcal{E}$ and $\mathcal{E}^\dagger$ now belong in the same sector $\mathsf{E}$. It forms a 27 dimensional irreducible representation of the $Sp(8)$ Lie algebra, and is invariant under the encompassing $E_6$, which includes the $\mathbb{Z}_2$ charges in $S$. The electric charge assignment is identical to the primary field $\mathcal{E}$ of $E_6$ at filling 16. The charge pattern of the $\mathsf{E}$ and $S$ sectors can be found in Table~\ref{tab:Sp8ChargeAssignment}.

In \S\ref{sec:SU3E6}, we noted the decomposition of the 248-dimensional $(E_8)_1$ algebra into the 8-dimensional $SU(3)_1$ algebra, the $78$-dimensional $(E_6)_1$ algebra, and the two tensor product spaces $\mathcal{E}_{SU(3)}\otimes\mathcal{E}_{E_6}$ and $\mathcal{E}_{SU(3)}^\dagger\otimes\mathcal{E}_{E_6}^\dagger$, each with dimension $3\times27=81$. Any $(E_8)_1$ current operator is a linear combination of a $SU(3)_1$ current, a $(E_6)_1$ current, and primary field pair product in $\mathcal{E}_{SU(3)}\otimes\mathcal{E}_{E_6}$ and $\mathcal{E}_{SU(3)}^\dagger\otimes\mathcal{E}_{E_6}^\dagger$. Here, in this section, the conformal embedding $(E_8)_1\supseteq Sp(8)_1\times SU(2)_4$ of orbifold theories splits the $E_8$ algebra into the direct sum: \begin{align}\begin{split}(E_8)_1&\supseteq Sp(8)_1\oplus SU(2)_4\\&\quad\oplus\left(\sigma\otimes[3/2]\right)\oplus\left(\tau\otimes[1/2]\right)\oplus\left(\mathsf{E}\otimes[1]\right).\end{split}\end{align} The first line embeds the $Sp(8)_1$ and $SU(2)_4$ algebras, which add up to dimension $36+3=39$. The tensor space $\mathsf{E}\otimes[1]$ pairs the spin $2/3$ and $1/3$ primary fields together and it contains $27\times3=81$ $E_8$ currents that have integral monodromy (i.e., trivial braiding) with the $\mathbb{Z}_2$ charges in $S$ and $[2]$. The twist fields (i.e., $\mathbb{Z}_2$ gauge fluxes) in the $Sp(8)_1$ sectors $\sigma$ and $\tau$ can be paired with those in $[3/2]$ and $[1/2]$ in $SU(2)_4$ to make up the remaining 128 $E_8$ currents that have $\pi$-monodromy with the $\mathbb{Z}_2$ charges. For instance, the spins of the paired twist fields add up to $1=3/8+5/8=7/8+1/8$, and the dimensions of the two tensor spaces of twist fields add up to $8\times4+48\times2=128$. (While not included in the $E_8$ algebra, the pair product of $\mathbb{Z}_2$ charges in $S\otimes[2]$ are spin-2 local bosons.) The 128 $E_8$ currents in $\left(\sigma\otimes[3/2]\right)\oplus\left(\tau\otimes[1/2]\right)$ are presented in Appendix~\ref{app:Sp8}. Knowing fields in $[1/2]$ and $[3/2]$ are electrically neutral allows us to read off the electric charge assignments of twist fields in $\sigma$ and $\tau$. The charge pattern is listed in Table~\ref{tab:Sp8ChargeAssignment}. The $\mathbb{Z}_2$ fluxes $\sigma$ and $\tau$ are referred to as metaplectic anyons~\cite{PhysRevB.87.165421,HastingsNayakWang14}.

\section{Discussion and Conclusion}
\label{discussionsection}

In this paper, we explored  the analogy between the integer quantum Hall state of electrons and the (bosonic) $E_8$ quantum Hall state.
In particular, we gave explicit constructions of the ``partially-filled" $E_8$ state---summarized in Fig.~\ref{fig:fillingcentralchargeplot} ---using the coupled-wire approach.
The topological orders of the various fractional $E_8$ states include Abelian and non-Abelian varieties.
Our approach in this paper for finding fractional $E_8$ states relied on the theory of conformal embeddings.
This theory details the various possible (bipartite) symmetry decompositions ${\cal G}_A \times {\cal G}_B \subseteq E_8$.
We then showed how the bulk-boundary correspondence (through the coupled-wire construction) allowed for the explicit construction of ${\cal G}_A$ and ${\cal G}_B$ fractional $E_8$ states.
These states have edge-state theories with ${\cal G}_A$ or ${\cal G}_B$ Kac-Moody symmetry.
Furthermore, the ${\cal G}_A$ and ${\cal G}_B$ states are related by a generalized ``particle-hole" symmetry, reviewed in Appendix \ref{PHconjugationappendix}, similar to particle-hole conjugate states $\nu$ and $1 - \nu$ in the lowest Landau level.

Our approach, using conformal embeddings, provides for a systematic understanding of the various possible fractional $E_8$ states.
This approach relies on the $E_8$ Kac-Moody symmetry of the edge-state theory.
This conformal embedding approach may be straightforwardly applied to construct fractional states of topological orders with edge-state theories possessing other Kac-Moody symmetries, e.g., $U(N)_1$ for $N$ filled Landau levels of electrons.

It is unclear whether the conformal embedding approach might provide a complementary avenue for understanding topological phases in the lowest Landau level.
What symmetry group can be embedded into $U(1)$?
One possibility is to consider a dual description \cite{Hsin:2016blu} of the $\nu=1$ integer quantum Hall state of electrons.
This dual description is in terms of a $U(N)_1 \times U(1)_{N+1}$ Chern-Simons gauge theory (see Appendix B of \cite{PhysRevB.99.125135}).
Here $N$ is an arbitrary integer; a physical interpretation of $N$ can be given in terms of partons \cite{PhysRevLett.66.802}.
We may then imagine applying the conformal embedding approach to the $U(N)_1$ Chern-Simons theory.
It would be interesting to know how such constructions might be related to the composite fermion/boson approach to fractional quantum Hall states (for a recent review and references therein, \cite{2020arXiv201113488J}).

Related to this is the question of whether there might be a gapless parent state for various fractional $E_8$ states found in this paper.
Here, we (again) have in mind an $E_8$ state analogy for the composite fermion theory at $\nu=1/2$.
Half-filling of the $E_8$ state occurs at $\nu=8$.
We have shown both Abelian and non-Abelian topological states can exist at $\nu=8$.
We do not yet have a candidate gapless parent state.

The bosonic topological states considered in this paper (summarized in Fig.~\ref{fig:familytree} and Eq.~\eqref{E8embeddings0}) do {\em not} exhaust all fractional states of $E_8$. Below are some possible examples not studied in this paper. (i) In this paper, we only considered bipartitions $\mathcal{G}_A\times\mathcal{G}_B\subset(E_8)_1$ where the WZW algebras $\mathcal{G}_A$ and/or $\mathcal{G}_B$ have level $k=1$. The $E_8$ state can be decomposed by into multi-partitions $\mathcal{G}_A\times\mathcal{G}_B\times\mathcal{G}_C\times\ldots$ of WZW theories with levels $k\geq1$. This may produce new bFQH states. (ii) In \S \ref{SOeven} and \ref{SOodd}, we constructed $SO(N)_1$ bFQH states, where $N=1,\ldots,15$. The coupled-wire model construction for $SO(15)_1$, in particular, that includes longer-range inter-wire interactions, may allow generalization to $SO(N)_1$ states with $N\geq16$. 

(iii) The exploration of orbifold bFQH states with twist liquid topological orders is incomplete. First, in our constructions of the $(E_8)_1$ wires, we do not find any $SO(16)_1$ subalgebras whose even and odd spinors are non-local. Therefore, the $SO(16)_1$ bFQH state, which is the orbifold state $(E_8)_1/\mathbb{Z}_2$ is missing in this paper. The construction of this state may be facilitated by introducing additional electron channels on each wire or allowing longer-range interactions. Second, in \S \ref{sec:Sp8}, we may not have exhausted all possible variations of the metaplectic orbifold embedding $SU(2)_4\times Sp(8)_1\subseteq(E_8)_1$. A more thorough investigation may discover more filling numbers with $\nu_{SU(2)_4}>0$ and $\nu_{Sp(8)_1}<16$. Third, in this paper, we only consider $\mathbb{Z}_2$ orbifold CFTs that are also WZW theories, such as $(E_6)_1/\mathbb{Z}_2=Sp(8)_1$ and $(E_7)_1/\mathbb{Z}_2=SU(8)_1$. There may be orbifold theories with higher-order gauge groups $G$ that are not WZW theories that decompose $(E_8)_1$. Fourth, it is very likely that there are $\mathbb{Z}_2$ orbifold CFTs that are also WZW theories but are not covered in this paper. One probable example is the orbifold state $SU(5)_1/\mathbb{Z}_2=SO(5)_2=Sp(4)_2$. Spin liquid or superconducting $SO(5)_2$ states have been constructed by coupled-wire models in ref.~\cite{TeoHu2021}. We anticipate charged $SO(5)_2$ bFQH states to occupy filling $\nu=16$. It would be interesting to see if such state can also occupy the particle-hole symmetric filling $\nu=8$. If so, such a metaplectic topological state can serve as the bosonic anolog of particle-hole symmetric Pfaffian state at $\nu=1/2$ that half-fills the Landau level.

Lastly, the $E_8$ state is not the only short-range entangled integer quantum Hall state. For example, the $D_{16}^+$ lattice \cite{conway2013sphere} at $c=16$, the Leech lattice and Niemeier lattices at $c=24$ are even unimodular lattices that can describe other bosonic short-range entangled integer quantum Hall states with unequal filling numbers and central charges, $\nu\neq c$. Just like the $E_8$, they also can be fractionally occupied by long-range entangled states. There can also be fermionic short-range entangled integer quantum Hall states corresponding to odd unimodular lattices, such as the $D_{12}^+$ lattice \cite{conway2013sphere} at $c=12$ and $(E_7\times E_7)^+$ lattice at $c=14$. They are postulated~\cite{HuSohalTeoToAppear} to occupy filling $\nu=4$ and $\nu=6$, which are much closer to experimentally observable regime than $\nu_{E_8}=16$.

\section*{Acknowledgments}
We thank Bo Han and Yichen Hu for the insightful discussion on conformal embeddings of $(E_8)_1$. 
We thank Hart Goldman and Yen-Wen Lu for useful discussions. In addition, we thank Pedro Lopes and Victor Quito for laying the foundation in ref.~\cite{PhysRevB.100.085116} that made this work possible.
This material is based upon work supported by the U.S. Department of Energy, Office of Science, Office of Basic Energy Sciences under Award No.~DE-SC0020007.
JCYT is supported by the National Science Foundation under Grant Number DMR-1653535.

\appendix

\section{Non-local Dirac fermion charge vectors}\label{Appendix:qIChargeParity}
The eight bosonic $E_8$ simple root operators $e^{i\tilde\Phi_I}$ (for each chiral sector $\sigma=R,L$ and on each wire $y$) are first constructed from integral combinations of electrons (see \eqref{eq:electronChevalleyConversion} and \eqref{E8simpleroots}). The $E_8$ root system consists of the 240 operators which are products of the simple root operators and have spin (scaling dimension) $h=1$. These boson operators and their electric charges are fixed. On the other hand, throughout this paper, we often use the non-local Dirac fermion presentation of the $E_8$ WZW currents. This presents the $E_8$ roots as bilinears and even spinor combinations of a set of 8 non-local Dirac fermions $d_j\sim e^{i\phi_j}$, where the ``Chevalley" and ``Cartan-Weyl" bosonized variables are related to that of the $E_8$ simple roots by the $R$-matrix, $\tilde\Phi_J=R_J^j\phi_j$ (see \eqref{nonlocaldiracdef} and \eqref{diracRmatrix}). These fermions are not fixed, and they depend on the choice of the $R$ matrix. In this appendix, we exhaust these choices and present the electric charges $q_j$ of the fermions.

The identification of the ``$p\dot{q}$" term of the Lagrangian density, \begin{align}\left(K_{E_8}^{-1}\right)^{IJ}\partial_{\mathsf{x}}\tilde\Phi_I\partial_t\tilde\Phi_J=\delta^{ij}\partial_{\mathsf{x}}\phi_i\partial_t\phi_j,\end{align} requires the $E_8$ Cartan matrix to agrees with the product $\left(K_{E_8}\right)_{IJ}=\delta_{ij}R_I^iR_J^j$. In this paper, we group $R_J^j$ in an $8\times8$ matrix whose rows $(R_J^{j=1},\ldots,R_J^{j=8})$ correspond to the $E_8$ simple roots. Therefore, $K_{E_8}=RR^T$. If there are two sets of non-local Dirac fermions $d_j$ and $d'_j$ presenting the $E_8$ simple roots according to $\tilde\Phi_J=R_J^j\phi_j={R'}_J^j\phi'_j$, then $RR^T=R'R'^T$ and thus $w=R^{-1}R'$ is an orthogonal matrix in $O(8)$ relating the two Cartan-Weyl bases. Moreover, since the $E_8$ root system is defined using the fixed simple roots and it does not depend on the choice of the fermions, the $w$ transformation leaves the $E_8$ root lattice invariant. In other words, $w$ belongs in the $E_8$'s automorphism group, which is identical to its Weyl group, $\mathrm{Aut}(E_8)=W(E_8)$ (see \eqref{E8Weylgroup} and ref.~\cite{conway2013sphere}). This group has finite order and its elements are generated by products of the following primitive linear manipulations: (i) permutations: $\phi_j\to\phi_{\sigma(j)}$ where $\sigma$ is a permutation of 8 elements, (ii) sign flips: $\phi_j\to(-1)^{s_j}\phi_j$, where $s_j=\pm1$, and (iii) Hadamard transformation: $\phi_j\to w_j^{j'}\phi_j$, where $w=(H_4\oplus H_4)/2$ and $H_4$ is the Hadamard matrix \eqref{hadamard}. Consequently, there are finitely many sets of non-local Dirac fermions that present the $E_8$. Starting with one such set $d_j\sim e^{i\phi_j}$, any other set $d'_j\sim e^{i\phi'_j}$ must be related to the first by a Weyl group transformation $\phi'_j=w_j^{j'}\phi_j$. Equivalently, any particular choice of the $R$ matrix can generate all possible ones by applying $R'=Rw$. 

The $E_8$ root operators carry electric charges. From \eqref{E8simlerootscharge}, the 8 simple roots have charges $\tilde{q}_I^\sigma = U^{\sigma\sigma'}_{Ia} t^{\sigma'}_a=(-4,2,0,0,-2,2,0,2)$, where $I=1,\ldots 8$, $a=1,\ldots,11$, $\sigma=R,L=+,-$, and $t^{\sigma'}_a=1$ is the charge of an electron (in units of $e$). The charges $\tilde{q}_I$ do not depend on the chiral sector, $\tilde{q}^R_I=\tilde{q}^J_R$, by the construction of the unimodular $U$ matrix in \eqref{Umatrix}. Given a $R$ matrix that defines a particular set of non-local Dirac fermions, the electric charges $q_j$ of $d_j$ can be calculated from the transformation $\tilde{q}_I=R_I^jq_j$. Below, we present some examples of $R$ matrices and their corresponding charge vectors ${\bf q}=(q_1,\ldots,q_8)$. 

We start with the following $R$ matrix:
\begin{align}
R=\left(
\begin{array}{cccccccc}
 1 & -1 & & & & & &\\
 & 1 & -1 & & & & &\\
 & & 1 & -1 & & & &\\
 & & & 1 & -1 & & &\\
 & & & & 1 & -1 & &\\
 & & & & & 1 & 1 &\\
 -\frac{1}{2} & -\frac{1}{2} & -\frac{1}{2} & -\frac{1}{2} & -\frac{1}{2} & -\frac{1}{2} & -\frac{1}{2} & -\frac{1}{2} \\
 & & & & & 1 & -1 &\\
\end{array}
\label{eq:2222Rmatrix}
\right).
\end{align}
The corresponding charge vector $q_j=(R^{-1})^I_j \tilde{q}_I$ is $(-2,2,0,0,0,2,0,-2)$.
By permuting the labels $j$ and flipping signs of some bosonized variables, other sets of Dirac fermions can be constructed with charge vectors of the form of $(2s_1,2s_2,2s_3,2s_4,0,0,0,0)$, where $s_{1,2,3,4}=\pm1$, or with any permuted entries.

The $R$-matrix, 
\begin{align}
R=\left(
\begin{array}{cccccccc}
 -1 & 0 & 0 & 1 & 0 & 0 & 0 & 0 \\
 \frac{1}{2} & \frac{1}{2} & \frac{1}{2} & -\frac{1}{2} & \frac{1}{2} & -\frac{1}{2} & \frac{1}{2}
   & -\frac{1}{2} \\
 0 & 0 & 0 & 0 & -1 & 0 & 0 & 1 \\
 0 & 0 & 0 & 0 & 0 & 0 & -1 & -1 \\
 -\frac{1}{2} & -\frac{1}{2} & \frac{1}{2} & -\frac{1}{2} & \frac{1}{2} & \frac{1}{2} &
   \frac{1}{2} & \frac{1}{2} \\
 \frac{1}{2} & \frac{1}{2} & -\frac{1}{2} & \frac{1}{2} & -\frac{1}{2} & \frac{1}{2} & \frac{1}{2}
   & -\frac{1}{2} \\
 0 & -1 & 0 & 0 & 0 & -1 & 0 & 0 \\
 \frac{1}{2} & \frac{1}{2} & -\frac{1}{2} & \frac{1}{2} & \frac{1}{2} & -\frac{1}{2} &
   -\frac{1}{2} & \frac{1}{2} \\
\end{array}
\right),
\label{eq:newRmatrixwith4}
\end{align} can be obtained by applying the primitive manipulations (i), (ii), followed by (iii) on \eqref{eq:2222Rmatrix}. The associated charge vector is ${\bf q}= (4,0,0,0,0,0,0,0)$. 
$R$ matrices that give rise to charge vectors with a single non-zero entry $q_j=\pm4$ can be obtained by further label permutations (i) and sign flips (ii).

The $R$ matrix,
\begin{align}
R= \left(
\begin{array}{cccccccc}
 1 & -1 & & & & & &\\
 & 1 & -1 & & & & &\\
 & & 1 & -1 & & & &\\
 & & & 1 & -1 & & &\\
 & & & & 1 & 1 & &\\
 & & & & & -1 & 1 &\\
 & & & & & & -1 & 1\\
 -\frac{1}{2} & -\frac{1}{2} & -\frac{1}{2} & -\frac{1}{2} & -\frac{1}{2} & -\frac{1}{2} & -\frac{1}{2} & -\frac{1}{2} 
\end{array}
\right),\label{eq:newRmatrixodd}
\end{align}
can be generated by rotating \eqref{eq:2222Rmatrix} with a Weyl group transformation $w$. Its charge vector is ${\bf q}=(-3,1,-1,-1,-1,-1,1,1)$. Any permutation and sign flip of such an odd charge vector, $(3s_1,s_2,s_3,s_4,s_5,s_6,s_7,s_8)$ for $s_{1,\ldots,8}=\pm1$ and $\prod_{j=1}^8s_j=-1$, can be obtained once again by (i) and (ii).

Lastly, the time-ordered correlations of the Cartan-Weyl bosonized variables are 
\begin{align}
\langle{\phi}_i(z){\phi}_j(w)\rangle=-\delta_{ij}\log(z-w)+\frac{i\pi}{2}{\rm sgn}(i-j),
\label{TOcorrelDirac}
\end{align}
where the sign function is $\mathrm{sgn}(x)=x/|x|$ for $x\neq0$ and $\mathrm{sgn}(0)=0$. The last term ensures that Dirac fermions anti-commute: $e^{i{\phi}_i}e^{i{\phi}_j}=-e^{i{\phi}_j}e^{i{\phi}_i}$, for $i\neq j$.

\section{Momentum conservation}\label{app:momentum}
We determine the Fermi momenta of the electron channels in all coupled-wire models. At the exactly-solvable fixed points, all backscattering interactions preserve momentum conservation when the electron channels' Fermi momenta take a set of specific values. Since the exactly-solvable models have a finite bulk excitation energy gap, weak perturbations that do not close the bulk gap will not alter the topological data of the quantum Hall state. Therefore, despite spoiling the exact-solubility, small deviations of the Fermi momenta away from the specific values presented below should not affect the topological phase.

The couple-wire models are constructed from an array of wires, each carrying $11$ non-chiral electron channels $c_{ya}(\mathsf{x})\sim c^R_{ya}(\mathsf{x})+c^L_{ya}(\mathsf{x})\sim e^{i(\Phi^R_{ya}(\mathsf{x})+k^R_{ya}\mathsf{x})}+e^{i(\Phi^L_{ya}(\mathsf{x})+k^L_{ya}\mathsf{x})}$ at the Fermi level. These electrons carry Fermi momenta, \begin{align}k^\sigma_{ya}=\frac{eBd}{\hbar c}y+\sigma k_{F,a},\label{appelecmomentum}\end{align} (see \eqref{electronmomentum} in \S\ref{sec:E8review}) in the presence of a magnetic field $B$. A many-electron interaction $(c_{ya}^\dagger c_{yb}^\dagger\ldots)(c_{y'a'}c_{y'b'}\ldots)$ splits near the Fermi level into multiple scattering processes $e^{i(-\Phi^{\sigma_1}_{ya}-\Phi^{\sigma_2}_{yb}+\Phi^{\sigma'_1}_{y'a'}+\Phi^{\sigma'_2}_{y'b'}+\ldots)}e^{i(-k^{\sigma_1}_{ya}-k^{\sigma_2}_{yb}+k^{\sigma'_1}_{y'a'}+k^{\sigma'_2}_{y'b'}+\ldots)\mathsf{x}}$ in a sum over all possible  $\sigma_1,\sigma_2,\ldots\sigma'_1,\sigma'_2,\ldots=R,L=+,-$. Terms with non-unit oscillation factors $e^{i(\sum k)\mathsf{x}}$ make vanishing contributions to the Hamiltonian after integrating over $\mathsf{x}$ in the thermodynamic limit where the wires have infinite length $l\to\infty$. The nonzero terms are the momentum conserving ones for which the oscillation factors vanish. In order for the backscattering interactions in the coupled-wire model to preserve momentum conservation at the Fermi level, the bare Fermi momenta $k_{F,a}$, for $a=1,\ldots,N$, must take specific values. In this appendix, we solve for these values.

There are three types of backscattering interactions considered in this paper: the intra-wire gapping of the integrated fermions $\mathcal{H}^f_{\mathrm{intra}}\sim\sum {f_{yn}^R}^\dagger f_{yn}^L+h.c.$ in \eqref{E8Hintra}; the intra-wire  and inter-wire current backscattering interactions $\mathcal{H}^B_{\mathrm{intra}}\sim\sum{\bf J}^R_{y,B}\cdot{\bf J}^L_{y,B}$ and $\mathcal{H}^A_{\mathrm{inter}}\sim\sum{\bf J}^R_{y,A}\cdot{\bf J}^L_{y+1,A}$, where ${\bf J}_A$ and ${\bf J}_B$ are the WZW current operators of the $\mathcal{G}_A\times\mathcal{G}_B$ decomposition of $E_8$. These interactions conserve momentum if the $L$ and $R$ operators in each term have identically cancelling momenta. We first consider the Abelian cases where the WZW affine Lie algebras are simply-laced. In these cases, all operators can be bosonized and they are either density operators of the form of $\partial_{\mathsf{x}}\Phi$, which carry zero momentum, or vertex operators $e^{i\sum m(\Phi+k\mathsf{x})}$. The bosonized variables and momentum both undergo a sequence of basis transformations, summarized below: \begin{align}\begin{split}\tilde{\Phi}_{yI}^{\sigma}&=\sum_{a,\sigma'}U_{Ia}^{\sigma\sigma'}\Phi_{ya}^{\sigma'},\quad I,a=1,\ldots,11,\\\phi^\sigma_{yj}&=\sum_J\left(R^{-1}\right)_j^J\tilde{\Phi}_{yJ}^{\sigma},\quad J,j=1,\ldots,8,\\\tilde\Phi^{A\sigma}_{yM}&=\boldsymbol\alpha^A_M\cdot\boldsymbol\phi^\sigma_y,\quad M=1,\ldots,r,\\\tilde\Phi^{B\sigma}_{yN}&=\boldsymbol\alpha^B_N\cdot\boldsymbol\phi^\sigma_y,\quad N=1,\ldots,8-r.\end{split}\label{basistransformationsummary}\end{align} First, the bosonized variables $\Phi^\sigma_{ya}$ of the electron operators $c^\sigma_{ya}\sim e^{i(\Phi^\sigma_{ya}+k^\sigma_{ya}\mathsf{x})}$, for $a=1,\ldots,11$, are transformed under the integral unimodular $U$ matrix in \eqref{Umatrix} into the Chevalley bosonized variables $\tilde\Phi^\sigma_{yI}$ of the $E_8$ simple root operators $\left[\mathtt{E}_{E_8}\right]^\sigma_{yJ}=e^{i(\tilde\Phi^\sigma_{yJ}+\tilde{k}^\sigma_{yJ}\mathsf{x})}$, for $J=1,\ldots,8$, and the three integrated fermions $f^\sigma_{yn}=e^{i(\tilde\Phi^\sigma_{y,I=8+n}+\tilde{k}^\sigma_{y,I=8+n}\mathsf{x})}$, for $n=1,2,3$. Next, using a $R$ matrix (examples include \eqref{eq:2222Rmatrix}, \eqref{eq:newRmatrixwith4} and \eqref{eq:newRmatrixodd} in Appendix~\ref{Appendix:qIChargeParity}) that specifies the $E_8$ simple root vectors in Euclidean 8-space, the Chevalley variables are transformed into the Cartan-Weyl variables, $\boldsymbol\phi=R^{-1}\widetilde{\boldsymbol\Phi}$. 
These bosons are associated with the 8 non-local Dirac fermions $d^\sigma_{yj}\sim e^{i(\phi^\sigma_{yj}+k^\sigma_{yj}\mathsf{x})}$ (see \eqref{nonlocaldiracdef}). Lastly, using a specific choice of simple roots vectors $\boldsymbol\alpha^A_M$ and $\boldsymbol\alpha^B_N$ of the $\mathcal{G}_A$ and $\mathcal{G}_B$ Lie algebras, the last two equations in \eqref{basistransformationsummary} give the bosonized variables of the root operators $\left[\mathtt{E}_A\right]^\sigma_{y\boldsymbol\alpha^A_M}=e^{i(\tilde\Phi^{A\sigma}_{yM}+\tilde{k}^{A\sigma}_{yM}\mathsf{x})}$ and $\left[\mathtt{E}_B\right]^\sigma_{y\boldsymbol\alpha^B_N}=e^{i(\tilde\Phi^{B\sigma}_{yN}+\tilde{k}^{B\sigma}_{yN}\mathsf{x})}$. The momenta $k^\sigma_{ya},\tilde{k}^\sigma_{yI},k^\sigma_{yj},\tilde{k}^{A\sigma}_{yM},\tilde{k}^{B\sigma}_{yN}$ of the various vertex operators are related by the same transformations in \eqref{basistransformationsummary}. The momenta of the non-local Dirac fermions $d^\sigma_{yj}$ are \begin{align}\begin{split}k^\sigma_{yj}&=\frac{eBd}{\hbar c}q_jy+\sigma k^d_{F,j},\\k^d_{F,j}&=\sum_{J=1}^8\left(R^{-1}\right)^J_j\sum_{a=1}^{11}\left(U^{++}_{Ja}-U^{+-}_{Ja}\right)k_{F,a},\end{split}\label{appDiracmomentum}\end{align} where the charge vector ${\bf q}=(q_1,\ldots,q_8)^T$ of the non-local Dirac fermions has entries, \begin{align}q_j=\sum_{J=1}^8\left(R^{-1}\right)^J_j\tilde{q}_J=\sum_{J=1}^8\left(R^{-1}\right)^J_j\sum_{a=1}^{11}\left(U^{++}_{Ja}+U^{+-}_{Ja}\right).\label{appchargevector}\end{align} The momenta of the integrated fermions $f^\sigma_{yn}$ are \begin{align}\begin{split}\tilde{k}^\sigma_{y,I=8+n}&=\frac{eBd}{\hbar c}\tilde{q}_ny+\sigma k^f_{F,n}\\k^f_{F,n}&=\sum_{a=1}^{11}\left(U^{++}_{I=8+n,a}-U^{+-}_{I=8+n,a}\right)k_{F,a},\end{split}\end{align} where the charges of the integrated fermions are \begin{align}\tilde{q}_{n=1,2,3}=\sum_{a=1}^{11}(U^{++}_{I=8+n,a}+U^{+-}_{I=8+n,a})=(3,1,1).\label{appchargeintegratedfermions}\end{align} The momenta of the simple roots in $\mathcal{G}_A$ and $\mathcal{G}_B$ are related to that of the non-local Dirac fermions by \begin{align}\begin{split}\tilde{k}^{A\sigma}_{yM}&=\boldsymbol\alpha^A_M\cdot{\bf k}^\sigma_y=\frac{eBd}{\hbar c}\boldsymbol\alpha^A_M\cdot{\bf q}y+\sigma\boldsymbol\alpha^A_M\cdot{\bf k}^d_F,\\\tilde{k}^{B\sigma}_{yN}&=\boldsymbol\alpha^B_N\cdot{\bf k}^\sigma_y=\frac{eBd}{\hbar c}\boldsymbol\alpha^B_N\cdot{\bf q}y+\sigma\boldsymbol\alpha^B_N\cdot{\bf k}^d_F,\end{split}\end{align} where $q^A_M=\boldsymbol\alpha^A_M\cdot{\bf q}$ and $q^B_N=\boldsymbol\alpha^B_N\cdot{\bf q}$ are the electric charges of the simple roots, and the entries of ${\bf k}^d_F=(k^d_{y,1},\ldots,k^d_{y,8})^T$ are the bare momenta computed in \eqref{appDiracmomentum}.

The coupled-wire Hamiltonian conserves momentum if the $L$ and $R$ vertex operators in all backscattering processes have cancelling momenta. 
(Here, we are ignoring possible Umklapp processes which allow momentum cancellation up to an integer multiple of $2 \pi$.)
This requires \begin{align}\begin{split}\tilde{k}^R_{y,I=8+n}-\tilde{k}^L_{y,I=8+n}&=\tilde{k}^{B,R}_{y,N}-\tilde{k}^{B,L}_{y,N}\\&=\tilde{k}^{A,R}_{y,M}-\tilde{k}^{A,L}_{y+1,M}=0,\end{split}\label{appmomentumconservation1}\end{align} for $n=1,2,3$, $M=1,\ldots,r$, and $N=1,\ldots,8-r$, where $r$ and $8-r$ are the ranks (as well as the central charges) of the $\mathcal{G}_A$ and $\mathcal{G}_B$ Lie algebras. Solving this system of 11 linear equations produces the solution to the 11 electron bare Fermi momenta $k_{F,a}$ in \eqref{appelecmomentum}. Using the basis transformations given above, the conditions in \eqref{appmomentumconservation1} become \begin{align}\begin{split}&\boldsymbol\alpha^A_M\cdot2{\bf k}^d_F=\frac{eBd}{\hbar c}\boldsymbol\alpha^A_M\cdot{\bf q}\\&\boldsymbol\alpha^B_N\cdot2{\bf k}^d_F=0,\quad 2k^f_{F,n=1,2,3}=0.\end{split}\label{appmomentumconservation2}\end{align}
Inter-wire backscattering interactions coupling $A$ currents in nearby wires acquire phases due to the magnetic field, when written the Landau gauge $A_\mathsf{x}=-B\mathsf{y}$. 
The intra-wire interactions that backscatter the $B$ currents and the integrated fermions $f_n$ within a wire are not affected by the magnetic field. 
\eqref{appmomentumconservation2} can be rewritten as a matrix equation in terms of the electron bare momenta ${\bf k}_F=(k_{F,1},\ldots,k_{F,11})^T$: \begin{align}X\left(U^{++}-U^{+-}\right)2{\bf k}_F=\frac{eBd}{\hbar c}\Lambda_AX\tilde{\bf q}_{\mathrm{tot}}.
\label{appmomentumconservation3}\end{align} Here, $X$ is the $11\times11$ matrix, \begin{align}X=\left(A_{\mathcal{G}_A\times\mathcal{G}_B}R^{-1}\right)\oplus\mathbb{1}_3=\begin{pmatrix}A_{\mathcal{G}_A\times\mathcal{G}_B}R^{-1}&0\\0&\mathbb{1}_3\end{pmatrix}.\end{align} $A_{\mathcal{G}_A\times\mathcal{G}_B}$ is the $8\times8$ matrix that combines the simple root vectors in  $\mathcal{G}_A$ and $\mathcal{G}_B$, \begin{align}A_{\mathcal{G}_A\times\mathcal{G}_B}=\begin{pmatrix}A_{\mathcal{G}_A}\\A_{\mathcal{G}_B}\end{pmatrix}=\begin{pmatrix}--&\boldsymbol\alpha^A_1&--\\\vdots&\vdots&\vdots\\--&\boldsymbol\alpha^A_r&--\\--&\boldsymbol\alpha^B_1&--\\\vdots&\vdots&\vdots\\--&\boldsymbol\alpha^B_{8-r}&--\end{pmatrix}_{8\times8}.\end{align} $\Lambda_A$ is responsible for projecting to the $A$ sector and is the $11\times11$ projection matrix, \begin{align}\Lambda_A=\mathbb{1}_r\oplus\mathbb{0}_{11-r}=\begin{pmatrix}\mathbb{1}_r&0\\0&0\end{pmatrix}_{11\times11}.\end{align} 
The charge vector $\tilde{\bf q}_{\mathrm{tot}}=(\tilde{q}_{J=1,\ldots,8},\tilde{q}_{n=1,2,3})^T$ contains the electric charges of the $E_8$ simple roots and the integrated fermions (see \eqref{appchargevector} and \eqref{appchargeintegratedfermions}). Since the $\mathcal{G}_A\times\mathcal{G}_B$ simple root vectors are linearly independent, the $A_{\mathcal{G}_A\times\mathcal{G}_B}$ matrix, and subsequently, the $X$ matrix are invertible. 
Using \begin{align}A_{\mathcal{G}_A\times\mathcal{G}_B}A_{\mathcal{G}_A\times\mathcal{G}_B}^T=K_{\mathcal{G}_A}\oplus K_{\mathcal{G}_B},\end{align}
we find
\begin{align}
    X^{-1}=\left(RA_{\mathcal{G}_A\times\mathcal{G}_B}^T\left(K_{\mathcal{G}_A}^{-1}\oplus K_{\mathcal{G}_B}^{-1}\right)\right)\oplus\mathbb{1}_3.
\end{align}
The $U$ matrix in \eqref{Umatrix} has unit determinant and is thus non-singular. 
In fact, $U^{-1}$ can be computed using the relation \begin{align}
\left(U^{++}-U^{+-}\right)\left(U^{++}+U^{+-}\right)^T=K_{E_8}\oplus\mathbb{1}_3,
\end{align} where 
$K_{E_8}=RR^T$ is the Cartan matrix of $E_8$ (see \eqref{KE8}). After simplifying, \eqref{appmomentumconservation3} has the unique solution: \begin{align}
k_{F,a}=\frac{1}{2}\frac{eBd}{\hbar c}\sum_{j,j',J=1}^8\left(U^{++}_{Ja}+U^{+-}_{Ja}\right)\left(R^{-1}\right)_j^JP_A^{jj'}q_{j'}.
\label{appmomentumAbelian}\end{align} Here, $q_{j'}$ is the charge of the non-local Dirac fermion $d_{j'}$ (see \eqref{appchargevector}) and \begin{align}P_A&=A_{\mathcal{G}_A}^TK_{\mathcal{G}_A}^{-1}A_{\mathcal{G}_A}\label{appAprojection}\end{align} is the projection matrix (obeying $P^2=P$) onto the $\mathcal{G}_A$ subspace in the Euclidean 8-space, where $K_{\mathcal{G}_A}=A_{\mathcal{G}_A}A_{\mathcal{G}_A}^T$ is the Cartan matrix of $\mathcal{G}_A$. 
For the $E_8$ integer quantum Hall state, $\mathcal{G}_A=E_8$ and $\mathcal{G}_B=0$ is the trivial algebra. 
The projection $P_A$ is the identity matrix $P_{E_8}=\mathbb{1}_8$ in the Euclidean 8-space, and \eqref{appmomentumAbelian} recovers the momentum solution in \eqref{E8momentum}.

In \S\ref{sec:E8review}, we saw that the filling number $\nu=N_e/N_B$ is expressed in terms of the electron Fermi momenta $k_{F,a}$ in \eqref{fillingnumber}. This equation holds for a general quantum Hall state constructed by any coupled-wire model. For an Abelian state, using the bare Fermi momentum solution \eqref{appmomentumAbelian} and the fermion charge vector ${\bf q}$ in \eqref{appchargevector}, the filling number of the $\mathcal{G}_A$ quantum Hall state is \begin{align}\begin{split}\nu&=\frac{\hbar c}{eBd}\sum_{a=1}^{11}2k_{F,a}=\sum_{j,j'=1}^8q_jP_A^{jj'}q_{j'}=\left|P_A{\bf q}\right|^2\\&=\sum_{M,M'=1}^rq^A_M\left(K_{\mathcal{G}_A}^{-1}\right)^{MM'}q^A_{M'}={\bf q}^A\cdot K_{\mathcal{G}_A}^{-1}{\bf q}^A.\end{split}\label{appAbelianfilling}\end{align} Here, ${\bf q}^A=(q^A_{M=1,\ldots,r})^T=A_{\mathcal{G}_A}{\bf q}$ is the vector containing the electric charges of the simple roots of $\mathcal{G}_A$. The second line of \eqref{appAbelianfilling} agrees with the electric Hall response of the $(2+1)D$ Chern-Simons theory $\mathcal{L}=\frac{1}{4\pi}K_{MM'}\alpha^M\wedge d\alpha^{M'}+\frac{e}{2\pi}q_MA\wedge d\alpha^M$. For the $E_8$ state, $P_A=\mathbb{1}_8$ and $\nu=|{\bf q}|^2=16$, which recovers \eqref{E8fillingnumber}.

Lastly, we address the issue of stability of the quantum Hall states. Momentum conservation only allows specific backscattering interactions to be present. In a coupled-wire model describing the $\mathcal{G}_A$ state, the magnetic field and commensurate electrons bare Fermi momenta $k_{F,a}$ force the $\mathcal{G}_A$ currents that carry finite electric charge to only scatter in between adjacent wires, and the charged $\mathcal{G}_B$ currents to only scatter within a wire. Other interaction terms that scatter charged currents in other directions violate momentum conservation and are therefore eliminated. The quantum Hall state is stable because most competing interactions that lead to a different topological phase are forbidden. However, momentum conservation does not eliminate all competing terms. Electrically neutral currents have uniform $\mathsf{x}$-momentum that does not couple to the magnetic field or depend on the vertical $y$ position. Both intra-wire and inter-wire backscattering interactions of neutral currents preserve momentum. Alternative gapping interactions in the neutral sector in general compete with the original and may drive a topological phase transition while keeping the filling number and electric Hall response unchanged. The instability of neutral sectors is most dominant when the quantum Hall state has trivial filling number. All $\mathcal{G}_A$ current operators in the $\nu=0$ $\mathcal{G}_A$ state have zero $\mathsf{x}$-momentum and can scatter in any direction. These neutral phases cannot be stabilized by the magnetic field alone, and their emergence may require separate mechanisms such as magnetism. 

\section{Particle-hole conjugation}
\label{PHconjugationappendix}

In this appendix, we discuss particle-hole symmetry with respect to the $E_8$ state, a symmetry first observed in \citep{PhysRevB.100.085116} as a relation between $G_2$ and $F_4$ Fibonacci states.
In general, the fillings of embedded theories
$\mathcal{G}_A\times \mathcal{G}_B\subseteq E_8$
are particle-hole conjugate: $\nu_{\mathcal{G}_B}=16-\nu_{\mathcal{G}_A}$.
Particle-hole symmetry is implemented by an anti-unitary operator ${\cal C}$.
Generally, on the electron operators $c_J$ in \eqref{electrons} (here, $J$ is a collective index for $\{\sigma, y, a \}$), it acts as $\mathcal{C}c_J\mathcal{C}^{-1}=\prod_{J'}(c_{J'})^{m^J_{J'}}$, where $m_{J'}^J$ are integers.
We are specifically interested in how particle-hole symmetry acts on the Chevalley basis bosons $\tilde{\Phi}^\sigma_{y,I}$, introduced in \eqref{eq:electronChevalleyConversion}, that generate the $E_8$ symmetry.
We define ${\cal C}$ to act as follows:
\begin{align}
\begin{split}
\mathcal{C}\tilde{\Phi}^R_{y,I}\mathcal{C}^{-1}&=\tilde{\Phi}^L_{y,I} - \frac{ \tilde{q}_I}{2}\mathsf{x},\\
\mathcal{C}\tilde{\Phi}^L_{y,I}\mathcal{C}^{-1}&=\tilde{\Phi}^R_{y-1,I} -\frac{ \tilde{q}_I}{2}\mathsf{x}.
\end{split}\label{eq:E8BosonVariableConjutation}
\end{align} 
According to the basis transformation \eqref{diracRmatrix}, particle-hole symmetry acts on the Cartan-Weyl bosons as
\begin{align}
\begin{split}
\mathcal{C}\phi^R_{y,j}\mathcal{C}^{-1}
&=\phi^L_{y,j}-q_j\frac{\mathsf{x}}{2},\\
\mathcal{C}\phi^L_{y,j}\mathcal{C}^{-1}
&=\phi^R_{y-1,j}-q_j\frac{\mathsf{x}}{2}.
\end{split}
\end{align}
Using these rules, the particle-hole transformation interchanges the inter and intra backscattering terms,
\begin{align}
\begin{split}
\mathcal{C}{[\mathtt{E}_{\mathcal{G}_A}]^R_{y,{\boldsymbol\alpha}}}^\dagger[\mathtt{E}_{\mathcal{G}_A}]^L_{y,{\boldsymbol\alpha}}\mathcal{C}^{-1}&= 
{[\mathtt{E}_{\mathcal{G}_A}]^R_{y-1,{\boldsymbol\alpha}}}^\dagger[\mathtt{E}_{\mathcal{G}_A}]^L_{y,{\boldsymbol\alpha}},\\
\mathcal{C}{[\mathtt{E}_{\mathcal{G}_B}]^R_{y,{\boldsymbol\alpha}}}^\dagger[\mathtt{E}_{\mathcal{G}_B}]^L_{y+1,{\boldsymbol\alpha}}\mathcal{C}^{-1}&= 
{[\mathtt{E}_{\mathcal{G}_B}]^R_{y,{\boldsymbol\alpha}}}^\dagger[\mathtt{E}_{\mathcal{G}_B}]^L_{y,{\boldsymbol\alpha}},
\end{split}
\end{align}
and thereby interchanges the Hamiltonians for the ${\cal G}_A$ and ${\cal G}_B$ states: 
$\mathcal{C}\mathcal{H}[\mathcal{G}_A]\mathcal{C}^{-1}=\mathcal{H}[\mathcal{G}_B]$ and
$\mathcal{C}\mathcal{H}[\mathcal{G}_B]\mathcal{C}^{-1}=\mathcal{H}[\mathcal{G}_A]$.

\begin{widetext}
\section{Topological order of \texorpdfstring{$SU(8)_1$}{SU(8)} state}
\label{topologicaldataSU8}
The primary field super-selection sectors of $SU(8)_1$ can be better represented using an alternative set of bosonized variables, different from those introduced in \S\ref{sec:abelianorbifolds}. Here, we consider the chiral $U(8)_1$ theory $\mathcal{L}_0=\frac{1}{4\pi}\delta^{jk}\partial_{\mathsf{x}}X_j \partial_tX_k-\mathcal{H}_0$ with 8 bosonized variables $X_{j=1,\ldots,8}$. The theory split into $U(1)_8\times SU(8)_1$. The $U(1)_8$ subalgebra is the diagonal sector generated by $\mathtt{H}_0=\partial_{\mathsf{x}}(X_1+\ldots+X_8)/\sqrt{8}$. The $SU(8)_1$ subalgebra is the off-diagonal sector spanned by the Cartan generators $\mathtt{H}_l=\left(\sum_{j=1}^l\partial_{\mathsf{x}}X_l-l\partial_{\mathsf{x}}X_{l+1}\right)/\sqrt{l(l+1)}$, for $l=1,\ldots,7$, and root operators $e^{\pm i(X_j-X_k)}$, for $1\leq j<k\leq8$. The compactification of the bosonized variables $X_j$ are defined by declaring the above $U(1)_8$ and $SU(8)_1$ currents, as well as the spin-4 bosons $e^{\pm i(X_1+\ldots,X_8)}$, to be primitive local bosons.

Using this representation, the $U(1)_8$ primary fields are $\mathcal{E}^m=e^{im(X_1+\ldots,X_8)/8}$, for $m=-3,\ldots,4$. The primary field super-selection sectors of $SU(8)_1$ are \begin{align}\mathcal{E}^m=\mathrm{span}\left\{e^{im(X_1+\ldots,X_8)/8-i(X_{j_1}+\ldots+X_{j_m})}:1\leq j_1<\ldots<j_m\leq8\right\},\label{appSU8primary}\end{align} for $m=0,\ldots,4$, and $\mathcal{E}^{m=-1,-2,-3}=(\mathcal{E}^{-m})^\dagger$. The $\mathcal{E}^m$ super-sector carries $C^8_{|m|}=8!/[|m|!(8-|m|)!]$ primary fields that irreducibly represent $SU(8)_1$ (c.f.~the current OPE \eqref{currentirrep}). The $X_j$ bosonized variables and the mixed variables $\tilde\phi_j$ defined in \eqref{mixedphidef} are related by a basis transformation, which identifies the primitive local bosons in the two representations. \begin{align}X_1+\ldots+X_8=2\tilde\phi_1-2\tilde\phi_2,\quad X_l-X_{l+1}=\alpha_l^j\tilde\phi_j,\label{Xtildephi}\end{align} where $\boldsymbol\alpha_{l=1,\ldots,7}$ are the rows of \eqref{ASU8}. The first equation equates the spin 4 primitive local boson in $U(1)_8$. The second equation identifies the 7 simple roots of $SU(8)$. The linear system of 8 equations in \eqref{Xtildephi} specifies a basis transformation $X_j=B_j^{j'}\tilde\phi_{j'}$ and allows the primary fields in \eqref{appSU8primary} to be expressed in terms of the mixed variables $\tilde\phi_j$. Their explicit forms are unimportant and will not be presented.

In Table~\ref{tab:AclassChargeAssgn2}, we see that there are two distinct charge patterns of the $SU(8)_1$ WZW currents at each of the fillings $\nu=14$ and $31/2$. They originate from the different choices of the non-local Dirac fermions $d_j\sim e^{i\phi_j}$ in constructing the coupled-wire models. In the last paragraph in subsection~\ref{sec:abelianorbifolds}, we claim that the two states belong to the same bFQH phase, despite the difference in the electric charges of the WZW currents. When the two states are juxtaposed, the $SU(8)_1$ currents cannot all be back-scattered from one edge to another because some carry unequal charges. On the other hand, the shared edge can be gapped while preserving charge symmetry by an alternative sine-Gordon interaction (c.f.~\eqref{SO101214sharededge} for $SO(10)_1$, $SO(10)_1$ and $SO(10)_1$). Here, we present such gapping interaction in terms of the transformed bosonized variables $X_j^{R/L}$ on the shared edge.

From the electric charges $q_j$ of $d_j$, the charges of the new fermions $e^{iX_j}$ can be determined by applying the basis transformations $\phi_j\to\tilde\phi_j$ in \eqref{mixedphidef} and $\tilde\phi_j\to X_j$ in \eqref{Xtildephi}. At filling $\nu_{SU(8)_1}=14$, the two WZW current charge patterns in Table~\ref{tab:AclassChargeAssgn2} correspond to the fermion charges $q\left(e^{iX_{j=1,\ldots,8}}\right)=(-3, 1, 1, 1, 1, 1, 1, 1)$ and $(-3, -1, -1, -1, -1, 1, 1, 1)$ up to permutations or an overall minus sign. At filling $\nu_{SU(8)_1}=31/2$, the two WZW current charge patterns in Table~\ref{tab:AclassChargeAssgn2} correspond to the fermion charges $q\left(e^{iX_{j=1,\ldots,8}}\right)=(1, 1, 1, 1, 1, 1, -3, -7)$ and $(1, 1, 1, 5, -3, -3, -3, -3)$ up to permutations or an overall minus sign. The gapping interaction on the shared edge can be chosen to be the sum of 7 local sin-Gordon terms \begin{align}\mathcal{U}=-u\sum_{l=1}^7\cos\left(\Theta_l\right),\quad\Theta_l=(N_R)^j_lX^R_j+(N_L)^j_lX^L_j.\label{SU8sharededge}\end{align}  The coefficients $(N_{R/L})^{j=1,\ldots,8}_{l=1,\ldots,7}$ form the entries of the $7\times16$ rectangular matrix $N=\begin{pmatrix}N_R&N_L\end{pmatrix}$. Charge conservation requires (i) the vertex operators $e^{i\Theta_l}$ to be electrically neutral, i.e., \begin{align}q\left(e^{i\Theta_l}\right)=\sum_{j=1}^8q\left(e^{iX^R_j}\right)(N_R)^j_l+\sum_{j=1}q\left(e^{iX^L_j}\right)(N_L)^j_l=0,\end{align} where $q\left(e^{iX^R_{j=1,\ldots,8}}\right)$ and $q\left(e^{iX^L_{j=1,\ldots,8}}\right)$ are the two distinct charge vectors written above at a given filling number. The $U(1)_8$ sectors generated by the diagonal combinations $X_1^{R/L}+\ldots+X_8^{R/L}$ are already gapped within a wire on individual sides. The sine-Gordon potentials in $\mathcal{U}$ operates solely on the off-diagonal $SU(8)$ sectors when (ii) \begin{align}\sum_{j=1}^8(N_{R/L})^j_l=0.\end{align} The sine-Gordon terms do not compete and simultaneously gap the shared edge if $[\Theta_l(x),\Theta_{l'}(x')]=0$. This requires (iii) $N$ is a null matrix, \begin{align}\sum_{j=1}^8\left[(N_R)^j_l(N_R)^j_{l'}-(N_L)^j_l(N_L)^j_{l'}\right]=0,\end{align} i.e.~$N\eta N^T=0$ where $\eta=\mathbb{1}_8\oplus(-\mathbb{1}_8)$ is the ``$K$-matrix" in the $X_j^{R/L}$ basis. The 7 sine-Gordon terms completely gap the $SU(8)_1$ edge if (iv) the null matrix has maximum rank, $\mathrm{rank}(N)=7$, so that the sine-Gordon variables are linearly independent. Lastly, the sine-Gordon potentials should all be local operators constructed by integral combinations of the $SU(8)_1$ current bosons. This requires $(N_{R/L})^j_lX_j$ to be integral combinations of the $SU(8)_1$ simple roots $X_l-X_{l+1}$. With condition (ii), locality simply requires (v) the entries $(N_{R/L})^j_l$ are integers. These five conditions (i)-(v) are satisfied by the following two null matrices \begin{align}N_{\nu=14}=\left(
\begin{smallmatrix}
 1 & 0 & 0 & 0 & 0 & -1 & 0 & 0 & -1 & 0 & 0 & 0 &
   0 & 1 & 0 & 0 \\
 1 & 0 & 0 & 0 & 0 & 0 & -1 & 0 & -1 & 0 & 0 & 0 &
   0 & 0 & 1 & 0 \\
 1 & 0 & 0 & 0 & 0 & 0 & 0 & -1 & -1 & 0 & 0 & 0 &
   0 & 0 & 0 & 1 \\
 0 & 1 & -1 & 0 & 0 & 0 & 0 & 0 & 0 & -1 & 1 & 0 &
   0 & 0 & 0 & 0 \\
 0 & 0 & 1 & -1 & 0 & 0 & 0 & 0 & 0 & 0 & -1 & 1 &
   0 & 0 & 0 & 0 \\
 0 & 0 & 0 & 1 & -1 & 0 & 0 & 0 & 0 & 0 & 0 & -1 &
   1 & 0 & 0 & 0 \\
 1 & 0 & -1 & -1 & 0 & 0 & 1 & 0 & 0 & -1 & 0 & 0 &
   -1 & 1 & 0 & 1 \\
\end{smallmatrix}
\right),\quad
N_{\nu=31/2}=\left(
\begin{smallmatrix}
 1 & -1 & 0 & 0 & 0 & 0 & 0 & 0 & 0 & 0 & 0 & 0 & 0
   & 0 & -1 & 1 \\
 0 & 1 & -1 & 0 & 0 & 0 & 0 & 0 & 0 & 0 & 0 & 0 & 0
   & -1 & 1 & 0 \\
 0 & 0 & 1 & -1 & 0 & 0 & 0 & 0 & 0 & 0 & 0 & 0 &
   -1 & 1 & 0 & 0 \\
 0 & 0 & 0 & 1 & 0 & 0 & 0 & -1 & 0 & 0 & 0 & -1 &
   1 & 0 & 0 & 0 \\
 0 & 0 & 0 & 0 & 0 & 0 & -1 & 1 & 0 & 0 & -1 & 1 &
   0 & 0 & 0 & 0 \\
 0 & 0 & 0 & 0 & 1 & -1 & 0 & 0 & -1 & 1 & 0 & 0 &
   0 & 0 & 0 & 0 \\
 0 & 0 & 0 & 1 & -2 & -1 & 1 & 1 & 2 & 1 & 0 & 0 &
   0 & -1 & -1 & -1 \\
\end{smallmatrix}
\right)\end{align} at the respective filling numbers.
% \end{widetext}

\section{Currents and primary fields of \texorpdfstring{$Sp(8)_1$}{Sp(8)} and \texorpdfstring{$SU(2)_4$}{SU(2)}}\label{app:Sp8}
In this appendix, we supplement the field operator presentations of the $Sp(8)_1$ and $SU(2)_4$ WZW algebras (in a given wire and chiral sector) in section~\ref{sec:Sp8}. The algebras were defined with the help of the basis transformation of bosonized variables $\phi\to\tilde\phi\to\bar\phi$ from \eqref{mixedphidef} and \eqref{dbar}. Here, we will present the field operators using the mixed variables $\tilde\phi_j$. The local $(E_8)_1$ WZW currents constructed by electrons are combinations of the 8 Cartan generators $\partial_{\mathsf{x}}\tilde\phi_j$ and the 240 roots $e^{i\tilde\alpha^j\tilde\phi_j}$. The $E_8$ root lattice (in the $\tilde\phi$ basis) consists of vectors $\tilde{\boldsymbol\alpha}$ of one of the following forms: (i) $\tilde{\boldsymbol\alpha}=\pm{\bf e}_i\pm{\bf e}_j$, where $i\equiv j$ mod 2, (ii) $\tilde{\boldsymbol\alpha}=\pm{\bf e}_{2p-1}+\sum_{q=1}^4\varepsilon^q{\bf e}_{2q}/2$, (iii) $\tilde{\boldsymbol\alpha}=\pm{\bf e}_{2q}+\sum_{p=1}^4\varepsilon^p{\bf e}_{2p-1}/2$, where $\varepsilon^l=\pm1$ and $\prod_{l=1}^4\varepsilon^l=+1$, or (iv) $\tilde{\boldsymbol\alpha}=\sum_{j=1}^8\varepsilon^j{\bf e}_j$, where $\varepsilon^j=\pm1$ and $\prod_{p=1}^4\varepsilon^{2p-1}=\prod_{q=1}^4\varepsilon^{2q}=-1$. The $SU(2)_4$ currents $J_1=2\sqrt{2}\cos\left(\bar\phi_2-\bar\phi_3\right)$, $J_2=2\sqrt{2}\cos\left(\bar\phi_3-\bar\phi_1\right)$, and $J_3=2\sqrt{2}\cos\left(\bar\phi_1-\bar\phi_2\right)$ can be re-expressed using bosonized variables $\tilde\phi_j$ or the fermions $\tilde{d}_j=e^{i\tilde\phi_j}=(\tilde\psi_{2j-1}+i\tilde\psi_{2j})/\sqrt{2}$: \begin{align}\begin{split}
J_1(z)&=\sqrt{2}i\left(\tilde\psi_4(z)\tilde\psi_7(z)+\tilde\psi_2(z)\tilde\psi_5(z)\right),\\
J_2(z)&=\sqrt{2}i\left(\tilde\psi_3(z)\tilde\psi_7(z)+\tilde\psi_1(z)\tilde\psi_5(z)\right),\\
J_3(z)&=\sqrt{2}i\left(\tilde\psi_3(z)\tilde\psi_4(z)+\tilde\psi_1(z)\tilde\psi_2(z)\right)=\sqrt{2}i\left(\partial_z\tilde\phi_1(z)+\partial_z\tilde\phi_2(z)\right),\\
J_-(z)&=\frac{J_1(z)-iJ_2(z)}{\sqrt{2}}=i\left(e^{i(\tilde\phi_1(z)+\tilde\phi_3(z))}-e^{i(\tilde\phi_1(z)-\tilde\phi_3(z))}+e^{i(\tilde\phi_2(z)+\tilde\phi_4(z))}-e^{i(\tilde\phi_2(z)-\tilde\phi_4(z))}\right),\end{split}\label{app:SU24current}\end{align} and $J_+=(J_1+iJ_2)/\sqrt{2}=J_-^\dagger$. The $SU(2)_4$ sectors decouples from the Majorana fermions $\eta_0=\tilde\psi_6$, $\chi_0=\tilde\psi_8$, and $\tilde\psi_{9,\ldots,16}$. In addition, the $SU(2)_4$ currents $J_{1,2,3}$ also have non-singular OPE with the following linear combinations of spinors: \begin{align}\begin{split}
\chi_+(z)&=\frac{i}{\sqrt{2}}\left(e^{i(\tilde\phi_1(z)-\tilde\phi_2(z)+\tilde\phi_3(z)-\tilde\phi_4(z))/2}-h.c.\right),\quad
\chi_-(z)=\frac{1}{\sqrt{2}}\left(e^{i(\tilde\phi_1(z)-\tilde\phi_2(z)+\tilde\phi_3(z)+\tilde\phi_4(z))/2}+h.c.\right),\\
\eta_+(z)&=\frac{i}{\sqrt{2}}\left(e^{i(\tilde\phi_1(z)-\tilde\phi_2(z)-\tilde\phi_3(z)+\tilde\phi_4(z))/2}-h.c.\right),\quad
\eta_-(z)=\frac{1}{\sqrt{2}}\left(e^{i(\tilde\phi_1(z)-\tilde\phi_2(z)-\tilde\phi_3(z)-\tilde\phi_4(z))/2}+h.c.\right).\end{split}\label{app:Sp8chieta}\end{align} Up to a sign, they are identical to the Majorana fermions $\chi_\pm$ and $\eta_\pm$ in \eqref{Sp8chieta}. The above field operator presentation using the $\tilde\phi_j$ variables shows that the 32 root operators of $Sp(8)_1$ in \eqref{Sp8roots1} are all linear combinations of the $E_8$ currents and are therefore local integral combinations of electrons.

Next, we present the 128 $E_8$ currents that belong in the tensor space $\left(\sigma\otimes[3/2]\right)\oplus\left(\tau\otimes[1/2]\right)$ of paired twist fields in $Sp(8)_1\times SU(2)_4$. These are current fields that carry non-trivial $\mathbb{Z}_2$ flux components in $Sp(8)_1=(E_6)_1/\mathbb{Z}_2$ and $SU(2)_4=SU(3)_1/\mathbb{Z}_2$. They correspond to the internal $\mathbb{Z}_2$ symmetry that flips the signs of $\tilde{d}_{1,3,5,7}$ while keeping $\tilde{d}_{2,4,6,8}$ unchanged. These $E_8$ current operators are \begin{align}e^{i\left(\sum_{l=1}^4\varepsilon^l\tilde\phi_{2l}/2\pm\tilde\phi_{2j-1}\right)},\quad e^{i\left(\sum_{l=1}^4\varepsilon^l\tilde\phi_{2l-1}/2\pm\tilde\phi_{2j}\right)},\end{align} where $j=1,2,3,4$,  $\varepsilon^{1,2,3,4}=\pm1$, and $\prod_{l=1}^4\varepsilon^l=+1$. We label these $E_8$ currents by $\mathtt{E}_{\gamma=1,\ldots,128}$. They rotate among each other under the $Sp(8)_1\times SU(2)_4$ current algebra. In particular, applying the $SU(2)_4$ currents $J_{i=1,2,3}$ gives the OPE \begin{align}J_i(z)\mathtt{E}_\gamma(w)=\frac{\sqrt{2}\left(S_i\right)_\gamma^{\gamma'}}{z-w}\mathtt{E}_{\gamma'}(w)+\ldots,\end{align} where $S_{i=1,2,3}$ are 128-dimensional matrices that represent the $SU(2)$ Lie algebra. The representation is reducible. The quadratic Casimir $S^2=S_1^2+S_2^2+S_3^2$ has eigenvalues $s(s+1)=3/4$ and $15/4$ with eigen-space dimensions 96 and 32, respectively. The two eigen-spaces are $\tau\otimes[1/2]$ and $\sigma\otimes[3/2]$. Since the $SU(2)_4$ sector is electrically neutral, this allows us to deduce the charge assignment pattern of the twisted sectors $\sigma$ and $\tau$ in $Sp(8)_1$. For example, in the 32-dimensional $\sigma\otimes[3/2]$, there are eight charge-4, eight charge-$(-2)$, and sixteen charge-neutral fields. 
Because $[3/2]$ contains four neutral fields, there must be two charge-4, two charge-$(-2)$, and four charge-neutral fields in $\sigma$.

\section{Topological data}
\label{topologicaldataappendix}

In this appendix, we collect the charge assignment tables for current operators and quasiparticle excitations for the conformal embeddings ${\cal G}_A \times {\cal G}_B \subset E_8$ studied in the main text.
Except for $SU(8)_1 \times U(1)_8 \subset (E_8)_1$, all conformal embeddings involve algebras at level 1.
Therefore, aside from this case, we will not specify the levels of the algebras in the tables below to simplify the notation.
In these tables, the numbers in each term $\ast\ (\star)$ specify the dimension $\ast$ of the subspace consisting of fields with charge $\star$.

\begin{table}[htbp]
\begin{center}
\caption{Current operator charge assignments for the conformal embeddings ${\cal G}_A \times {\cal G}_B \subset E_8$, where ${\cal G}_A$ or ${\cal G}_B$ is an exceptional group.
} \label{tab:ExceptionalChargeAssignmentRoot}
\begin{tabular}{ c|c|c|c }
$\nu_{G_2}$ & $\nu_{F_4}$ & $\#Q_{G_2}$ & $\#Q_{F_4}$ \\
\hline
$0$ & $16$
&$14(0)$
&$22(0),8(\pm2),7(\pm4)$
\\
$8$ & $8$
&$4(0),4(\pm2),1(\pm4)$
&$22(0),14(\pm2), 1(\pm4)$\\
\hline
$\nu_{SU(2)}$ & $\nu_{E_7}$ & $\#Q_{SU(2)}$ & $\#Q_{E_7}$ \\
\hline
$0$ & $16$
&$3(0)$
&$49(0),32(\pm2),10(\pm4)$
\\
$2$ & $14$
&$1(0),1(\pm2)$
&$49(0),35(\pm2),7(\pm4)$
\\
$8$ & $8$
&$1(0),1(\pm4)$
&$67(0),32(\pm2),1(\pm4)$
\\
\hline
$\nu_{SU(3)}$ & $\nu_{E_6}$ & $\#Q_{SU(3)}$ & $\#Q_{E_6}$ \\
\hline
$0$ & $16$
&$8(0)$&$30(0),16(\pm2),8(\pm4)$
\\
$8/3$ & $40/3$
&$4(0),2(\pm2)$
&$28(0),20(\pm2),5(\pm4)$
\\
$8$ & $8$
&$2(0),2(\pm2),1(\pm4)$
&$36(0),20(\pm2), 1(\pm4)$
\\
$32/3$ & $16/3$
&$4(0),2(\pm4)$
&$46(0),16(\pm2)$ \\
\hline
\end{tabular}
\end{center}
\end{table}

\begin{table}[htbp]
\begin{center}
\caption{Quasiparticle charge assignments for the conformal embedding $G_2 \times F_4 \subset E_8$.
Below, $\tau$ refers to the nontrivial super-selection sector of the $G_2$ topological phase and $\bar\tau$ refers to the nontrivial super-selection sector of the $F_4$ topological phase.}
 \label{tab:G2F4SuperselectionsectorChargeAssgn}
\begin{tabular}{ c|c|c|c }
$\nu_{G_2}$ & $\nu_{F4}$ & $\#(Q_{\tau})_{G_2}$ & $\#(Q_{\bar\tau})_{F_4}$ \\
\hline
$0$ & $16$
&$7(0)$
&$8(0),8(\pm2),1(\pm4)$
\\
$8$ & $8$
&$3(0),2(\pm2)$
&$14(0),6(\pm2)$ \\
\hline
\end{tabular}
\end{center}
\end{table}
\begin{table}[htbp]
\begin{center}
\caption{Quasiparticle charge assignments for the conformal embedding $SU(2) \times E_7 \subset E_8$.
Below, $\cal{S}$ refers to the nontrivial superselection sectors of the $SU(2)$ and $E_7$ topological phases.}
\label{tab:SU2E7SuperselectionsectorChargeAssgn}
\begin{tabular}{ c|c|c|c }
$\nu_{SU(2)}$ & $\nu_{E_7}$ & $\#(Q_{\mathcal{S}})_{SU(2)}$ & $\#(Q_{{\mathcal{S}}})_{E_7}$ \\
\hline
$0$ & $16$
&$2(0)$
&$20(0), 16(\pm2), 2(\pm4)$
\\
$2$ & $14$
&$1(\pm1)$
&$21(\pm1),7(\pm3)$
\\
$8$ & $8$
&$1(\pm2)$
&$32(0),12(\pm4)$\\
\hline
\end{tabular}
\end{center}
\end{table}

\begin{table}[htbp]
\begin{center}
\caption{Quasiparticle charge assignments for the conformal embedding $SU(3) \times E_6 \subset E_8$.
Below, $\cal{E}$ and $\overline{\cal{E}}$ refer to the nontrivial superselection sectors of the $SU(3)$ and $E_6$ topological phases.}
\label{tab:SU3E6SuperselectionsectorChargeAssgn}
\begin{tabular}{ c|c|c|c|c|c }
$\nu_{SU(3)}$ & $\nu_{E_6}$ 
& $\#(Q_{\mathcal{E}})_{SU(3)}$ & $\#(Q_{\mathcal{E}})_{E_6}$ 
& $\#(Q_{\bar{\mathcal{E}}})_{SU(3)}$ & $\#(Q_{\bar{\mathcal{E}}})_{E_6}$   \\
\hline
$0$ & $16$ 
&$3(0)$& $9(0),8(\pm2),1(\pm4)$
&$3(0)$& $9(0),8(\pm2),1(\pm4)$
\\
$8/3$ & $40/3$ 
&$1(-\frac{4}{3}),2(\frac{2}{3})$& $5(-\frac{8}{3}),10(-\frac{2}{3}),10(\frac{4}{3}),2(\frac{10}{3})$
&$2(-\frac{2}{3}),1(\frac{4}{3})$& $2(-\frac{10}{3}),10(-\frac{4}{3}),10(\frac{2}{3}),5(\frac{8}{3})$
\\
$8/3$ & $40/3$ 
&$2(-\frac{2}{3}),1(\frac{4}{3})$& $2(-\frac{10}{3}),10(-\frac{4}{3}),10(\frac{2}{3}),5(\frac{8}{3})$
&$1(-\frac{4}{3}),2(\frac{2}{3})$& $5(-\frac{8}{3}),10(-\frac{2}{3}),10(\frac{4}{3}),2(\frac{10}{3})$
\\
$8$ & $8$ 
&$1(0),1(\pm2)$& $15(0),6(\pm2)$
&$1(0),1(\pm2)$& $15(0),6(\pm2)$
 \\
$32/3$ & $16/3$ 
&$1(-\frac{8}{3}),2(\frac{4}{3})$&
$10(-\frac{4}{3}),16(\frac{2}{3}),1(\frac{8}{3})$
&$2(-\frac{4}{3}), 1(\frac{8}{3})$& 
$1(-\frac{8}{3}),16(-\frac{2}{3}),10(\frac{4}{3})$
 \\
$32/3$ & $16/3$ 
&$2(-\frac{4}{3}), 1(\frac{8}{3})$& $1(-\frac{8}{3}),16(-\frac{2}{3}),10(\frac{4}{3})$
&$ 1(-\frac{8}{3}),2(\frac{4}{3})$& $10(-\frac{4}{3}),16(\frac{2}{3}),1(\frac{8}{3})$ \\
\hline
\end{tabular}
\end{center}
\end{table}

\begin{table}[htbp]
\begin{center}
\caption{Current operator charge assignments for the conformal embedding $SU(5)^A \times SU(5)^B \subseteq E_8$.}\label{tab:SU5ChargeAssignmentRoot}
\begin{tabular}{ c|c|c|c}
$\nu_{SU(5)^A}$ & $\nu_{SU(5)^B}$ & $\#(Q)_{SU(5)^A}$ & $\#(Q)_{SU(5)^B}$ \\
\hline
$0$ & $16$ 
&$24(0)$  
&$8(0),4(\pm2),4(\pm4)$ 
 \\
$16/5$ & $64/5$
&$16(0),4(\pm2)$
&$16(0),4(\pm4)$
\\
$16/5$ & $64/5$
&$16(0),4(\pm2)$
&$10(0),4(\pm2),3(\pm4)$
\\
$24/5$ & $56/5$
&$12(0),6(\pm2)$
&$8(0),6(\pm2),2(\pm4)$
\\
$8$ & $8$
&$10(0),6(\pm2),1(\pm4)$
&$10(0),6(\pm2),1(\pm4)$
\\\hline
\end{tabular}
\end{center}
\end{table}

 \begin{table}[htbp]
\begin{center}
\caption{Charge assignments of the $SU(5)_1$ primary fields in $\mathcal{E}^{m=-2,-1,0,1,2}$. Here, $\mathcal{E}^0$ is the vacuum and $\mathcal{E}^{-m}=(\mathcal{E}^m)^\dagger$. We only list the charge assignments of fields in $\mathcal{E}^m$ for $m=1,2$.}
\label{tab:SU5ChargeAssignmentPrimary1to2}
\begin{tabular}{ c|c|c}
$\nu_{SU(5)}$&
$\#(Q_{\mathcal{E}^1})_{SU(5)}$&
$\#(Q_{\mathcal{E}^2})_{SU(5)}$
\\\hline
0&$5(0)$&$10(0)$\\
$16/5$&$1(-8/5),4(2/5)$&$4(-6/5),6(4/5)$\\
$16/5$&$4(-2/5),1(8/5)$&$6(-4/5),4(6/5)$\\
$24/5$&$2(-6/5),3(4/5)$&$1(-12/5),6(-2/5),3(8/5)$\\
$24/5$&$3(-4/5),2(6/5)$&$3(-8/5),6(2/5),1(12/5)$\\
$8$&$3(0),1(\pm2)$&$4(0),3(\pm2)$\\
$56/5$&$1(-12/5),2(-2/5),2(8/5)$&$2(-14/5),3(-4/5),4(6/5),1(16/5)$\\
$56/5$&$2(-8/5),2(2/5),1(12/5)$&$1(-16/5),4(-6/5),3(4/5),2(14/5)$\\
$64/5$&$1(-16/5),4(4/5)$&$4(-12/5),6(8/5)$\\
$64/5$&$4(-4/5),1(16/5)$&$6(-8/5),4(12/5)$\\
$64/5$&$1(-14/5),1(-4/5),3(6/5)$&$1(-18/5),3(-8/5),3(2/5),3(12/5)$\\
$64/5$&$3(-6/5),1(4/5),1(14/5)$&$3(-12/5),3(-2/5),3(8/5),1(18/5)$\\
$16$&$1(0),2(\pm2)$&$4(0),2(\pm2),1(\pm4)$\\\hline
\end{tabular}
\end{center}
\end{table}

\begin{table}[h]
\begin{center}
\caption{Current operator charge assignments for the conformal embeddings $SU(8)_1 \times U(1)_8 \subseteq (E_8)_1$ and $Sp(8)_1\times SU(2)_4\subseteq(E_8)_1$ that involve the orbifold theories.
The charge assignment in $U(1)_8$ is obtained from $b=\exp[\pm2i(\tilde\phi_1-\tilde\phi_2)]=\exp[\pm i(2\phi_1-\phi_2-\phi_4-\phi_6-\phi_8)]$.
} \label{tab:AclassChargeAssgn2}
\begin{tabular}{ c|c|c|c }
$\nu^{}_{SU(8)_1}$ & $\nu^{}_{U(1)_8}$ & $\#Q_{SU(8)_1}$ & $\#Q_{U(1)_8}$ \\
\hline
$7/2$ & $25/2$
&$49(0),7(\pm2)$
&$1(\pm10)$
\\
$8$ & $8$
&$31(0),16(\pm2)$
&$1(\pm8)$
\\
$23/2$ & $9/2$
&$29(0),15(\pm2),2(\pm4)$
&$1(\pm6)$
\\
$14$ & $2$
&$25(0),16(\pm2),3(\pm4)$
&$1(\pm4)$
\\
$14$ & $2$
&$49(0),7(\pm4)$
&$1(\pm4)$
\\
$31/2$ & $1/2$
&$37(0),7(\pm2),6(\pm4)$
&$1(\pm2)$
\\
$31/2$ & $1/2$
&$25(0),15(\pm2),4(\pm4)$
&$1(\pm2)$
\\
$16$ & $0$
&$23(0),16(\pm2),4(\pm4)$
&$2(0)$ \\
\hline
$\nu^{}_{Sp(8)_1}$ & $\nu^{}_{SU(2)_4}$ & $\#Q_{Sp(8)_1}$ & $\#Q_{SU(2)_4}$ \\\hline
$16$ & $0$
&$14(0),8(\pm2),3(\pm4)$
&$3(0)$\\\hline
\end{tabular}
\end{center}
\end{table}

\begin{table}[htbp]
\begin{center}
\caption{Quasiparticle charge assignments for the conformal embedding $SU(8)_1 \times U(1)_8 \subset (E_8)_1$.
We denote the super-selection sectors of the $SU(8)_1$ and $U(1)_8$ topological phases by $\mathcal{E}^m$ with $m = -3,\ldots,4$.
Here, $\mathcal{E}^0$ is the vacuum.
Because the charge assignments for $\mathcal{E}^m$ and $\mathcal{E}^{-m}=(\mathcal{E}^m)^\dagger$ are opposite, we only list the charge assignments of $\mathcal{E}^m$ sectors with $m = 1, \ldots, 4$.}
\label{tab:SU8ChargeAssignmentPrimary1to2}
\begin{tabular}{ c|c|c|c|c|c}
$\nu_{SU(8)}$&$\nu_{U(1)_8}$&
$\#(Q_{\mathcal{E}^1})_{SU(8)}$&$\#(Q_{\mathcal{E}^1})_{U(1)_8}$&
$\#(Q_{\mathcal{E}^2})_{SU(8)}$&$\#(Q_{\mathcal{E}^2})_{U(1)_8}$ 
\\
\hline
$7/2$ & $25/2$   
& $7(-\frac{1}{4}), 1(\frac{7}{4})$
&$1(-\frac{5}{4})$
& $21(-\frac{1}{2}),7(\frac{3}{2})$&$1(-\frac{5}{2})$ 
 \\
$7/2$ & $25/2$   
& $1(-\frac{7}{4}), 7(\frac{1}{4})$
&$1(\frac{5}{4})$
& $7(-\frac{3}{2}), 21(\frac{1}{2})$&$1(\frac{5}{2})$ 
 \\
$8$ & $8$   
& $4(\pm1)$
&$1(1)$ 
& $6(\pm2),16(0)$&$1(2)$ 
\\
$8$ & $8$   
& $4(\pm1)$
&$1(-1)$
& $6(\pm2),16(0)$&$1(-2)$ 
\\
$23/2$ & $9/2$  
& $1(-\frac{9}{4}),5(-\frac{1}{4}),2(\frac{7}{4})$
&$1(\frac{3}{4})$
& $5(-\frac{5}{2}),12(-\frac{1}{2}),10(\frac{3}{2}),1(\frac{7}{2})$&$1(\frac{3}{2})$
\\
$23/2$ & $9/2$  
& $2(-\frac{7}{4}),5(\frac{1}{4}),1(\frac{9}{4})$
&$1(-\frac{3}{4})$
& $1(-\frac{7}{2}),10(-\frac{3}{2}),12(\frac{1}{2}),5(\frac{5}{2})$&$1(-\frac{3}{2})$
 \\
$14$ & $2$  
& $1(-\frac{5}{2}),4(-\frac{1}{2}),3(\frac{3}{2})$
&$1(-\frac{1}{2})$
& $4(-3),9(-1),12(1),3(3)$&$1(-1)$
 \\
$14$ & $2$  
& $3(-\frac{3}{2}),4(\frac{1}{2}),1(\frac{5}{2})$
&$1(\frac{1}{2})$
& $3(-3),12(-1),9(1),4(3)$&$1(1)$
 \\
$14$ & $2$  
& $7(-\frac{1}{2}),1(\frac{7}{2})$
&$1(-\frac{1}{2})$
& $21(-1),7(3)$&$1(-1)$
 \\
$14$ & $2$  
& $1(-\frac{7}{2}),7(\frac{1}{2})$
&$1(\frac{1}{2})$ 
& $7(-3),21(1)$&$1(1)$
 \\
$31/2$ & $1/2$ 
& $6(-\frac{3}{4}),1(\frac{5}{4}),1(\frac{13}{4})$
&$1(\frac{1}{4})$
& $15(-\frac{3}{2}),6(\frac{1}{2}),6(\frac{5}{2}),1(\frac{9}{2})$&$1(\frac{1}{2})$
\\
$31/2$ & $1/2$ 
& $1(-\frac{13}{4}),1(-\frac{5}{4}),6(\frac{3}{4})$
&$1(-\frac{1}{4})$
& $1(-\frac{9}{2}),6(-\frac{5}{2}),6(-\frac{1}{2}),15(\frac{3}{2})$&$1(-\frac{1}{2})$
\\
$31/2$ & $1/2$ 
& $4(-\frac{5}{4}),3(\frac{3}{4}),1(\frac{11}{4})$
&$1(-\frac{1}{4})$
& $6(-\frac{5}{2}),12(-\frac{1}{2}),7(\frac{3}{2}),3(\frac{7}{2})$&$1(-\frac{1}{2})$
\\
$31/2$ & $1/2$ 
& $1(-\frac{11}{4}),3(-\frac{3}{4}),4(\frac{5}{4})$
&$1(\frac{1}{4})$
& $3(-\frac{7}{2}),7(-\frac{3}{2}),12(\frac{1}{2}),6(\frac{5}{2})$&$1(\frac{1}{2})$ 
\\
$16$ & $0$ 
& $2(\pm2),4(0)$
&$1(0)$
& $1(\pm4),8(\pm2),10(0)$&$1(0)$
\\
\hline
$\nu_{SU(8)}$&$\nu_{U(1)_8}$&
$\#(Q_{\mathcal{E}^3})_{SU(8)}$&$\#(Q_{\mathcal{E}^3})_{U(1)_8}$&
$\#(Q_{\mathcal{E}^4})_{SU(8)}$&$\#(Q_{\mathcal{E}^4})_{U(1)_8}$
\\
\hline
$7/2$ & $25/2$   
& $35(-\frac{3}{4}), 21(\frac{5}{4})$
&$1(-\frac{15}{4})$
& $35(\pm1)$&$1(-5)$
 \\
$7/2$ & $25/2$   
& $21(-\frac{5}{4}), 35(\frac{3}{4})$
&$1(\frac{15}{4})$
& $35(\pm1)$&$1(5)$
 \\
$8$ & $8$   
& $4(\pm3),24(\pm1)$
&$1(3)$ 
& $1(\pm4),16(\pm2),36(0)$&$1(4)$
\\
$8$ & $8$   
& $4(\pm3),24(\pm1)$
&$1(-3)$
& $1(\pm4),16(\pm2),36(0)$&$1(-4)$
\\
$23/2$ & $9/2$  
& $10(-\frac{11}{4}),20(-\frac{3}{4}),21(\frac{5}{4}),5(\frac{13}{4})$
&$1(\frac{9}{4})$
& $10(\pm3),25(\pm1)$&$1(3)$
\\
$23/2$ & $9/2$  
& $5(-\frac{13}{4}),21(-\frac{5}{4}),20(\frac{3}{4}),10(\frac{11}{4})$
&$1(-\frac{9}{4})$
& $10(\pm3),25(\pm1)$&$1(-3)$
 \\
$14$ & $2$  
& $6(-\frac{7}{2}),16(-\frac{3}{2}),21(\frac{1}{2}),12(\frac{5}{2}),1(\frac{9}{2})$
&$1(-\frac{3}{2})$
& $4(\pm4),19(\pm2),24(0)$&$1(-2)$
 \\
$14$ & $2$  
& $1(-\frac{9}{2}),12(-\frac{5}{2}),21(-\frac{1}{2}),16(\frac{3}{2}),6(\frac{7}{2})$
&$1(\frac{3}{2})$
& $4(\pm4),19(\pm2),24(0)$&$1(2)$
 \\
$14$ & $2$  
& $35(-\frac{3}{2}),21(\frac{5}{2})$
&$1(-\frac{3}{2})$
& $35(\pm2)$&$1(-2)$
 \\
$14$ & $2$  
& $21(-\frac{5}{2}),35(\frac{3}{2})$
&$1(\frac{3}{2})$
& $35(\pm2)$&$1(2)$
 \\
$31/2$ & $1/2$ 
& $20(-\frac{9}{4}),15(-\frac{1}{4}),15(\frac{7}{4}),6(\frac{15}{4})$
&$1(\frac{3}{4})$
& $15(\pm3),20(\pm1)$&$1(1)$
\\
$31/2$ & $1/2$ 
& $6(-\frac{15}{4}),15(-\frac{7}{4}),15(\frac{1}{4}),20(\frac{9}{4})$
&$1(-\frac{3}{4})$ 
& $15(\pm3),20(\pm1)$&$1(-1)$
\\
$31/2$ & $1/2$ 
& $4(-\frac{15}{4}),18(-\frac{7}{4}),18(\frac{1}{4}),13(\frac{9}{4}),3(\frac{17}{4})$
&$1(-\frac{3}{4})$
& $1(\pm5),12(\pm3),22(\pm1)$&$1(-1)$
\\
$31/2$ & $1/2$ 
& $3(-\frac{17}{4}),13(-\frac{9}{4}),18(-\frac{1}{4}),18(\frac{7}{4}),4(\frac{15}{4})$
&$1(\frac{3}{4})$ 
& $1(\pm5),12(\pm3),22(\pm1)$&$1(1)$ 
\\
$16$ & $0$ 
& $4(\pm4),14(\pm2),20(0)$
&$1(0)$
& $6(\pm4),16(\pm2),26(0)$&$1(0)$ \\
\hline
\end{tabular}
\end{center}
\end{table}

\begin{table}[htbp]
\begin{center}
\caption{Quasiparticle charge assignments for $Sp(8)_1$.}\label{tab:Sp8ChargeAssignment}
\begin{tabular}{ c|c|c|c|c }
$\nu_{Sp(8)_1}$&$\#Q_\sigma$&$\#Q_\mathsf{E}$&$\#Q_\tau$&$\#Q_S$\\\hline
16&$4(0),2(\pm2)$&$9(0),8(\pm2),1(\pm4)$&$16(0),12(\pm2),4(\pm4)$&$16(0),8(\pm2),5(\pm4)$\\\hline
\end{tabular}
\end{center}
\end{table}

\begin{table}[htbp]
\begin{center}
\caption{Current operator charge assignments for the conformal embeddings $SO(2r+1) \times SO(15 - 2r) \subset E_8$ for $r = 1, 2, 3$ and $\text{Ising} \times SO(15) \subset E_8$.} 
\label{tab:SOoddChargeAssignmentRoot}
\begin{tabular}{ c|c|c|c }
$\nu_{SO(7)}$ & $\nu_{SO(9)}$ & $\#Q_{SO(7)}$ & $\#Q_{SO(9)}$ \\
\hline
$0$ & $16$ 
&$21(0)$
&$22(0),7(\pm4)$
\\
$0$ & $16$ 
&$21(0)$
&$16(0),4(\pm2),6(\pm4)$
\\
$4$ & $12$ 
&$11(0), 5(\pm2)$
&$12(0), 9(\pm2),3(\pm4)$
\\
$8$ & $8$ 
& $7(0),6(\pm2),1(\pm4)$ 
& $14(0),10(\pm2),1(\pm4)$ 
\\
$12$ & $4$ 
& $9(0),3(\pm2),3(\pm4)$
& $22(0),7(\pm2)$
\\
$16$ & $0$ 
& $11(0), 5(\pm4)$
& $36(0)$
\\
\hline
$\nu_{SO(5)}$ & $\nu_{SO(11)}$ & $\#Q_{SO(5)}$ & $\#Q_{SO(11)}$ \\
\hline
$0$ & $16$ 
&$10(0)$
&$37(0),9(\pm4)$
\\
$0$ & $16$ 
&$10(0)$
&$19(0),12(\pm2),6(\pm4)$
\\
$4$ & $12$ 
&$4(0),3(\pm2)$
&$19(0), 15(\pm2),3(\pm4)$
\\
$8$ & $8$ 
&$4(0),2(\pm2),1(\pm4)$ 
&$25(0), 14(\pm2),1(\pm4)$
\\
$16$ & $0$ 
&$4(0),3(\pm4)$ 
&$55(0)$
\\
\hline
$\nu_{SO(3)}$ & $\nu_{SO(13)}$ & $\#Q_{SO(3)}$ & $\#Q_{SO(13)}$ \\
\hline
$0$ & $16$
&$3(0)$
&$56(0),11(\pm4)$
\\
$0$ & $16$
&$3(0)$
&$26(0),20(\pm2),6(\pm4)$
\\
$4$ & $12$
&$1(0),1(\pm4)$
&$30(0),21(\pm2),3(\pm4)$
\\
$16$ & $0$
&$1(0),1(\pm8)$
&$78(0)$
\\
\hline
$\nu_{\rm Ising}$ & $\nu_{SO(15)}$ & $\#Q_{\text{Ising}}$ & $\#Q_{SO(15)}$ \\
\hline
$0$ & $16$
&N/A
&$79(0),13(\pm4)$
\\
$0$ & $16$
&N/A
&$37(0),28(\pm2),6(\pm4)$ \\
\hline
\end{tabular}
\end{center}
\end{table}

\begin{table}[htbp]
\begin{center}
\caption{Quasiparticle charge assignments for the conformal embeddings $SO(2r+1) \times SO(15 - 2r) \subset E_8$ for $r = 1, 2, 3$ and $\text{Ising} \times SO(15) \subset E_8$.
Below, $f$ and $\sigma$ denote the nontrivial super-selection sectors of the $SO(2r+1)$ and $SO(15 - 2r)$ topological phases.}
\label{tab:SOoddChargeAssignmentPrimary}
\begin{tabular}{ c|c|c|c|c|c }
$\nu_{SO(7)}$ & $\nu_{SO(9)}$ & $\#(Q_f)_{SO(7)}$ & $\#(Q_f)_{SO(9)}$ & $\#(Q_\sigma)_{SO(7)}$ & $\#(Q_\sigma)_{SO(9)}$  \\
\hline
$0$ & $16$ 
&$7(0)$&$7(0),1(\pm4)$
& $8(0)$&$8(\pm2)$
 \\
$0$ & $16$ 
&$7(0)$&$1(0),4(\pm2)$
& $8(0)$&$6(0),4(\pm2),1(\pm4)$
 \\
$4$ & $12$ 
&$5(0),1(\pm2)$&$3(0),3(\pm2)$
& $4(\pm1)$&$6(\pm1),2(\pm3)$  
\\
$8$ & $8$ 
&$3(0),2(\pm2)$&$5(0),2(\pm2)$
&$4(0),2(\pm2)$&$ 8(0),4(\pm2)$
\\
$12$ & $4$ 
& $1(0),3(\pm2)$ & $7(0),1(\pm2)$
& $3(\pm1),1(\pm3)$ & $8(\pm1)$
\\
$16$ & $0$
&$5(0),1(\pm4)$&$9(0)$ 
&$4(\pm2)$&$16(0)$
\\
\hline
$\nu_{SO(5)}$ & $\nu_{SO(11)}$ & $\#(Q_f)_{SO(5)}$ & $\#(Q_f)_{SO(11)}$ & $\#(Q_\sigma)_{SO(5)}$ & $\#(Q_\sigma)_{SO(11)}$  \\
\hline
$0$ & $16$ 
&$5(0)$&$9(0),1(\pm4)$
&$4(0)$&$16(\pm2)$
\\
$0$ & $16$ 
&$5(0)$&$3(0),4(\pm2)$
&$4(0)$&$12(0),8(\pm2),2(\pm4)$ 
\\
$4$ & $12$ 
&$3(0),1(\pm2)$&$5(0),3(\pm2)$
&$2(\pm1)$&$12(\pm1),4(\pm3)$ 
\\
$8$ & $8$ 
&$1(0),2(\pm2)$&$7(0),2(\pm2)$
&$2(0),1(\pm2)$&$16(0),8(\pm2)$
\\
$16$ & $0$ 
&$3(0),1(\pm4)$&$11(0)$
& $2(\pm2)$&$32(0)$
\\
\hline
$\nu_{SO(3)}$ & $\nu_{SO(13)}$ & $\#(Q_f)_{SO(3)}$ & $\#(Q_f)_{SO(13)}$ & $\#(Q_\sigma)_{SO(3)}$ & $\#(Q_\sigma)_{SO(13)}$  \\
\hline
$0$ & $16$
&$3(0)$&$11(0),1(\pm4)$
&$2(0)$&$32(\pm2),$
\\
$0$ & $16$
&$3(0)$&$5(0),4(\pm2)$
&$2(0)$&$24(0),16(\pm2),4(\pm4)$ 
\\
$4$ & $12$
&$1(0),1(\pm2)$&$7(0),3(\pm2)$
&$1(\pm1)$&$24(\pm1),8(\pm3)$ 
\\
$16$ & $0$
&$1(0),1(\pm4)$&$13(0)$
&$1(\pm2)$&$64(0)$ 
\\
\hline
$\nu_{\rm Ising}$ & $\nu_{SO(15)}$ & $\#(Q_f)_{\rm Ising}$ & $\#(Q_f)_{SO(15)}$ & $\#(Q_\sigma)_{\rm Ising}$ & $\#(Q_\sigma)_{SO(15)}$  \\
\hline
$0$ & $16$
&$1(0)$&$13(0),1(\pm4)$
&$1(0)$&$64(\pm2),$
\\
$0$ & $16$
&$1(0)$&$7(0),4(\pm2)$
&$1(0)$&$48(0),32(\pm2),8(\pm4)$ \\
\hline
\end{tabular}
\end{center}
\end{table}

\begin{table}[htbp]
\begin{center}
\caption{Current operator charge assignments for the conformal embeddings $SO(2r) \times SO(16 - 2r) \subset E_8$ for $r = 4, 3, 2,1$. }\label{tab:SOevenChargeAssignmentRoot}
\begin{tabular}{ c|c|c|c}
$\nu_{SO(8)}$ & $\nu_{SO(8)}$ & $\#(Q)_{SO(8)}$ & $\#(Q)_{SO(8)}$ \\
\hline
$0$ & $16$ 
&$28(0)$  
&$16(0),6(\pm4)$ 
 \\
$4$ & $12$ 
&$16(0)$,$6(\pm2)$  
&$10(0),6(\pm2),3(\pm4)$ 
 \\
$8$ & $8$ 
&$10(0)$,$8(\pm2)$,$1(\pm4)$
&$10(0),8(\pm2),1(\pm4)$
\\
$12$ & $4$ 
&$10(0),6(\pm2),3(\pm4)$ 
&$16(0)$,$6(\pm2)$ 
\\
$16$ & $0$ 
&$16(0),6(\pm4)$  
&$28(0)$ 
\\
\hline
$\nu_{SO(6)}$ & $\nu_{SO(10)}$ & $\#(Q)_{SO(6)}$ & $\#(Q)_{SO(10)}$ \\
\hline
$0$ & $16$ 
&$15(0)$
&$29(0),8(\pm4)$  
\\
$0$ & $16$ 
&$15(0)$
&$17(0),8(\pm2),6(\pm4)$  
\\
$3$ & $13$ 
&$9(0),3(\pm2)$
&$17(0),10(\pm2),4(\pm4)$  
\\
$4$ & $12$
&$7(0),4(\pm2)$
&$15(0),12(\pm2),3(\pm4)$  
\\
$8$ & $8$ 
&$5(0),4(\pm2),1(\pm4)$
&$19(0),12(\pm2),1(\pm4)$  
\\
$11$ & $5$ 
&$5(0),3(\pm2),2(\pm4)$
&$25(0),10(\pm2)$    
\\
$12$ & $4$ 
&$9(0),3(\pm4)$
&$29(0),8(\pm2)$  
\\
$16$ & $0$ 
&$7(0),4(\pm4)$
&$45(0)$
\\
\hline
$\nu_{SO(4)}$ & $\nu_{SO(12)}$ & $\#(Q)_{SO(4)}$ & $\#(Q)_{SO(12)}$ \\
\hline
$0$ & $16$ 
&$6(0)$
&$46(0),10(\pm4)$
\\
$0$ & $16$ 
&$6(0)$
&$22(0),16(\pm2),6(\pm4)$
\\
$2$ & $14$ 
&$4(0),1(\pm2)$
&$26(0),15(\pm2),5(\pm4)$
\\
$4$ & $12$ 
&$2(0),2(\pm2)$
&$24(0),18(\pm2),3(\pm4)$
\\
$8$ & $8$ 
&$4(0),1(\pm4)$
&$32(0),16(\pm2),1(\pm4)$
\\
$10$ & $6$ 
&$2(0),1(\pm2),1(\pm4)$
&$36(0),15(\pm2)$
\\
$16$ & $0$ 
&$2(0),2(\pm4)$
&$66(0)$
\\
\hline
$\nu_{SO(2)}$ & $\nu_{SO(14)}$ & $\#(Q)_{SO(2)}$ & $\#(Q)_{SO(14)}$ \\
\hline
$0$ & $16$ 
& $2(0)$
& $67(0),12(\pm4)$
\\
$0$ & $16$ 
&$2(0)$
&$31(0),24(\pm2),6(\pm4)$  
\\
$1$ & $15$ 
&$1(\pm2)$
&$37(0),21(\pm2),6(\pm4)$ 
\\
$4$ & $12$ 
&$1(\pm4)$
&$37(0),24(\pm2),3(\pm4)$  
\\
$9$ & $7$ 
&$1(\pm6)$
&$49(0),21(\pm2)$  
\\
$16$ & $0$ 
&$1(\pm8)$
&$91(0)$  \\
\hline
\end{tabular}
\end{center}
\end{table}

\begin{table}[htbp]
\begin{center}
\scriptsize
\caption{Quasiparticle charge assignments for the conformal embeddings $SO(2r) \times SO(16 - 2r) \subset E_8$ for $r = 4, 3, 2,1$.
Below, $f$, $s+$, and $s-$ denote the nontrivial super-selection sectors of the $SO(2r)$ and $SO(16 - 2r)$ topological phases.}
\label{tab:SOevenChargeAssignmentPrimary}
\resizebox{18.2cm}{!}{$
\begin{tabular}{ c|c|c|c|c|c|c|c}
$\nu_{SO(8)}$&$\nu_{SO(8)}$&$\#(Q_f)_{SO(8)}$&$\#(Q_f)_{SO(8)}$&$\#(Q_{s+})_{SO(8)}$&$\#(Q_{s+})_{SO(8)}$ &$\#(Q_{s-})_{SO(8)}$&$\#(Q_{s-})_{SO(8)}$\\
\hline
$0$ & $16$   
& $8(0)$&$6(0),1(\pm4)$
& $8(0)$&$4(\pm2)$ 
& $8(0)$&$4(\pm2)$
 \\
$0$ & $16$   
& $8(0)$&$4(\pm2)$ 
& $8(0)$&$6(0),1(\pm4)$
& $8(0)$&$4(\pm2)$
\\
$0$ & $16$   
& $8(0)$&$4(\pm2)$ 
& $8(0)$&$4(\pm2)$
& $8(0)$&$6(0),1(\pm4)$
\\
$4$ & $12$  
& $6(0), 1(\pm2)$&$2(0), 3(\pm2)$ 
& $4(\pm1)$&$3(\pm1),1(\pm3)$
& $4(\pm1)$&$3(\pm1),1(\pm3)$
 \\
$4$ & $12$  
& $4(\pm1)$&$3(\pm1),1(\pm3)$ 
& $6(0),1(\pm2)$&$2(0),3(\pm2)$
& $4(\pm1)$&$3(\pm1),1(\pm3)$
 \\
$4$ & $12$  
& $4(\pm1)$&$3(\pm1),1(\pm3)$ 
& $4(\pm1)$&$3(\pm1),1(\pm3)$
& $6(0),1(\pm2)$&$2(0),3(\pm2)$
 \\
$8$ & $8$ 
& $4(0),2(\pm2)$&$4(0),2(\pm2)$
&  $4(0), 2(\pm2)$&$4(0), 2(\pm2)$
&  $4(0), 2(\pm2)$&$4(0), 2(\pm2)$
\\
$12$ & $4$ 
&  $2(0),3(\pm2)$&$6(0),1(\pm2)$
&  $3(\pm1),1(\pm3)$&$4(\pm1)$
&  $3(\pm1),1(\pm3)$&$4(\pm1)$
 \\
$12$ & $4$  
&$3(\pm1),1(\pm3)$& $4(\pm1)$
&$2(0),3(\pm2)$&$6(0),1(\pm2)$
&$3(\pm1),1(\pm3)$&$4(\pm1)$
\\
$12$ & $4$ 
& $3(\pm1),1(\pm3)$&$4(\pm1)$
& $3(\pm1),1(\pm3)$&$4(\pm1)$
& $2(0),3(\pm2)$&$6(0),1(\pm2)$
\\
$16$ & $0$   
&$6(0),1(\pm4)$& $8(0)$
&$4(\pm2)$& $8(0)$
&$4(\pm2)$& $8(0)$
 \\
$16$ & $0$   
& $4(\pm2)$&$8(0)$
& $6(0),1(\pm4)$&$8(0)$
& $4(\pm2)$&$8(0)$
 \\
$16$ & $0$   
& $4(\pm2)$&$8(0)$
& $4(\pm2)$&$8(0)$
& $6(0),1(\pm4)$&$8(0)$
 \\
\hline
$\nu_{SO(6)}$&$\nu_{SO(10)}$&$\#(Q_f)_{SO(6)}$&$\#(Q_f)_{SO(10)}$&$\#(Q_{s+})_{SO(6)}$&$\#(Q_{s+})_{SO(10)}$ &$\#(Q_{s-})_{SO(6)}$&$\#(Q_{s-})_{SO(10)}$\\
\hline
$0$ & $16$ 
&  $6(0)$&$8(0),1(\pm4)$
&  $4(0)$&$8(\pm2)$
&  $4(0)$&$8(\pm2)$
\\
$0$ & $16$ 
&  $6(0)$&$2(0),4(\pm2)$
&  $4(0)$&$6(0),4(\pm2),1(\pm4)$
&  $4(0)$&$6(0),4(\pm2),1(\pm4)$
\\
$3$ & $13$ 
&  $3(\pm1)$&$4(\pm1),1(\pm3)$
& $1(-\frac{3}{2}),3(\frac{1}{2})$ &$4(-\frac{5}{2}),5(-\frac{1}{2}),6(\frac{3}{2}),1(\frac{7}{2})$
&$3(-\frac{1}{2}),1(\frac{3}{2})$ &$1(-\frac{7}{2}),6(-\frac{3}{2}),5(\frac{1}{2}),4(\frac{5}{2})$
\\
$3$ & $13$ 
&  $3(\pm1)$&$4(\pm1),1(\pm3)$
& $3(-\frac{1}{2}),1(\frac{3}{2})$
&$1(-\frac{7}{2}),6(-\frac{3}{2}),5(\frac{1}{2}),4(\frac{5}{2})$
&$1(-\frac{3}{2}),3(\frac{1}{2})$ &$4(-\frac{5}{2}),5(-\frac{1}{2}),6(\frac{3}{2}),1(\frac{7}{2})$
\\
$4$ & $12$
&  $4(0),1(\pm2)$&$4(0),3(\pm2)$
&  $2(\pm1)$&$6(\pm1),2(\pm3)$
&  $2(\pm1)$&$6(\pm1),2(\pm3)$
\\
$8$ & $8$ 
&  $2(0),2(\pm2)$&$6(0),2(\pm2)$
&  $2(0),1(\pm2)$&$8(0),4(\pm2)$
&  $2(0),1(\pm2)$&$8(0),4(\pm2)$
\\
$11$ & $5$ 
&  $2(\pm1),1(\pm3)$&$5(\pm1)$
&  $1(-\frac{5}{2}),1(-\frac{1}{2}),2(\frac{3}{2})$
&$5(-\frac{3}{2}),10(\frac{1}{2}),1(\frac{5}{2})$
&  $2(-\frac{3}{2}),1(\frac{1}{2}),1(\frac{5}{2})$
&$1(-\frac{5}{2}),10(-\frac{1}{2}),5(\frac{3}{2})$
\\
$11$ & $5$ 
&  $2(\pm1),1(\pm3)$&$5(\pm1)$
&  $2(-\frac{3}{2}),1(\frac{1}{2}),1(\frac{5}{2})$&$1(-\frac{5}{2}),10(-\frac{1}{2}),5(\frac{3}{2})$
&  $1(-\frac{5}{2}),1(-\frac{1}{2}),2(\frac{3}{2})$&$5(-\frac{3}{2}),10(\frac{1}{2}),1(\frac{5}{2})$
\\
$12$ & $4$ 
&  $3(\pm2)$&$8(0),1(\pm2)$
&  $1(-3),3(1)$&$8(\pm1)$
&  $3(-1),1(3)$&$8(\pm1)$
\\
$12$ & $4$ 
&  $3(\pm2)$&$8(0),1(\pm2)$
&  $3(-1),1(3)$&$8(\pm1)$
&  $1(-3),3(1)$&$8(\pm1)$
\\
$16$ & $0$ 
&  $4(0),1(\pm4)$&$10(0)$
&  $2(\pm2)$&$16(0)$
&  $2(\pm2)$&$16(0)$
\\
\hline
$\nu_{SO(4)}$&$\nu_{SO(12)}$&$\#(Q_f)_{SO(4)}$&$\#(Q_f)_{SO(12)}$&$\#(Q_{s+})_{SO(4)}$&$\#(Q_{s+})_{SO(12)}$ &$\#(Q_{s-})_{SO(4)}$&$\#(Q_{s-})_{SO(12)}$\\
\hline
$0$ & $16$ 
&  $4(0)$&$10(0),1(\pm4)$
&  $2(0)$&$16(\pm2)$
&  $2(0)$&$16(\pm2)$
\\
$0$ & $16$ 
&  $4(0)$&$4(0),4(\pm2)$
&  $2(0)$&$12(0),8(\pm2),2(\pm4)$
&  $2(0)$&$12(0),8(\pm2),2(\pm4)$
\\
$2$ & $14$ 
&  $2(\pm1)$&$5(\pm1),1(\pm3)$
&  $1(\pm1)$&$11(\pm1),5(\pm3)$
&  $2(0)$&$10(0),10(\pm2),1(\pm4)$
\\
$2$ & $14$ 
&  $2(\pm1)$&$5(\pm1),1(\pm3)$
&  $2(0)$&$10(0),10(\pm2),1(\pm4)$
&  $1(\pm1)$&$11(\pm1),5(\pm3)$
\\
$4$ & $12$ 
&  $2(0),1(\pm2)$&$6(0),3(\pm2)$
&  $1(\pm1)$&$12(\pm1),4(\pm3)$
&  $1(\pm1)$&$12(\pm1),4(\pm3)$
\\
$8$ & $8$ 
&  $2(\pm2)$&$8(0),2(\pm2)$
&  $1(\pm2)$&$16(0),8(\pm2)$
&  $2(0)$&$16(0),8(\pm2)$
\\
$8$ & $8$ 
&  $2(\pm2)$&$8(0),2(\pm2)$
&  $2(0)$&$16(0),8(\pm2)$
&  $1(\pm2)$&$16(0),8(\pm2)$
\\
$10$ & $6$ 
&  $1(\pm1),1(\pm3)$&$6(\pm1)$
&  $1(\pm2)$&$20(0),6(\pm2)$ 
&  $1(\pm1)$&$15(\pm1),1(\pm3)$
\\
$10$ & $6$ 
&  $1(\pm1),1(\pm3)$&$6(\pm1)$
&  $1(\pm1)$&$15(\pm1),1(\pm3)$
&  $1(\pm2)$&$20(0),6(\pm2)$ 
\\
$16$ & $0$ 
&  $2(0),1(\pm4)$&$12(0)$
&  $1(\pm2)$&$32(0)$
&  $1(\pm2)$&$32(0)$
\\
\hline
$\nu_{SO(2)}$&$\nu_{SO(14)}$&$\#(Q_f)_{SO(2)}$&$\#(Q_f)_{SO(14)}$&$\#(Q_{s+})_{SO(2)}$&$\#(Q_{s+})_{SO(14)}$ &$\#(Q_{s-})_{SO(2)}$&$\#(Q_{s-})_{SO(14)}$\\
\hline
$0$ & $16$ 
&  $2(0)$&$12(0),1(\pm4)$
&  $1(0)$&$32(\pm2)$
&  $1(0)$&$32(\pm2)$
\\
$0$ & $16$ 
&  $2(0)$&$6(0),4(\pm2)$
&  $1(0)$&$24(0),16(\pm2),4(\pm4)$
&  $1(0)$&$24(0),16(\pm2),4(\pm4)$
\\
$1$ & $15$ 
&  $1(\pm1)$&$6(\pm1),1(\pm3)$
&  $1(-\frac{1}{2})$&$6(-\frac{7}{2}),21(-\frac{3}{2}),21(\frac{1}{2}),15(\frac{5}{2}),1(\frac{9}{2})$
&  $1(\frac{1}{2})$&$1(-\frac{9}{2}),15(-\frac{5}{2}),21(-\frac{1}{2}),21(\frac{3}{2}),6(\frac{7}{2})$
\\
$1$ & $15$ 
&  $1(\pm1)$&$6(\pm1),1(\pm3)$
&  $1(\frac{1}{2})$&$1(-\frac{9}{2}),15(-\frac{5}{2}),21(-\frac{1}{2}),21(\frac{3}{2}),6(\frac{7}{2})$
&  $1(-\frac{1}{2})$&$6(-\frac{7}{2}),21(-\frac{3}{2}),21(\frac{1}{2}),15(\frac{5}{2}),1(\frac{9}{2})$
\\
$4$ & $12$ 
&  $1(\pm2)$&$8(0),3(\pm2)$
&  $1(-1)$&$24(\pm1),8(\pm3)$
&  $1(1)$&$24(\pm1),8(\pm3)$
\\
$4$ & $12$ 
&  $1(\pm2)$&$8(0),3(\pm2)$
&  $1(1)$&$24(\pm1),8(\pm3)$
&  $1(-1)$&$24(\pm1),8(\pm3)$
\\
$9$ & $7$ 
&  $1(\pm3)$&$7(\pm1)$
&  $1(-\frac{3}{2})$&$7(-\frac{5}{2}),35(-\frac{1}{2}),21(\frac{3}{2}),1(\frac{7}{2})$
&  $1(\frac{3}{2})$&$1(-\frac{7}{2}),21(-\frac{3}{2}),35(\frac{1}{2}),7(\frac{5}{2})$
\\
$9$ & $7$ 
&  $1(\pm3)$&$7(\pm1)$
&  $1(\frac{3}{2})$&$1(-\frac{7}{2}),21(-\frac{3}{2}),35(\frac{1}{2}),7(\frac{5}{2})$
&  $1(-\frac{3}{2})$&$7(-\frac{5}{2}),35(-\frac{1}{2}),21(\frac{3}{2}),1(\frac{7}{2})$
\\
$16$ & $0$ 
&  $1(\pm4)$&$14(0)$
&  $1(-2)$&$64(0)$
&  $1(2)$&$64(0)$
\\
$16$ & $0$ 
&  $1(\pm4)$&$14(0)$
&  $1(2)$&$64(0)$
&  $1(-2)$&$64(0)$ \\
\hline
\end{tabular}
$}
\end{center}
\end{table}
\end{widetext}

\end{document}